\documentclass{aa}  

\usepackage{graphicx}
\usepackage{txfonts}
\usepackage[usenames,dvipsnames]{xcolor}
\usepackage{adjustbox}
\usepackage{hyperref}
\usepackage{natbib}
\usepackage{makecell}
\usepackage{gensymb}
\usepackage{subcaption}
\usepackage{multirow}
\usepackage{pdflscape}
\usepackage{pgffor}
\usepackage{multicol}
\usepackage{lscape}
\usepackage{placeins}
\usepackage{rotating, threeparttable}
\usepackage{array}
\usepackage{float}
\usepackage[morefloats=500]{morefloats}
\usepackage{placeins}
\usepackage[normalem]{ulem}

\definecolor{Blue}{rgb}{0,0,1}
\hypersetup{colorlinks, citecolor=Blue, linkcolor=Blue, urlcolor=Blue}

\DeclareUnicodeCharacter{2212}{-}

\defcitealias{Unnikrishnan_et_al_2024}{UK24}

\begin{document} 

   \title{Charting circumstellar chemistry of carbon-rich \\asymptotic giant branch stars}

   \subtitle{II. Abundances and spatial distributions of CS}

   \author{R. Unnikrishnan
          \inst{1}
          \and
          M. Andriantsaralaza\inst{1,2}
          \and
          E. De Beck\inst{1}
          \and
          L.-\AA. Nyman\inst{1,3,4}
          \and
          H. Olofsson\inst{1}
          \and
          W. H. T. Vlemmings\inst{1}
          \and
          M. Maercker\inst{1}
          \and
          M. Van de Sande\inst{5}
          \and
          T. Danilovich\inst{6,7}
          \and
          T. J. Millar\inst{8}
          \and
          S. B. Charnley\inst{9}
          \and
          M. G. Rawlings\inst{10}
          }

   \institute{Department of Space, Earth and Environment, Chalmers University of Technology, {SE-412 96 Gothenburg}, Sweden\\\email{ramlal.unnikrishnan@chalmers.se}
      \and
      Theoretical Astrophysics, Division for Astronomy and Space Physics, Department of Physics and Astronomy, Uppsala University, Box 516, 751 20 Uppsala, Sweden
      \and
      Joint ALMA Observatory (JAO), Alonso de Córdova 3107, Vitacura 763-0355, Casilla 19001, Santiago, Chile
      \and
      European Southern Observatory (ESO), Alonso de Córdova 3107, Vitacura 763-0355, Santiago, Chile
      \and
      Leiden Observatory, Leiden University, PO Box 9513, NL-2300 RA Leiden, The Netherlands
      \and
      School of Physics \& Astronomy, Monash University, Wellington Road, Clayton 3800, Victoria, Australia
      \and
      Institute of Astronomy, KU Leuven, Celestijnenlaan 200D, 3001 Leuven, Belgium
      \and
      Astrophysics Research Centre, School of Mathematics and Physics, Queen’s University Belfast, University Road, Belfast, BT7 1NN, UK
      \and
      NASA Goddard Space Flight Center, 8800 Greenbelt Road, Greenbelt, MD 20771, USA
      \and
      Gemini Observatory / NSF’s NOIRLab, 670 N. A’ohoku Place, Hilo, Hawai’i, 96720, USA
      }

   \date{Received 1 April 2025 / Accepted 16 May 2025}


  \abstract
   {The circumstellar envelopes (CSEs) of asymptotic giant branch (AGB) stars harbour a rich variety of molecules and are sites of complex chemistry. Our current understanding of the circumstellar chemical processes of carbon-rich AGB stars is predominantly based on observations of a single star, IRC~+10\,216, often regarded as an archetypical carbon star.}
   {We aim to estimate stellar and circumstellar properties for five carbon stars, and constrain their circumstellar CS abundances. This study compares the CS abundances among the sources, informs circumstellar chemical models, and helps to assess if IRC+10~216 is a good representative of the physics and chemistry of carbon star CSEs.}
   {We modelled the spectral energy distributions (SEDs) and CO line emission to derive the stellar and outflow properties. Using these, we then retrieved CS abundance profiles with detailed radiative transfer modelling, imposing spatial and excitation constraints from ALMA and single-dish observations.}
   {We obtain good fits to the SEDs and CO lines for all sources and reproduce the CS line emission across various transitions and apertures, yielding robust estimates of the CS abundance profiles. Peak CS fractional abundances range from 1$\times$10$^{-6}$ - 4$\times$10$^{-6}$, with e-folding radii of 1.8$\times$10$^{16}$ - 6.8$\times$10$^{16}$ cm. We also derive reliable $^{12}$C/$^{13}$C and $^{32}$S/$^{34}$S ratios from CS isotopologue modelling.}
  {Our results refine previous single-dish CS abundance estimates and improve the relative uncertainty on the CS e-folding radius for IRAS 07454$-$7112 by a factor of $\sim$2.5. Chemical models reproduce our estimates of the CS radial extent, corroborating the CS photodissociation framework used therein. We find no significant differences between the derived CS abundance profiles for IRC~$+$10$\,$216 and the rest of the sample, apart from the expected density-driven variations.}


   \titlerunning{CS abundances in carbon stars}
   \authorrunning{R. Unnikrishnan et al.}

   \maketitle
%
\section{Introduction}
\label{sec:introduction}
After the main-sequence and the first red-giant phase of their evolution, low- to intermediate-mass stars ($\sim$0.8$M_\odot$ < $M_\mathrm{*}$ < 8$M_\odot$) become cool, luminous giants \citep[$T_\mathrm{eff} \lesssim3000$\,K, $L_{*}\sim10^{3}-10^{4}\,L_\odot$, $R_\mathrm{*} \sim 100 R_\odot$;][]{Herwig_2005} known as asymptotic giant branch (AGB) stars. These stars are characterised by dust-driven stellar winds \citep{Habing_and_Olofsson_2004, Hofner_and_Olofsson_2018} that cause extensive mass loss \citep[$10^{-8}$ {to} $10^{-5}\,M_\odot\,\mathrm{yr}^{-1}$;][]{Olofsson_1999, Hofner_and_Olofsson_2018}, forming expanding circumstellar envelopes (CSEs) of dust and gas around them. The CSEs harbour a large number of molecular species \citep[e.g.][]{Cernicharo_et_al_2000, Patel_et_al_2011, Unnikrishnan_et_al_2024, Decin_2021} and are also sites of dust formation \citep{Habing_and_Olofsson_2004}. AGB stars contribute significantly to the chemical enrichment of the interstellar medium (ISM) and thereby to overall galactic chemical evolution \citep[e.g.][]{Tielens_2005, Matsuura_et_al_2009, Kobayashi_et_al_2011}.

AGB stars with a photospheric carbon-to-oxygen abundance ratio (C/O) of greater than 1 are termed C-type (carbon-rich) stars, whereas the ones with C/O $<$ 1 are called M-type (oxygen-rich). The CSEs of C-type AGB stars exhibit richer chemistry compared to M-type stars, featuring a wide variety of molecular species, including long carbon chains \citep[][]{Olofsson_1993, Cernicharo_et_al_2000, Gong_et_al_2015, Woods_et_al_2003, Unnikrishnan_et_al_2024} and likely also polycyclic aromatic hydrocarbons \citep[PAHs, e.g.][]{Cherchneff_2012, Tielens_2008, Zeichner_et_al_2023, Anand_et_al_2023}. More than 100 molecular species have been detected so far in carbon star CSEs \citep{McGuire_2022, Cernicharo_et_al_2023b, Cernicharo_et_al_2023a, Tuo_et_al_2024}.

Some of these species, such as CO, C$_2$H$_2$, HCN, CS, SiO, and SiS, are of photospheric origin, influenced by non-equilibrium shock chemistry in the extended atmosphere \citep[e.g.][]{Cherchneff_2006}, and are injected into the circumstellar gas from the warm, dense inner regions. They are commonly referred to as parent molecules \citep[e.g.][]{Woods_et_al_2003, Agundez_et_al_2020}, in contrast to the daughter species \citep[e.g. CN, HNC, HC$_3$N, see][]{Agundez_et_al_2020} that are formed farther out in the CSE, mostly by photodissociation of the parent species. The complex interplay of various processes including gas-phase reactions, dust-gas interactions, and photodissociation determines the radial variation in molecular abundances across AGB CSEs. Therefore, observations of molecular species and dust at multiple wavelengths are needed to obtain a comprehensive picture of the physical conditions and chemical processes influencing circumstellar molecular abundances.

Most observational studies of circumstellar carbon chemistry so far, the interferometric ones in particular, have almost exclusively focussed on the nearby ($\sim$190 pc), high mass-loss rate (1.5$\times$10$^{-5}$ $M_\odot$ yr$^{-1}$, MLR), C-type AGB star IRC~+10~216, which is generally regarded as an archetype carbon star and has reasonably well-determined molecular abundances \citep[e.g.][]{Agundez_et_al_2012, Agundez_et_al_2017, Cernicharo_et_al_2000, Agundez_et_al_2015, Pardo_et_al_2022feb, Siebert_et_al_2022, Velilla-Prieto_et_al_2015}. Abundances derived from single-dish (SD) line observations are, however, available for a few other carbon stars as well \citep[e.g.][]{Woods_et_al_2003, Schoier_et_al_2006, Schoier_et_al_2007, Schoier_et_al_2013, Danilovich_et_al_2018, Massalkhi_et_al_2024}. In contrast to SD observations, spatially resolved interferometric maps of circumstellar line emission \citep[e.g.][]{Agundez_et_al_2017, Danilovich_et_al_2019, Velilla-Prieto_et_al_2019} offer the possibility to directly determine the spatial distribution and radial extent of the molecular gas. The first spatially resolved spectral survey of a set of carbon stars other than IRC~+10~216 was presented by \citet[][hereafter \citetalias{Unnikrishnan_et_al_2024}]{Unnikrishnan_et_al_2024}, aimed at observationally testing the archetypal status often attributed to IRC~+10~216. This study revealed similar complexity in both morphology and chemistry in the observed sources as was previously found towards IRC~+10~216 \citep[e.g.][]{Agundez_et_al_2017}. 

\citetalias{Unnikrishnan_et_al_2024} derived molecular abundances assuming local thermodynamic equilibrium (LTE) conditions in the circumstellar gas. Though such LTE calculations are regularly used to obtain reasonable first-order abundance estimates \citep[e.g.][]{Smith_et_al_2015, Pardo_et_al_2022feb, Velilla-Prieto_et_al_2015_OH_231.8+4.2}, non-LTE radiative transfer (RT) modelling is required to calculate precise radial abundance profiles \citep[e.g.][]{Agundez_et_al_2012, Brunner_et_al_2018, Van_de_Sande_et_al_2018_R_Dor, Danilovich_et_al_2018, Danilovich_et_al_2019, Massalkhi_et_al_2024}. Most of the existing RT models of molecular line emission from carbon star CSEs are limited to either sampling a range of excitation conditions using SD spectra \citep[e.g.][]{Woods_et_al_2003, Schoier_et_al_2007, Schoier_et_al_2013, Agundez_et_al_2012, Danilovich_et_al_2018, Massalkhi_et_al_2019, Massalkhi_et_al_2024} or focussing on the spatial distribution of the emission using interferometric observations of a few selected lines \citep[e.g.][]{Schoier_et_al_2006, Agundez_et_al_2017, Velilla-Prieto_et_al_2019}. RT modelling with strict constraints based on the combination of the above, i.e. both excitation information from a broad range of SD data, and morphological information from spatially resolved images, as presented in this work, is the necessary next step in advancing the modelling of circumstellar line emission to estimate precise molecular abundances, as has been noted by several authors including \citet{Danilovich_et_al_2018} and \citet{Massalkhi_et_al_2024}. This has been done for both M-type stars \citep[e.g.][]{Danilovich_et_al_2019} and S-type stars \citep[e.g.][]{Brunner_et_al_2018}, yielding good estimates of fractional abundances of several molecular species including CS.

Among the parent species, CS is easily formed in carbon-star envelopes under thermodynamical equilibrium conditions due to the availability of large amounts of C \citep[e.g.][]{Agundez_et_al_2020, Danilovich_et_al_2018}, while SiS and SiO formation may require shock-induced chemistry \citep[e.g.][]{Cherchneff_2012}. Being refractory species, they are expected to play a major role in circumstellar dust formation and growth as well \citep[see][and references therein]{Gail_et_al_2013}. Factors like these affect the distribution of such species in the CSE, and can complicate their RT modelling, making it difficult to obtain simultaneous good fits to a large number of observed lines \citep[e.g.][]{Velilla-Prieto_et_al_2019, Schoier_et_al_2007}. Therefore, the modelling of such species, particularly for the high-MLR C-type stars, demands a more detailed analysis than in the case of CS, including the use of non-Gaussian abundance profiles, especially when using both spatial and excitation constraints together. In the current work, we focus on CS and its isotopologues, and will present the modelling of the remaining parent species in a forthcoming paper. We present, for the first time, CS RT models constrained using spatially resolved observations for carbon stars other than IRC~+10~216.

This work is the second in a series of planned publications, starting with \citetalias{Unnikrishnan_et_al_2024}, that aim to leverage the high-angular-resolution of ALMA to expand spatially resolved spectral line observations to other carbon stars beyond the well-studied IRC~+10~216. The paper is organised as follows. The source sample is presented in Sect.~\ref{sec:The_Sources}. The observations used in this work are described in Sect.~\ref{sec:Observations}. Sect.~\ref{sec:Radiative_Transfer_Modelling} explains the RT modelling method employed, with Sect.~\ref{subsec:Envelope_model} describing the adopted envelope model, and the dust, CO, and CS modelling procedures detailed in sections~\ref{subsec:Dust_Modelling}, \ref{subsec:CO_Modelling}, and \ref{subsec:CS_Modelling}, respectively. We present the results in Sect.~\ref{sec:Results}, and compare and discuss them in light of existing literature and chemical models in Sect.~\ref{sec:Discussion}.

\section{The sources}
\label{sec:The_Sources}
Our carbon star sample includes IRAS 15194$-$5115 (II Lup), IRAS 15082$-$4808 (V358 Lup), IRAS 07454$-$7112 (AI Vol), AFGL 3068 (LL Peg), and IRC~+10\,216 (CW Leo). The reasoning behind the source selection and the properties of the Mira variables IRAS 15194$-$5115, IRAS 15082$-$4808, and IRAS 07454$-$7112 were already discussed by \citetalias{Unnikrishnan_et_al_2024}. In the current work, we also include the extreme carbon star AFGL 3068 along with the archetypical carbon star IRC~+10\,216, to enable direct comparisons with the rest of the sample. The basic parameters of the five stars are summarised in Table~\ref{tab:source_properties}, and the two additions to the original sample are briefly described below.

\begin{table*}[t]
   \caption{Source properties}
   \label{tab:source_properties}
   \centering
      \begin{tabular}{l l r r c c c c}
      \hline\hline & \\[-2ex]
      Source & \makecell{Common\\name} & RA$^{(a)}$ & DEC$^{(b)}$ & \makecell{Distance$^{(c)}$\\(pc)} &  \makecell{$\varv_\mathrm{sys}$\\(km s$^{-1}$)} & \makecell{$\varv_\mathrm{exp}$\\(km s$^{-1}$)} & \makecell{Period$^{(d)}$\\(days)}\\
      \hline & \\[-2ex]
      IRAS 15194$-$5115 & II Lup   & 15:23:05.07 & $-$51:25:58.76 & 696 (+129/$-$93) & $-$15.0 & 21.5 & 576 \\
      IRAS 15082$-$4808 & V358 Lup & 15:11:41.44 & $-$48:19:58.95 & 1050 ($\pm$60)   & $-$3.3  & 19.5 & 632 \\
      IRAS 07454$-$7112 & AI Vol   & 07:45:02.41 & $-$71:19:45.86 & 583 (+70/$-$56)  & $-$38.7 & 13.0 & 511 \\
      AFGL 3068         & LL Peg   & 23:19:12.60 & +17:11:33.13   & 1220 (+70/$-$60) & $-$30.0 & 14.0 & 696 \\
      IRC~+10\,216      & CW Leo   & 09:47:57.40 & +13:16:43.56   & 190 ($\pm$20)    & $-$26.5 & 14.5 & 630 \\
      \hline
      \end{tabular}
      \tablefoot{$^{(a)}$J2000 h:m:s;
      $^{(b)}$J2000 d:m:s;
      $^{(c)}$\citet{Andriantsaralaza_et_al_2022}; 
      $^{(d)}$\citet{Whitelock_et_al_2006} or references therein.}
\end{table*}

AFGL 3068 is a Mira variable with a pulsation period of 696 days \citep{Whitelock_et_al_2006}. It is an extreme carbon star, expected to be at the tip of the AGB phase. It has a well-defined large-scale spiral pattern in its CSE, thought to be caused by an embedded binary companion \citep[e.g.][]{Kim_et_al_2017, Guerrero_et_al_2020}. Similar arc and shell patterns have been observed towards IRC~+10\,216 \citep{Agundez_et_al_2017, Mauron_and_Huggins_1999, Leao_et_al_2006, Trung_and_Lim_2008}, and also in molecular emission around IRAS 15194$-$5115 \citep[][\citetalias{Unnikrishnan_et_al_2024}]{Lykou_et_al_2018}. 

IRC~+10\,216 is also a Mira variable with a period of $\sim$630-678 days \citep{Menten_et_al_2012, Groenewegen_et_al_2012, Kim_et_al_2015, Pardo_et_al_2018, Dharmawardena_et_al_2019}. It is the brightest astronomical source at 5\,$\mu$m excluding solar system objects \citep{Becklin_et_al_1969}, and is one of the most observed evolved stars at mm/sub-mm/radio wavelengths \citep[see e.g.][and references therein]{Tuo_et_al_2024}. All five sources in our sample are expected to be in their high mass-loss, late AGB phase \citep{Vassiliadis_and_Wood_1993}.

\section{Observations}
\label{sec:Observations}
\subsection{Photometric fluxes}
\label{subsec:SED_Observations}
To characterise the stellar parameters and circumstellar dust properties, we need to model the stellar and dust emission contributing to the spectral energy distributions (SEDs) of our sources. We used photometric flux densities compiled from the literature to constrain our dust RT models. High-quality fluxes were obtained from Gaia DR3 \citep{Gaia_collaboration_et_al_2023} in the $B_\text{p}$, $G$, and $R_\text{p}$ bands, where available, and from the 2MASS $J$, $H$, and $K_\text{s}$ bands \citep{Cutri_et_al_2003} with quality flags ranging from 1 to 3. Additional fluxes were included from the DIRBE $L$ and $M$ bands \citep{Price_et_al_2010, Smith_et_al_2004}, the IRAS Point Source Catalogue at 12, 25, 60, and 100 $\mu$m \citep{Helou_and_Walker_1988} with quality flag 3, and from Akari at 8.6–160 $\mu$m \citep{Ishihara_et_al_2010, Yamamura_et_al_2010}, also with quality flag 3, where available. Data from the Planck catalogue \citep{Planck2011} and the SCUBA Legacy Catalogues \citep{scuba2008} were also included when available. WISE data were excluded due to well-documented saturation issues with AGB stars, which led to significant flux uncertainties \citep[e.g.][]{Lian_et_al_2014}. We note that the WISE bands $W1$ to $W4$ anyway overlap in wavelength with the DIRBE $L$ and $M$ bands, and IRAS at 12 and 25 $\mu$m.

The uncertainty in flux measurements can reflect the variability of the star, which is more pronounced at shorter wavelengths and diminishes progressively at longer wavelengths \citep[e.g.][]{Pardo_et_al_2018, Ramstedt_et_al_2008}. Based on this, we assigned uncertainties of 75\% for the optical measurements from Gaia, 60\%, 50\%, and 40\% for the 2MASS $J$, $H$, and $K_\text{s}$ band fluxes, respectively, 30\% for the $L$ and $M$ band fluxes, and 20\% for fluxes beyond 5 $\mu$m, in line with the estimates of \citet{Ramstedt_et_al_2008}. We corrected the fluxes below 3.5 $\mu$m for interstellar extinction following the method described in \citet{Andriantsaralaza_et_al_2022}. The photometric flux densities used in this work for all sources are listed in Table~\ref{tab:photometric_fluxes_SED}. 

\subsection{CO line emission}
\label{subsec:CO_Observations} 
We used archival SD CO observations, along with Atacama Compact Array (ACA) interferometric observations of CO $J = 2-1$ and $ J = 3 - 2$ lines from the DEATHSTAR project \citep{Ramstedt_et_al_2020, Andriantsaralaza_et_al_2021} for IRAS 07454$-$7112 and AFGL~3068, to constrain our gas RT models. We refer to \citet{Ramstedt_et_al_2020} and \citet{Andriantsaralaza_et_al_2021} for details of the DEATHSTAR ACA observations and data processing. The majority of the archival SD CO data used in this work were sourced from \citet{Olofsson_1993}, \citet{Schoier_and_Olofsson_2001}, 
\citet{Woods_et_al_2003}, \citet{Ramstedt_et_al_2008}, \citet{De_Beck_et_al_2010}, \citet{Danilovich_et_al_2015}, the APEX pointing catalogue\footnote{\url{https://www.apex-telescope.org/observing/pointing/spectra/}}, and the James Clerk Maxwell Telescope (JCMT) catalogue\footnote{\url{https://www.cadc-ccda.hia-iha.nrc-cnrc.gc.ca/en/jcmt/}}. While most of these observations are of the CO  $J = 1 - 0$, $J = 2 - 1$, and $J = 3 - 2$ transitions, IRAS 15194$-$5115 and IRAS 07454$-$7112 also have several higher $J$ transitions up to $J = 16 - 15$ and $J = 14 - 13$, respectively, observed using \textit{Herschel}/HIFI as part of the SUCCESS \citep{Danilovich_et_al_2015} project. The archival SD CO line intensities used in this work are listed in Table~\ref{tab:CO_int_ints_SD}.

Multiple spectra were extracted from progressively increasing apertures from the ACA cubes by convolving them to larger and larger circular beam sizes, starting at the major-axis size of the original synthesised beam. The beam size was gradually increased in steps of half the initial circular beam until the measured line flux plateaued, indicating that all detected emission is retrieved. This spectral extraction approach was employed to account for flux contributions across various spatial scales, and captures the variation in line intensities with distance from the centre, which serves as an additional constraint to the RT models. The integrated intensities of the CO spectra extracted from various apertures are listed in Table~\ref{tab:CO_int_ints_ACA}.

The measurement errors on the CO line intensities were dominated by telescope calibration uncertainties, which we assumed to be 20\% of the line intensity for SD data \citep[see e.g.][]{Schoier_et_al_2007, Danilovich_et_al_2018}. In cases with low signal-to-noise ratio, like the archival SEST $J = 3 − 2$ data, the overall uncertainty on the intensity was estimated to be around $\sim$30\%, taking into account the high noise in the spectra as well.
For the ALMA lines, we set the uncertainties to 10\%, following \citet{Ramstedt_et_al_2020}, \citet{Andriantsaralaza_et_al_2021}, and \citetalias{Unnikrishnan_et_al_2024}.

\subsection{CS line emission}
\label{subsec:CS_Observations}
We have modelled the CS, $^{13}$CS, and C$^{34}$S lines from our ALMA, APEX, and \textit{Herschel}/HIFI spectral surveys (see Sections~\ref{subsubsec:ALMA_spectral_survey},~\ref{subsubsec:APEX_spectral_survey},~\ref{subsubsec:HIFI_spectral_survey}) in this work. We could not model the C$^{33}$S emission due to the lack of line detections having good signal-to-noise ratios. In some cases, we also used supplementary spectral data obtained from the literature (see Sect.~\ref{subsubsec:Supplementary_CS_observations}). Our ALMA line survey of three C-type AGB stars (Sect.~\ref{subsubsec:ALMA_spectral_survey}) has already been presented in detail in \citetalias{Unnikrishnan_et_al_2024}, along with selected APEX molecular line observations that are also part of an unbiased SD spectral survey. This APEX survey and the HIFI spectral scans are discussed below, along with the supplementary archival observations used. The integrated intensities of the CS lines used in this work are listed in Table~\ref{tab:CS_line_intensities}, and those of the $^{13}$CS and C$^{34}$S lines are given in Tables~\ref{tab:13CS_line_intensities} and \ref{tab:C34S_line_intensities}, respectively.

\begin{table*}[ht]
   \caption{CS lines used in this work}
   \label{tab:CS_line_intensities}
   \centering
      \begin{adjustbox}{width=18cm}
      \begin{tabular}{c r c c c c c c c}
      \hline\hline & \\[-2ex]
      \makecell{Transition} & \makecell{Rest frequency\\\ [GHz]} & \makecell{E$_{up}$\\\ [K]} & \makecell{Telescope} & \multicolumn{5}{c}{\makecell{Integrated intensity [Jy km s$^{-1}$ for ALMA; K km s$^{-1}$ for others]}} \\
      \cline{5-9} & \\[-2ex]  
      & & & & 15194$-$5115 & 15082$-$4808 & 07454$-$7112 & AFGL 3068 & IRC~+10\,216 \\
      \hline & \\[-2ex]
      $1 - 0$ & 48.9909549 & 2.35 & Yebes & $-$ & $-$ & $-$ & $-$ & 49.5$^{(d)}$\\

      $2 - 1$ & 97.9809533 & 7.05 & ALMA & 281.8 & 130.3 & 58.6 & $-$ & 1695.2\\
              &            &      & OSO & $-$ & $-$ & $-$ & 3.9$^{(a)}$ & 118.9$^{(a)}$\\

      $3 - 2$ & 146.9690287 & 14.11 & IRAM & $-$ & $-$ & $-$ & $-$ & 363.8$^{(d)}$\\

      $4 - 3$ & 195.9542109 & 23.51 & APEX & 21.7 & 16.0 & 8.4$^{(b)}$ & $-$ & $-$\\
              &             &       & IRAM & $-$ & $-$ & $-$ & $-$ & 443.9$^{(d)}$\\

      $5 - 4$ & 244.9355565 & 35.27 & APEX & 32.4 & 14.6 & 10.6 & 5.9 & 208.9\\
              &             &       & IRAM & $-$ & $-$ & $-$ & $-$ & 592.2$^{(d)}$\\

      $6 - 5$ & 293.9120865 & 49.37 & APEX & 27.6 & $-$ & 12.8$^{(b)}$ & $-$ & $-$\\
              &             &       & IRAM & $-$ & $-$ & $-$ & $-$ & 882.8$^{(d)}$\\

      $7 - 6$ & 342.8828503 & 65.83 & ALMA & $-$ & $-$ & 425.0$^{(c)}$ & 248.0$^{(c)}$ & $-$\\
              &             &       & APEX & 20.3 & $-$ & $-$ & $-$ & $-$\\\\

      $10 - 9$ & 489.750921 & 129.29 & HIFI & 2.8 & $-$ & $-$ & $-$ & $-$\\
      $11 - 10$ & 538.6889972 & 155.14 & HIFI & 3.0 & $-$ & $-$ & $-$ & $-$\\
      $12 - 11$ & 587.616485 & 183.35 & HIFI & 2.8 & $-$ & $-$ & $-$ & $-$\\
      $13 - 12$ & 636.53246 & 213.89 & HIFI & 2.5 & $-$ & $-$ & $-$ & $-$\\
      $14 - 13$ & 685.4359238 & 246.79 & HIFI & 3.0 & $-$ & $-$ & $-$ & $-$\\
      $15 - 14$ & 734.325929 & 282.03 & HIFI & 2.9 & $-$ & $-$ & $-$ & $-$\\
      $16 - 15$ & 783.201514 & 319.62 & HIFI & 1.8 & $-$ & $-$ & $-$ & $-$\\
      $17 - 16$ & 832.0617177 & 359.55 & HIFI & 2.1 & $-$ & $-$ & $-$ & $-$\\
      $18 - 17$ & 880.9055792 & 401.83 & HIFI & 1.3 & $-$ & $-$ & $-$ & $-$\\
      $19 - 18$ & 929.732125 & 446.45 & HIFI & 1.9 & $-$ & $-$ & $-$ & $-$\\
      $20 - 19$ & 978.5404329 & 493.41 & HIFI & 1.4 & $-$ & $-$ & $-$ & $-$\\
      $21 - 20$ & 1027.3295036 & 542.71 & HIFI & 0.9 & $-$ & $-$ & $-$ & $-$\\
      $22 - 21$ & 1076.0983895 & 594.36 & HIFI & 1.6 & $-$ & $-$ & $-$ & $-$\\
      \hline
      \end{tabular}
      \end{adjustbox}
      \tablefoot{The ALMA and APEX lines reported are from \citetalias{Unnikrishnan_et_al_2024}, unless otherwise specified. All HIFI lines listed are from our \textit{Herschel}/HIFI spectral survey of IRAS 15194$-$5115 (see Sect.~\ref{subsubsec:HIFI_spectral_survey}). The Yebes 40 m $J = 1 - 0$ line and all IRAM 30 m lines are from \citet{Massalkhi_et_al_2024}. For the ALMA (and ACA) lines, the integrated intensity values reported are in units of Jy km s$^{-1}$ and are calculated using spectra extracted from apertures large enough to encompass all detected line emission. For all other SD observations (APEX, IRAM, Yebes, OSO, HIFI), the integrated intensities are given in units of K km s$^{-1}$, in the main beam ($T_\mathrm{MB}$) temperature scale. The relevant beam size ranges and main beam efficiencies for the different telescopes are described in Sect.~\ref{subsec:CS_Observations}, and the individual beam sizes at each transition frequency are indicated in Fig.~\ref{fig:15194_CS_line_fits} and Figs.~\ref{fig:15082_and_07454_CS_line_fits} - \ref{fig:10216_CS_line_fits} alongside the corresponding line spectra. \textbf{References.} $^{(a)}$ \citet{Woods_et_al_2003}; $^{(b)}$ \citet{Danilovich_et_al_2018}; $^{(c)}$ The DEATHSTAR ACA survey \citep{Ramstedt_et_al_2020, Andriantsaralaza_et_al_2021}; $^{(d)}$ \citet{Massalkhi_et_al_2024}.}
\end{table*}

\subsubsection{ALMA line survey}
\label{subsubsec:ALMA_spectral_survey}
ALMA band 3 ($85 - 116$\,GHz) spectral surveys (project code: 2013.1.00070.S; PI: Nyman, L. \AA.) were presented by \citetalias{Unnikrishnan_et_al_2024} for IRAS 15194$-$5115, IRAS 15082$-$4808, and IRAS 07454$-$7112, and include the $J = 2 - 1$ line of CS and its isotopologues $^{13}$CS and C$^{34}$S. AFGL 3068 was also originally included in our ALMA survey, but the observations for this source were not completed, and yielded no relevant lines, leading us to rely only on SD observations (see Sections~\ref{subsubsec:APEX_spectral_survey},~\ref{subsubsec:Supplementary_CS_observations}) for this source. We extracted spectra from multiple, progressively increasing apertures for the spatially resolved ALMA CS lines, using the same procedure as presented above for the ACA CO lines (Sect.~\ref{subsec:CO_Observations}). We see no evidence of resolved-out flux in our CS $J = 2 - 1$ maps. The multi-aperture line profiles are used along with the SD spectra to constrain our RT models (see Sect.~\ref{subsubsec:CS_modelling_procedure}), ensuring that both spatial and excitation constraints are in place for the determination of the best-fit models. Following \citetalias{Unnikrishnan_et_al_2024}, we use a standard 10\% calibration uncertainty on the CS integrated intensities for lines observed with ALMA. For more details of the ALMA survey, including the specifics of the observations, data processing, and availability of processed data, we refer the reader to \citetalias{Unnikrishnan_et_al_2024}.

\subsubsection{APEX line survey}
\label{subsubsec:APEX_spectral_survey}
All five sources in our sample were observed using the Atacama Pathfinder Experiment \citep[APEX, ][]{Gusten_et_al_2006}, in the frequency ranges 159-211 GHz (band 5) and 200-270 GHz (band 6). Additionally, IRAS 15194$-$5115 was also observed at the frequencies 272-376 GHz (band 7). We used only the CS (and isotopologues) lines (Table~\ref{tab:CS_line_intensities}) in the present work. Band 5 observations were conducted using the SEPIA receiver \citep{Billade_et_al_2012, Belitsky_et_al_2018} for all sources. IRAS 15194$-$5115 observations in bands 6 and 7 were carried out using the SHeFI receiver \citep{Vassilev_et_al_2008}, while the PI230\footnote{PI230 is a collaboration between the European Southern Observatory (ESO) and the Max-Planck-Institut für Radioastronomie (MPIfR). See \url{https://www.eso.org/public/teles-instr/apex/pi230/}} instrument was used for band 6 observations of the rest of the sources. The SHeFI bands 6 and 7 observations of IRAS 15194$-$5115 were carried out in October 2014. The SEPIA Band 5 observations were performed in October 2016 for IRAS 15194$-$5115, and in June-July 2021 for the other four sources. The PI230 observations were done in July and October 2019. 

All observations were performed in wobbler switching mode, with a standard wobbler throw of 1$\arcmin$. The SEPIA band 5 and PI230 observations were sideband separated (2SB). Single-sideband (SSB) observations with SHeFI for IRAS 15194$-$5115 were, when possible, optimised to avoid image sideband leakage from bright emission lines. Pointing, focus, and calibration were regularly checked with the standard catalogues employed at APEX, and we note that the science targets themselves are part of these catalogues.

The APEX spectra were originally delivered in the antenna temperature \citep[$T_\mathrm{A}^\mathrm{*}$,][]{Ulich_and_Haas_1976} scale, which has already been corrected for atmospheric attenuation \citep[see e.g.][and references therein]{Pardo_et_al_2022aug, Pardo_et_al_2025}. We converted all spectra used in this work to the main beam temperature ($T_\mathrm{MB}$) scale, using $T_\mathrm{MB} = T_\mathrm{A}^\mathrm{*}/\eta_\mathrm{MB}$, where $\eta_\mathrm{MB}$ is the main beam efficiency. The adopted values for APEX main beam efficiency\footnotemark[\value{footnote}] are 0.8 for SEPIA band 5, 0.75 for PI230 and SHeFI band 6, and 0.73 for SHeFI band 7. The APEX beam\footnote{\url{https://www.apex-telescope.org/telescope/efficiency/}} varies from $\sim$39$\arcsec$ at 159 GHz to $\sim$16$\arcsec$ at 376 GHz. Data reduction was performed using the GILDAS/CLASS\footnote{\url{https://www.iram.fr/IRAMFR/GILDAS}} software. A first-order polynomial baseline, calculated on the line-free regions, was subtracted from the spectra. As in \citetalias{Unnikrishnan_et_al_2024}, we assume a standard 20\% calibration uncertainty for the APEX data.

\subsubsection{HIFI survey of II Lup}
\label{subsubsec:HIFI_spectral_survey}
We present observations of CS, $^{13}$CS, and C$^{34}$S line emission towards IRAS 15194−5115 from the high-sensitivity, high-resolution spectral scan (PI: De Beck, E.) carried out between 5 and 19 September 2012 using the Heterodyne Instrument for the Far Infrared \citep[HIFI,][]{de_Graauw_et_al_2010} instrument on board the Herschel Space Observatory \citep{Pilbratt_et_al_2010}. The observations cover the frequency range 479.5 - 1121.9 GHz, spanning bands 1a, 1b, 2a, 2b, 3a, 3b, 4a, and 4b. All observations were carried out in spectral scan mode, using a dual beam switch calibration scheme with off-source positions 3\arcmin\/ away from the science target's position. To improve baseline stability in bands 3 and 4, the fast chop mode was selected; this was not needed for bands 1 and 2. As in the case of the APEX data, we used a first-order polynomial for baseline subtraction.

For a detailed description of the different observing modes, the in-orbit performance of HIFI, and the data processing schemes, we refer to \citet{Roelfsema_et_al_2012}, \citet{Mueller_et_al_2014}, and \citet{Shipman_et_al_2017}. The conversion of the HIFI spectral scans from double-sideband spectra to single-sideband spectra follows the conjugate gradient method described by \citet{Comito_and_Schilke_2002} taking into consideration all relevant sideband gain ratios. The data used here were extracted from the highly processed data product (HPDP\footnote{\url{https://www.cosmos.esa.int/web/herschel/highly-processed-data-products}}\textsuperscript{,}\footnote{\url{https://archives.esac.esa.int/hsa/legacy/HPDP/HIFI/HIFI_spectral_scans/HIFI_Spectral_Scans_HPDP_Release_Notes.pdf}}) made available in the Herschel Science Archive as part of an effort to uniformly process 500 spectral scan observations  towards different objects. The HPDP consists of one spectrum per polarisation (horizontal and vertical) for each band (1a through 4b). For our purposes, we averaged the spectra obtained for the horizontal and vertical polarisations.

The nominal spectral resolution of HIFI is 1.1\,MHz, corresponding to a velocity resolution of $0.3-0.7$ km s$^{-1}$ across the survey. Beam widths vary from 44\farcs2 at 479.5\,GHz to 18\farcs9 at 1121.9\,GHz \cite[][see their Eq. 3]{Roelfsema_et_al_2012}. The absolute pointing accuracy is $\sim$0\farcs8, with a pointing stability of 0\farcs3\footnote{\url{https://www.cosmos.esa.int/web/herschel/pointing-performance}}. HIFI main-beam efficiencies ($\sim$0.54 - 0.66 in our frequency range) for $T_\mathrm{A}^\mathrm{*}$ to $T_\mathrm{MB}$ conversion, and calibration uncertainty estimates ($\sim$10\%) are adopted from \citet{Mueller_et_al_2014}. The overall uncertainty on the line intensities used in this work can be larger, especially for the higher $J$ lines due to relatively low signal-to-noise ratios, as can be seen from Fig.~\ref{fig:15194_CS_line_fits}. The full molecular inventory from the APEX and HIFI surveys will be presented in a future work.

\subsubsection{Supplementary CS observations}
\label{subsubsec:Supplementary_CS_observations}
In addition to the ALMA, APEX, and HIFI data described above, we used archival CS lines available from the literature to provide better constraints to our RT models. For IRAS 07454$-$7112, we used APEX spectra of CS $J = 4-3$ and $6-5$ lines from \citet{Danilovich_et_al_2018}. We also obtained the same lines for IRAS 15194$-$5115, for which we already had these lines from our APEX survey. In this case, we compared the line profiles from both APEX datasets and found them to be identical within calibration uncertainties. 

For AFGL 3068, we used an ACA CS $J = 7-6$ data cube from the DEATHSTAR project \citep{Ramstedt_et_al_2020, Andriantsaralaza_et_al_2021}, and an OSO 20 m CS $J = 2-1$ integrated intensity reported by \citet{Woods_et_al_2003}. For IRC~+10\,216, we obtained an interferometric data cube of the CS $J = 2 - 1$ line, produced by combining ALMA data with an IRAM 30 m on-the-fly map, from \citet{Velilla-Prieto_et_al_2019}. This helped to recover all the emission for this very extended source \citep[see e.g.][]{Cernicharo_et_al_2015}, without being restricted by the maximum recoverable scale of the ALMA observations. We also received a Yebes 40 m spectrum of the CS $J = 1-0$ line, along with IRAM 30 m spectra of the $J = 2-1$, $3-2$, $4-3$, $5-4$, $6-5$, and $7-6$ lines from \citet{Massalkhi_et_al_2024}. From these, we used all lines except the CS $J = 7-6$, for which \citet{Massalkhi_et_al_2024} mention potential observational issues, in our RT modelling. The calibration uncertainty for the IRAM 30 m varies from $\sim$10\% - 30\% depending on the observation frequency of observation \citep[see][]{Massalkhi_et_al_2024}. The integrated intensities of the supplementary CS lines used in this work are listed in Table~\ref{tab:CS_line_intensities}.

\section{Radiative transfer modelling}
\label{sec:Radiative_Transfer_Modelling}

\subsection{Envelope model}
\label{subsec:Envelope_model}
We adopted a spherically symmetric, smooth CSE model, assuming an isotropic, constant mass-loss rate (MLR, $\dot{M}$) and a constant dust-to-gas mass ratio ($\Psi$). Though some of our sources display asymmetric features in their CSEs (see Sect.~\ref{sec:The_Sources}), we were still able to reproduce the overall line profiles and radial intensity profiles from the ALMA observations with these assumptions, similar to those regularly used in the literature to model similar sources \citep[e.g.][]{Massalkhi_et_al_2024, Danilovich_et_al_2015, Danilovich_et_al_2018, Danilovich_et_al_2019, Velilla-Prieto_et_al_2019, Schoier_et_al_2007, Schoier_et_al_2013, Agundez_et_al_2012, Agundez_et_al_2017}. A full treatment of the circumstellar asymmetries and density variations is beyond the scope of the modelling presented in this paper.

The central AGB star is characterised by its luminosity ($L_{\star}$) and temperature ($T_{\star}$). Dust is assumed to be present only beyond the dust condensation radius ($R_\mathrm{c}$), which is assumed to be at three times the stellar radius \citep[e.g.][]{Hofner_2009, Bladh_and_Hofner_2012, Hofner_and_Olofsson_2018} in our models (see Sect.~\ref{subsec:Dust_Modelling}). We assume a smoothly accelerating wind with a terminal radial expansion velocity $\varv_\infty$ (Table~\ref{tab:source_properties}). Between the surface of the star and the dust condensation radius, the gas expansion velocity ($\varv_{\text{exp}}$) is assumed to be a constant ($\varv_0$ = 3 km s$^{-1}$), approximately equal to the sound speed at $R_\mathrm{c}$ \citep[e.g.][]{Danilovich_et_al_2015}. As the gas volume in this region (1-3 $R_\star$) is very low compared to the rest of the CSE, any variation in the velocity profile here will not significantly affect our results. Beyond $R_\mathrm{c}$, the velocity profile can be expressed as \citep{Hofner_and_Olofsson_2018}
\begin{equation}
    \varv_{\text{exp}}(r) = \varv_0 + (\varv_\infty - \varv_0) \left( 1 - \frac{R_\text{c}}{r} \right)^\beta,
    \label{eqn:v_exp}
\end{equation}
where the parameter $\beta$ controls the wind acceleration. We also adopted a constant turbulent velocity of 1.5 km s$^{-1}$ and assumed that dust and gas have the same velocity. The circumstellar gas density was assumed to decrease as $r^{-2}$, given by
\begin{equation}
    n_{\text{H}_2} = \frac{\dot{M}_g}{4 \pi r^2 \varv(r)} \frac{1}{m_{\text{H}_2}},
    \label{eqn:n_H2}
\end{equation}
where $\dot{M}_g$ is the gas MLR, $\varv(r)$ the wind velocity at radius $r$, and $m_{\text{H}_2}$ is the molecular mass of hydrogen. The circumstellar gas temperature profile was assumed to be given by a power law of the format
\begin{equation}
    T(r) = T_0 \left( \frac{r}{R_\text{in}} \right)^\alpha,
    \label{eq:T_profile}
\end{equation}
where $T_0$ is the gas temperature at the inner radius ($R_\text{in}$) of the models, and the slope, $\alpha$, is a free parameter. We tried setting a lower limit of 10 K for the gas temperature in the outermost regions of the models, but it made no impact on the results. The fractional abundance of molecular species is defined with respect to molecular hydrogen (H$_2$), as 
\begin{equation}
f_X(r) = \frac{n_X(r)}{n_{\text{H}_2}(r)},
\label{eq:frac_abund_formula}
\end{equation}
where $n_\text{X}(r)$ and $n_{\text{H}_2}$ are the number density distributions of the molecule and H$_2$, respectively. It was implicitly assumed here that there is no photodissociation of H$_2$ within the spatial extents of the other molecules, which is valid since H$_2$ has the most spatially extended abundance distribution among all circumstellar species \citep{Huggins_and_Glassgold_1982}.

\subsection{Dust modelling}
\label{subsec:Dust_Modelling}
We modelled the SEDs of our sources using MCMax \citep{Min_et_al_2009}, a Monte Carlo RT code that takes into account absorption, re-emission, and scattering to solve the radiative transfer of dust. 
In addition to standard Monte Carlo RT, where photon packets and their interaction with dust grains are traced across the medium, MCMax includes an implementation of the diffusion approximation \citep{Min_et_al_2009} to incorporate energy transport by radiative diffusion \citep[e.g.][]{Wehrse_Et_al_2000} as well, which improves the accuracy of the estimated temperature profile in regions of high optical depth, and can also increase computational speed. The code assumes axial symmetry for the medium.

Key input parameters include the radiation field from the central star, the dust density distribution, and the optical properties of the dust. We used the spherical, hydrostatic COMARCS model atmospheres \citep{Aringer_et_al_2019} as input stellar spectra for IRAS 15194$-$5115, IRAS 15082$-$4808, and IRAS 07454$-$7112, with a surface gravity (log~$g$) of $−$0.5, and a microturbulence velocity of 0.5 km s$^{-1}$. 
The COMARCS programme generates synthetic spectra, assuming local thermodynamic and chemical equilibrium, taking into account atomic and molecular lines \citep{Aringer_et_al_2016, Aringer_et_al_2019}. These models thus serve as better representations of the atmospheric spectra of AGB stars than simple black bodies.
The input COMARCS spectra were extrapolated and their resolution was reduced following the same procedure as described by \citet{Andriantsaralaza_et_al_2022}. For AFGL 3068 and IRC~+10\,216, we had to resort to black body spectra instead of the COMARCS models, as we could not obtain good fits using models in the available temperature range of COMARCS (2500 - 3300 K) and had to assume cooler stars.

We modelled the dusty envelope with a fixed inner radius of 3$R_{\star}$, in line with both dynamical models and interferometric measurements, that place the dust condensation radius ($R_\mathrm{c}$) for carbon stars in the range 2-3 $R_{\star}$ \citep[e.g.][]{ Andersen_et_al_2003, Nowotny_et_al_2011, Wittkowski_et_al_2007, Karovicova_et_al_2011, Norris_et_al_2012, Bladh_and_Hofner_2012, Hofner_and_Olofsson_2018}. We used amorphous carbon dust grains with a grain density of 2 g cm$^{-3}$, typical of carbon stars \citep{Hofner_and_Olofsson_2018}. The optical properties of the dust were adopted from \citet{Suh_2000}. 

Testing different assumed grain sizes usually expected for AGB circumstellar dust \citep[see e.g.][and references therein]{Mattsson_et_al_2010, Hofner_and_Olofsson_2018}, we concluded that only 1 $\mu$m grains led to satisfying results. Models with smaller grains all gave poor fits to the short-wavelength part of the SEDs, and/or led to unreasonably high dust condensation temperatures ($T_\text{c}$ $>2000$ K), as long as $R_\mathrm{c}$ was fixed at 3$R_{\star}$. When we attempted to leave $R_\mathrm{c}$ as a free parameter instead of fixing it to 3$R_{\star}$, models with grains smaller than 1 $\mu$m all returned unrealistically large $R_\mathrm{c}$ values of up to $\sim$16 - 31 $R_{\star}$, incompatible with dust-driven wind models. This is further discussed in Sect.~\ref{subsec:Dust_and_CO_discussion}. 

The stellar luminosity ($L_\star$), stellar temperature ($T_\star$), and dust-MLR ($\dot{M}_\text{d}$) were the free parameters in the SED fitting. For each star, an initial grid was constructed based on previously reported values of the free parameters from the literature, when available. The grids were then refined in the vicinity of local minima. A dust-MLR step of 0.5 dex was employed. The luminosity step was initially set to 1000 \(L_\odot\) and subsequently refined to 500 \(L_\odot\) when necessary. Distance estimates for all five stars were obtained from \citet{Andriantsaralaza_et_al_2022}. MCMax computes the SED, the optical depth at \(10 \, \mu\text{m}\) (\(\tau_{10}\)), and the temperature distribution of the dust for each grid point.
The best-fit model for each source was identified by minimising the reduced chi-squared (\(\chi^2_{\text{red}}\)), using
\begin{equation}
    \chi_{\text{red}}^2 = \frac{1}{(N-p)} \sum_{i=1}^{N} \frac{(F_{\text{mod},\lambda} - F_{\text{obs},\lambda})^2}{\sigma_\lambda^2},
\label{eq:chi2}
\end{equation}
where $N$ is the number of flux measurements used, $F_{\text{mod},\lambda}$ is the modelled flux density at wavelength $\lambda$, $F_{\text{obs},\lambda}$ denotes the corresponding observed flux density (Table~\ref{tab:photometric_fluxes_SED}) with uncertainty $\sigma_\lambda$ (see Sect.~\ref{subsec:SED_Observations}), and $p$ is the number of free parameters, equal to 3 in our models. The observed SEDs and the corresponding best-fit models are shown in Fig.~\ref{fig:best_fit_SEDs}.

\subsection{CO modelling}
\label{subsec:CO_Modelling}
To ensure consistency in our modelling across the sources, we derived gas mass-loss rates by CO emission line modelling for each star. We used the one-dimensional (1D), non-LTE, Monte Carlo programme (MCP) RT code \citep{Schoier_and_Olofsson_2001} to determine the CO molecular level populations across the CSEs. MCP has been extensively used for RT modelling of AGB CSEs \citep[e.g.][]{Danilovich_et_al_2014, Danilovich_et_al_2015, Ramstedt_et_al_2008, Schoier_et_al_2002, Schoier_et_al_2006, Schoier_et_al_2007, Schoier_et_al_2013}. With the results from the dust modelling ($T_{\star}$, $L_{\star}$, $\tau_{10}$ and $T_\text{dust}(r)$; see Sect.~\ref{subsec:Dust_Modelling}) as input, along with basic source parameters (Table~\ref{tab:source_properties}) and molecular data, MCP calculates the molecular excitation across different radial bins by using the Monte Carlo method to solve the equations of statistical equilibrium for each model in a grid of gas MLRs and temperatures.

We included rotational energy levels from $J = 0 - 40$ in both the ground ($\varv = 0$) and first vibrationally excited ($\varv = 1$) states of CO in our modelling, covering upper-level energies ($E_\text{up}$) from 5.5 K ($\varv = 0, J = 1$) up to $\sim$7578 K ($\varv = 1, J = 40$). We use rate coefficients for inelastic CO-H$_2$ collisions provided by \citet{Yang_et_al_2010}, weighted assuming the statistical ortho-to-para H$_2$ ratio of 3 \citep{Danilovich_et_al_2018, Massalkhi_et_al_2024}. The inner radius of the models was set to the dust condensation radius ($R_\text{c}$), equal to 3$R_\star$. The fractional abundance of CO was calculated using
\begin{equation}
f(r) = f_0 \exp\left(-\text{ln}\,2\left(\frac{r}{R_{1/2}}\right)^s\right),
\label{eq:CO_abundance_profile}
\end{equation}
where $f_0$ is the initial photospheric abundance of CO with respect to H$_2$, $R_{1/2}$ is the photodissociation radius defined as the radius where the CO abundance has dropped to $f_0/2$, and $s$ is the slope of the profile. We adopted a typical canonical value of $f_0 = 1\times10^{−3}$ for all stars \citep[e.g.][]{Danilovich_et_al_2015}. For the sake of consistency, we did not leave the CO radial extent as a free parameter, as ALMA data that can constrain this parameter were only available for two of the sources in our sample. Instead, we used the values of $R_{1/2}$ and $s$ tabulated by \citet{Saberi_et_al_2019}, which are dependent on the gas mass-loss rate ($\dot{M}_\text{g}$) and the terminal expansion velocity ($\varv_\infty$, see Table~\ref{tab:source_properties}). The gas expansion velocity is described by Eq.~\ref{eqn:v_exp}, with $\varv_\infty$ taken from \citet{Woods_et_al_2003}. We assumed no drift velocity between the gas and the dust.

The kinetic temperature profile of the gas, assumed to be a power law (Eq.~\ref{eq:T_profile}), is varied in the grid with $T_0$ and $\alpha$ as free parameters, along with the gas-MLR, $\dot{M}_\text{g}$. We created individual grids for each source, with $T_0$ in the range 500$-$2250 K, with a step-size of 250 K, $\alpha$ varying from -1.4 to -0.4, in steps of 0.2. We centred the $\dot{M}_\text{g}$ grids around previously determined gas-MLR values for the respective sources from the literature, scaled to the new distances from \citet{Andriantsaralaza_et_al_2022}, with a step of 0.2 dex, overall spanning over 1.6 units in log scale.

Model line profiles were generated for all the SD observations and for all beam sizes used to extract spectra from the ALMA line cubes (see Sect.~\ref{subsec:CO_Observations}). 
For each source, models with $\chi^2_\mathrm{red}$ $\le$ 5 were first identified. From among these, the best-fit models were then determined by $\chi^2$ minimisation coupled with visual inspection of the modelled line profile shapes, as commonly done in the literature \citep[see e.g.][]{Ramstedt_et_al_2008}. This is done since the $\chi^2$ values are based exclusively on the integrated line intensities and do not take into account the shape of the line profiles. The observed CO line profiles along with the best-fit modelled spectra are shown in Figs.~\ref{fig:CO_line_profiles_15194} - \ref{fig:CO_line_profiles_10216}. We estimate an overall uncertainty of a factor of $\sim$2-3 in our reported MLR values (see Sect.~\ref{subsec:Dust_and_CO_discussion}). 

\subsection{CS modelling}
\label{subsec:CS_Modelling}

\subsubsection{Molecular data}
\label{subsubsec:CS_molecular_data}
The CS molecular description employed in this work is the same as that used by \citet{Danilovich_et_al_2018}, comprising of rotational energy levels ranging from $J = 0 - 40$ within both the ground ($\varv = 0$) and first excited ($\varv = 1$) vibrational states, amounting to a total of 82 energy levels. 
The $\varv = 0$ levels include 820 collisional transitions, and the $\varv = 0$ and $\varv = 1$ levels collectively account for 160 radiative transitions. The energy levels and radiative transition data were sourced from the Cologne Database for Molecular Spectroscopy \citep[CDMS\footnote{\url{https://cdms.astro.uni-koeln.de}}; ][]{Muller_et_al_2005, Endres_et_al_2016}. The collisional rate coefficients were adapted from CO-H$_2$ collision calculations presented by \citet{Yang_et_al_2010}, assuming an H$_2$ ortho-to-para ratio of 3. We also tested the more recent CS-H$_2$ collisional rates from \citet{Denis-Alpizar_et_al_2018}, and found no significant difference between the modelled line intensities in the two cases. For the isotopologue models of $^{13}$CS and C$^{34}$S, we used energy levels and transitions for the same range of states ($J = 0 - 40$, $\varv = 0, 1$) as $^{12}$CS, taken from the CDMS. We employed the same set of collisional rates as for the main CS isotopologue for both $^{13}$CS and C$^{34}$S. 

\subsubsection{Modelling procedure}
\label{subsubsec:CS_modelling_procedure}
We used a 1D accelerated lambda iteration (ALI) non-LTE spectral line RT code to model CS line emission. The code is based on the ALI method \citep{Rybicki_and_Hummer_1991}, and is described in detail by \citet{Maercker_et_al_2008}. ALI provides reliable convergence even at high optical depths, and has been widely used to model circumstellar line emission \citep[e.g.][]{Justtanont_et_al_2005, Maercker_et_al_2008, Maercker_et_al_2009, Schoier_et_al_2011, Brunner_et_al_2018, Danilovich_et_al_2016, Danilovich_et_al_2017, Danilovich_et_al_2018, Danilovich_et_al_2019}. The code has been well benchmarked against other similar non-LTE line RT codes \citep[see][]{van_Zadelhoff_et_al_2002, Maercker_et_al_2008}. In this work, we use the latest available version of the ALI code, the same used by \citet{Danilovich_et_al_2018, Danilovich_et_al_2019} to model circumstellar molecular line emission.

ALI takes the results from the dust and CO RT models (see Sections~\ref{subsec:Dust_Modelling},~\ref{subsec:CO_Modelling}), i.e. the stellar luminosity and temperature, the kinetic temperature profiles of the circumstellar dust and gas, the dust optical depth at 10 $\mu$m, and the gas mass-loss rate, as inputs, and for each modelled line, calculates the molecular excitation at each radial point in the CSE. We used 40 radial points distributed logarithmically between $R_\text{in}$ and $R_\text{out}$. New level populations are calculated at each iteration, and we assume that the model has converged once the average population change at each radial point between two successive iterations is less than $10^{-3}$. The above choices of the number of radial shells and the convergence threshold are well justified from the literature \citep[see e.g.][]{Maercker_et_al_2008}. Considering that CS is a parent species originating close to the stellar surface, we assumed a centrally-peaked Gaussian distribution to model the circumstellar CS fractional abundance (Eq.~\ref{eq:frac_abund_formula}) profile, given by
\begin{equation}
f(r) = f_0 \exp\left(-\left(\frac{r}{R_e}\right)^2\right),
\label{eq:gaussian_abundance_profile}
\end{equation}
where $f_0$ is the peak abundance at the inner edge of the envelope, and $R_\text{e}$ is the $e$-folding radius, at which the CS abundance has dropped to $f_0/e$. $f_0$ and $R_\text{e}$ as the only free parameters in our modelling. The outer radius ($R_\text{out}$) of the models was set to five times the respective $R_\text{e}$, ensuring that the model covers a sufficient volume to recover all observed emission.

We designed individual grids for each source. We periodically checked the model outputs to ensure the grid coverage was sufficiently refined and extended over an adequate region of the parameter space. We used integrated line intensities from a number of progressively increasing apertures for the ALMA lines (see Fig.~\ref{fig:ALMA_radial_intensity_plots}), along with line intensities from SD observations (Table~\ref{tab:CS_line_intensities}) to constrain the models, as we did for the CO modelling. This way, we employed both the excitation information (from the different $J$ transitions) and the azimuthally averaged radial distribution of the emission (from the ALMA maps) to constrain the CS models. The small apertures from the ALMA maps and the high-$J$ SD lines give constraints on the inner wind, while the larger ALMA apertures and low-$J$ SD lines constrain more the radial extent.

{\renewcommand{\arraystretch}{1.35}
\begin{table*}[h]
   \caption{Modelling results}
   \label{tab:RT_modelling_results}
   \centering
   \begin{adjustbox}{width=18.25cm}
   \begin{tabular}{c@{\extracolsep{10pt}} cccccccc@{\extracolsep{10pt}} cccc@{\extracolsep{10pt}} cc}
      \hline\hline \\[-3.5ex]
      \makecell{\\\\Source} & 
      \multicolumn{6}{c}{Dust modelling} & 
      \multicolumn{5}{c}{CO modelling} & 
      \multicolumn{2}{c}{CS modelling} \\
      \cline{2-7} \cline{8-12} \cline{13-14} \\[-1.5ex]
      & \makecell{{$L_\star$}\\{[$L_\odot$]}} & \makecell{{$T_\star$}\\{[K]}} & \makecell{{$R_\text{in}$}\\{[cm]}} & \makecell{{$\dot{M}_\text{d}$}\\{[$M_\odot$ yr$^{-1}$]}} & \makecell{{$T_\text{c}$}\\{[K]}} & $\tau_\text{10}$ & 
      \makecell{{$\dot{M}_\text{g}$}\\{[$M_\odot$ yr$^{-1}$]}} & \makecell{{$T_0$}\\{[K]}} & $\alpha$ & \makecell{{{$R_\text{1/2}$}}\\{[cm]}} & \makecell{{$\Psi$}\\{($\times10^{-3}$)}} &
      $f_0$ & \makecell{{$R_\text{e}$}\\{[cm]}} \\[2ex]
      \hline \\[-1.5ex]
        15194$-$5115  & 11000 & 2500 & $1.20\times10^{14}$ & $4.20\times10^{-8}$ & 1510 & 1.81 & $2.20\times10^{-5}$ & 1750 & -0.8 & $3.37\times10^{17}$ & 1.91 & $3.1^{+1.6}_{-1.6}\times10^{-6}$ & $6.8^{+5.5}_{-3.9}\times10^{16}$ \\
        15082$-$4808  & 11500 & 2500 & $1.20\times10^{14}$ & $7.00\times10^{-8}$ & 1780 & 3.39 & $2.20\times10^{-5}$ & 1000 & -0.6 & $3.37\times10^{17}$ & 3.18 & $4.0^{+3.5}_{-2.4}\times10^{-6}$ & $5.2^{+1.6}_{-1.9}\times10^{16}$ \\
        07454$-$7112  & 5000 & 3000 & $5.98\times10^{13}$ & $1.26\times10^{-8}$ & 1780 & 1.93 & $8.30\times10^{-6}$ & 2250 & -0.8 & $1.96\times10^{17}$ & 1.52 & $2.0^{+1.5}_{-0.9}\times10^{-6}$ & $2.4^{+0.6}_{-0.5}\times10^{16}$ \\
        AFGL 3068     & 10000 & 2100 & $1.50\times10^{14}$ & $1.26\times10^{-7}$ & 1825 & 6.87 & $4.20\times10^{-5}$ & 1750 & -0.8 & $4.79\times10^{17}$ & 3.00 & $3.6^{+6.4}_{-3.0}\times10^{-6}$ & $1.8^{+6.7}_{-0.9}\times10^{16}$ \\
        IRC~$+$10\,216 & 12000 & 2200 & $1.50\times10^{14}$ & $8.00\times10^{-8}$ & 1675 & 4.07 & $1.50\times10^{-5}$ & 1250 & -0.8 & $2.67\times10^{17}$ & 5.33 & $1.4^{+0.5}_{-0.7}\times10^{-6}$ & $6.0^{+1.6}_{-1.6}\times10^{16}$ \\[0.5ex]
      \hline \\[-2ex]
   \end{tabular}
   \end{adjustbox}
   \tablefoot{$L_\star$: stellar luminosity, $T_\star$: stellar temperature, $R_\text{in}$: inner radius of the models ($R_\text{in}$ = 3$R_\star$), $\dot{M}_\text{d}$: dust-MLR, $T_\text{c}$: dust condensation temperature, $\tau_\text{10}$: dust optical depth at 10 $\mu$m, $\dot{M}_\text{g}$: gas-MLR, $T_0$: gas temperature at $R_\text{in}$, $\alpha$: exponent of gas temperature profile power law (see Eq.~\ref{eq:T_profile}),  $R_{1/2}$: photodissociation radius, $\Psi$: dust-to-gas mass ratio, $f_0$: peak-abundance of CS, $R_\text{e}$: e-folding radius of CS abundance profile.}
\end{table*}}

CS can be photodissociated by radiation having a broad range of frequencies from across the broadband continuum, unlike the case of CO where photodissociation can occur only due to radiation from specific UV lines \citep{van_Dishoeck_1988}. Hence, CS has no differential self-shielding between its isotopologues \citep[e.g.][]{Hrodmarsson_and_van_Dishoeck_2023}, there is no reason to expect that the radial extent of the abundance profiles, which is determined by photodissociation in the outer envelope, should vary between the different isotopologues. So, for $^{13}$CS and C$^{34}$S, we used the same modelling procedure as described above for CS, with the exception that we now have only $f_0$ as a free parameter, with $R_\text{e}$ being fixed to that of the corresponding best-fit CS model. The $f_0$ grids for isotopologue modelling were generated by scaling down the peak-abundance grids used for $^{12}$C$^{32}$S with the expected isotopic ratios for each source (see Sect.~\ref{sec:Discussion}).

For each star, the uncertainties on the line intensities of spectra extracted from different ALMA apertures (see Sect.~\ref{subsubsec:ALMA_spectral_survey}) are inherently correlated as they are taken from the same data cube. To account for this in the uncertainty estimates for the fit parameters ($f_0$ and $R_\text{e}$), we included these correlations in the $\chi^2$ calculations, treating the ALMA spectra as correlated and the SD lines as independent observations. If we sort the lines with all the ALMA spectra first, followed by the SD ones, the resulting covariance matrix, \(\mathbf{C}\), is defined as
\[
\begin{aligned}
C_{ij} = 
\begin{cases} 
\sigma_i^2, & i = j, \\ 
\sigma_i \sigma_j \rho, &  i \neq j, \text{ with } i \text{ and } j \leq n_{\text{ALMA}}, \\
0, &  i \neq j, \text{ with } i \text{ or } j > n_{\text{ALMA}},
\end{cases} \\
\end{aligned}
\]
where $\sigma_i$ represents the uncertainty in the integrated line intensities, $\rho$ is the correlation coefficient, and \(n_{\text{ALMA}}\) denotes the number of spectra extracted from the ALMA cube. We assume a correlation coefficient of 0.9, indicating that the spectra from different ALMA apertures are strongly correlated. We did not set $\rho$ exactly equal to 1, since the radial substructure and complex morphology of the line emission would lead to a small degree of decorrelation among the spectra extracted from different apertures. We checked that minor changes to the correlation coefficient around the chosen value do not affect the results significantly, and only lead to marginal changes in the estimated uncertainties as well. The off-diagonal elements of \(\mathbf{C}\) were set to zero for the SD lines. It was also ensured that the spatial information from the ALMA spectra and the excitation information from the different $J$ lines both had the same total weight, to prevent the $\chi^2$ minimisation from being strongly biased towards the spatial information simply due to the larger number of ALMA apertures compared to the number of available SD lines.  We note that while such equal weighting is a sensible scheme in our context, there is no one true way to assign these relative weights in a general case, and an informed choice should be made based on the number of available ALMA and SD spectra.

For each model in the grid, the residual vector, \(\mathbf{r}\), is defined as the difference between the modelled (\(\mathbf{m}\)) and observed (\(\mathbf{o}\)) integrated line intensity vectors ($\mathbf{r} = \mathbf{m} - \mathbf{o}$). The inverted covariance matrix (\(\mathbf{C^{-1}}\)) was then used to calculate the $\chi^2$ values as
\begin{equation}
\chi^2 = \mathbf{r}^\top \mathbf{C}^{-1} \mathbf{r}.
\end{equation}
Line profiles from models with the lowest $\chi^2$ values in the grid were visually inspected before finalising the choice of the best-fit model, for the same reasons as described in Sect.~\ref{subsec:CO_Modelling}. For a fit with two free parameters, as is the case here, the 1$\sigma$, 2$\sigma$, and 3$\sigma$ uncertainty levels for the best-fit parameters are given by $\chi_\text{min}^2$ + 2.3, $\chi_\text{min}^2$ + 6.18, and $\chi_\text{min}^2$ + 11.83, respectively, where $\chi_\text{min}^2$ is the minimum value of $\chi^2$. We note here that the uncertainties estimated for $f_0$ and $R_\text{e}$ are assuming the gas MLRs listed in Table~\ref{tab:RT_modelling_results}, and depend on the choice of relative weighting between the different lines. The results from the CS modelling are presented in sections~\ref{subsec:CS_Modelling_Results} and \ref{subsec:isotopologue_modelling_results}.

\section{Results}
\label{sec:Results}

\subsection{Dust}
\label{subsec:Dust_Modelling_Results}
For each star, the best-fit model corresponds to the model with the lowest reduced chi-square ($\chi^2_\mathrm{red}$) values. Furthermore, to ensure consistency with a dust-driven wind scenario (see Sect.~\ref{subsec:Dust_Modelling}), we only considered models that satisfy an upper limit constraint on the condensation temperature ($T_{\text{c}} < 1900$\,K) of amorphous carbon dust \citep[see][]{Nanni_et_al_2016, Hofner_and_Olofsson_2018}. The SEDs of our sources peak at approximately 10 $\mu$m, as shown in Fig.~\ref{fig:best_fit_SEDs}. The models reproduce the photometric data reasonably well for all stars. The most notable deviation between data and model occurs at the shortest wavelengths, corresponding to the Gaia data, which are consistently overestimated by the models for IRAS~15194-5115, IRAS~15082-4808, and IRAS~07457-7112, for which we used the COMARCS models as input stellar spectra (Sect.~\ref{subsec:Dust_Modelling}). However, due to the intrinsic variability of these stars, which is most prominent at short wavelengths, and the large contribution of dust emission in the SEDs at such relatively high dust mass-loss rates, the Gaia data points are not the most critical constraints. This is reflected in the high uncertainties assigned to them (see Sect.~\ref{subsec:SED_Observations}). 

The three free parameters ($\dot{M}_\text{d}$, $T_\star$, and $L_\star$) are well constrained for all stars except for IRAS~15194-5115, where $T_\star$ and $L_\star$ remain poorly constrained,  resulting in degenerate viable models for that star. The dust mass-loss rates ($\dot{M}_\text{d}$) derived across our sample span over an order of magnitude, from 1.26~$\times 10^{-8}$\,$M_\odot$ yr$^{-1}$ for IRAS 07454$-$7112 to 1.26~$\times 10^{-7}$\,$M_\odot$ yr$^{-1}$ for AFGL~3068, with an average value of 6.60~$\times 10^{-8}$\,$M_\odot$ yr$^{-1}$. The results of our dust modelling are listed in Table~\ref{tab:RT_modelling_results}. 

\begin{figure*}[h]
    \centering
     \includegraphics[width=\linewidth]{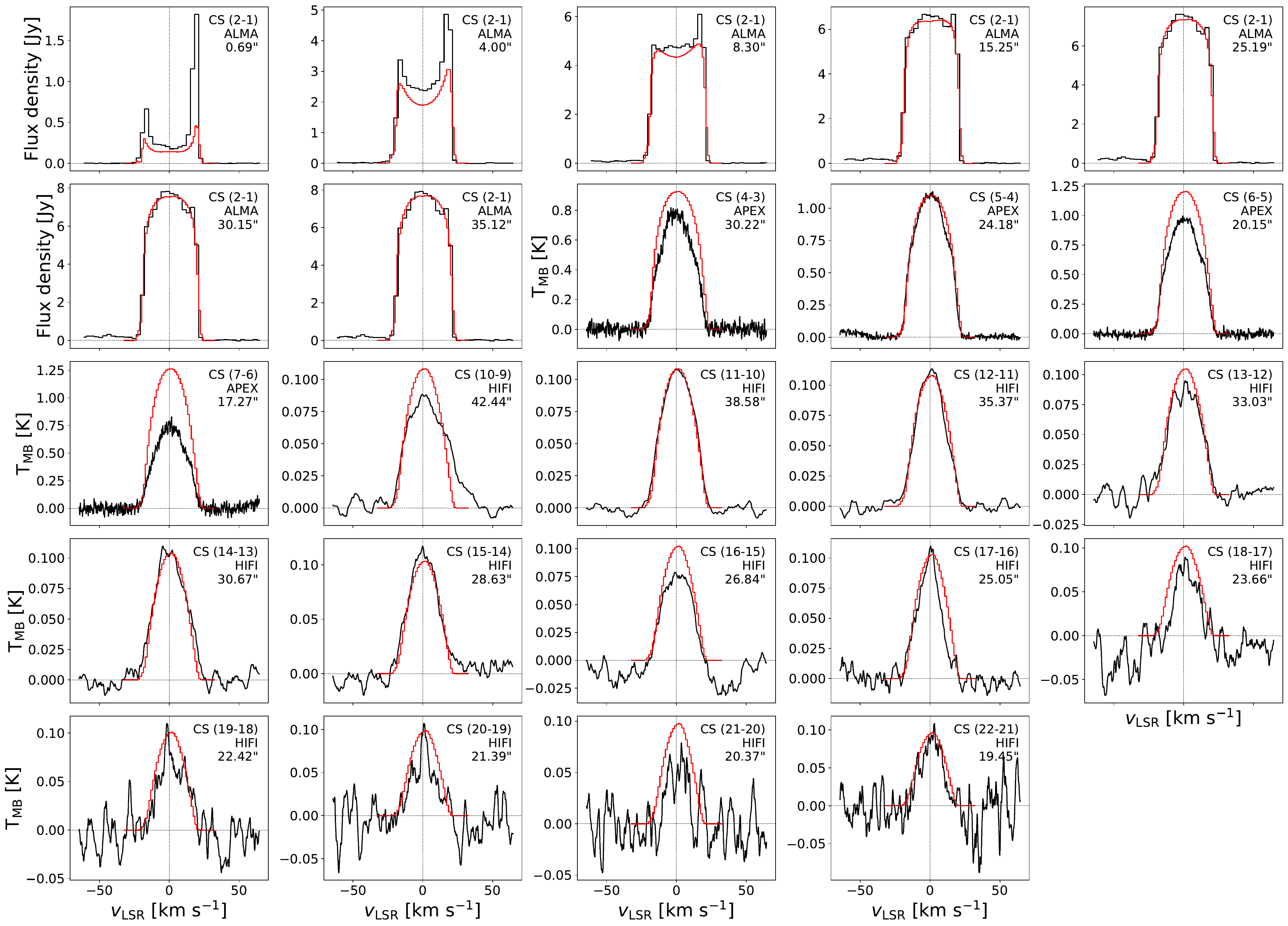}
    \caption{Observed (black) and modelled (red) CS line profiles for IRAS 15194$-$5115. The transition quantum numbers, telescope used for the observation, and the corresponding beam size in arcseconds are listed at the top right corner of each panel. The line fluxes are in units of Jansky for the ALMA spectra, whereas they are given in the main-beam temperature ($T_\mathrm{MB}$ [K]) scale for all SD spectra shown.}
    \label{fig:15194_CS_line_fits}
\end{figure*}

\begin{figure*}[h]
    \centering
    \includegraphics[width=\linewidth]{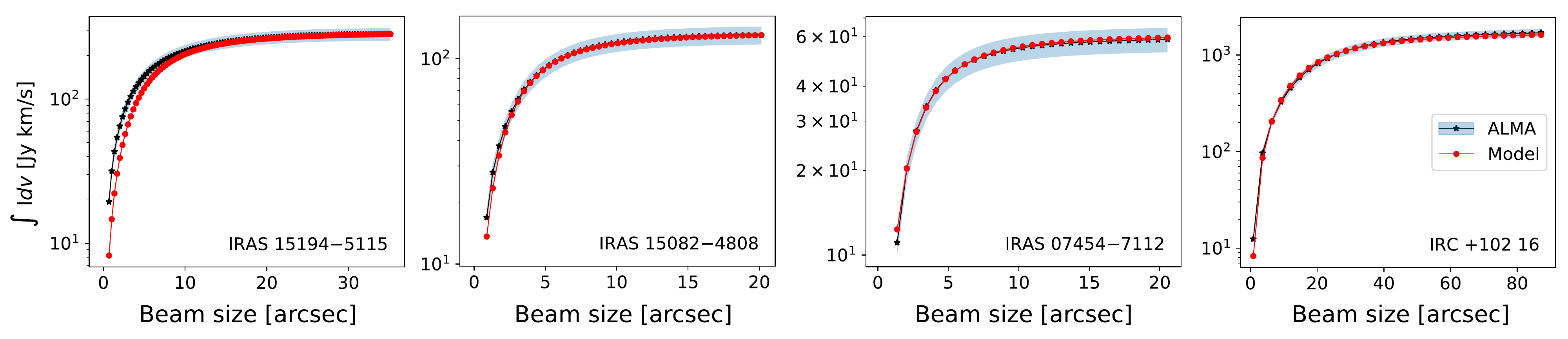}
    \caption{Comparison of modelled (red) and observed (black) integrated intensities of CS $J = 2 - 1$ spectra extracted from successively larger beams (see Sect.~\ref{subsubsec:CS_modelling_procedure}). The blue-shaded region represents the uncertainty in the observed line intensities.}
    \label{fig:ALMA_radial_intensity_plots}
\end{figure*}

\subsection{CO}
\label{subsec:CO_Modelling_Results}

Our best-fit models to the CO line emission (see Figs.~\ref{fig:CO_line_profiles_15194} - \ref{fig:CO_line_profiles_3068}) yielded well-constrained mass-loss rates for the studied sources, spanning from $8.3 \times 10^{-6}$ to $4.2 \times 10^{-5}$ M$_{\odot}$ yr$^{-1}$ (Table~\ref{tab:RT_modelling_results}). To define the accelerating gas velocity profiles (Eq.~\ref{eqn:v_exp}), we initially used $\beta$ = 1 for all sources, as is commonly done in the literature \citep[e.g.][]{Massalkhi_et_al_2024, Danilovich_et_al_2015}. This led to good models for all stars except IRAS 15194$-$5115, for which $\beta$ = 1 resulted in very weak modelled lines for the high-$J$ CO transitions $J = 9-8$ and $J = 10-9$. Hence, we tested different $\beta$ values for this source, and found that only $\beta$ = 5, implying slower acceleration compared to $\beta$ = 1, produced significant emission for the high-$J$ transitions ($J \geq 6-5$). We then tested $\beta$ = 5 for the other sources as well, and found no significant difference in the line intensities beyond the calibration uncertainties of the observations, compared to the $\beta$ = 1 case.

For IRAS 15194$-$5115, only SD observations were available, and a total of 11 transitions were used to constrain the models. The best-fit model for this source (with $\beta$ = 5, see above) fits the data reasonably well overall, though some lines are slightly overestimated while others are underestimated (Fig.~\ref{fig:CO_line_profiles_15194}). The most problematic case is the APEX $J = 7-6$ line, where the intensity ratio between model and data is 1.77. This line might be affected by calibration issues, and is also less reliable overall because of the relatively high noise level. High J-lines are systematically underestimated, indicating that the inner wind is not well constrained.

Similarly, only SD observations were available for IRAS 15082$-$4808, covering four transitions in total, up to the CO $J = 4-3$ line. In general, the fit is reasonable (Fig.~\ref{fig:CO_line_profiles_15082}), as the APEX data and the SEST $J = 1-0$ lines match well with the model, although two of the SEST lines are not well reproduced. Since the APEX data are expected to be more reliable compared to SEST, we prioritised achieving better agreement with APEX rather than SEST.

IRAS 07454$-$7112 has the lowest gas MLR in the sample. Both SD and ALMA data were available for this star. The analysis includes transitions up to $J = 14-13$, with a total of six lines. The best model successfully reproduces the general shape of the lines (Fig.~\ref{fig:CO_line_profiles_07454}). Although the fits are generally acceptable for the ALMA lines, with some lines being underestimated ($J = 3-2$) and others overestimated ($J = 2-1$), no model satisfactorily reproduces the $J = 2-1$ APEX line. For the CO $J = 3-2$ ALMA line, the integrated intensity decreases beyond 14.3$\arcsec$, likely due to resolved-out flux. For this star, high-J transitions are also systematically underestimated, suggesting that the inner wind is not very well constrained.

For IRC$+$10$,$216, only SD data were used as constraints, and the analysis included four transitions. While the line shapes are well reproduced, the modelled line strengths do not agree consistently with observations. For example, the model overestimates the NRAO CO 1$-$0 line intensity while reproducing the OSO and SEST CO 1$-$0 lines well. We note that these 1$-$0 SD observations should, however, be treated with caution as the telescope intensity calibrations can be highly uncertain for that transition (Fig.~\ref{fig:CO_line_profiles_10216}).

Both ALMA and SD data were available for AFGL 3068. A total of five transitions were used, with the highest being $J = 6-5$. This star has the highest MLR in the sample, and its CO line profiles exhibit an irregular triangular shape, characteristic of high-MLR AGB stars. The best-fit model reproduces the line widths well. The CO $J = 3-2$ ALMA line shows signs of resolved-out flux, as beyond 15$\arcsec$, the integrated intensity decreases with radial distance. While ALMA lines show reasonably good agreement with the model, some APEX lines could not be reconciled with ALMA observations (see Fig.~\ref{fig:CO_line_profiles_3068}).

\subsection{CS}
\label{subsec:CS_Modelling_Results}

{\renewcommand{\arraystretch}{1.5}
\begin{table*}
   \caption{Peak abundances of $^{13}$CS and C$^{34}$S and CS isotopic ratios}
   \label{tab:isotopic_peak_abundances_and_ratios}
   \centering
      \begin{tabular}{l c c c c}
      \hline\hline & \\[-3ex]
      Source & $f_0$ ($^{13}$CS) & $f_0$ (C$^{34}$S) & $^{12}$CS/$^{13}$CS & C$^{32}$S/C$^{34}$S \\
      \hline & \\[-3ex]
      IRAS 15194$-$5115    & $4.2\times10^{-7}$ & $1.1\times10^{-7}$ & $7.4^{+2.1}_{-3.8}$ & $29.5^{+8.6}_{-15.2}$  \\
      IRAS 15082$-$4808    & $6.7\times10^{-8}$ & $1.4\times10^{-7}$ & $60^{+43}_{-36}$    & $28.0^{+20.3}_{-16.8}$ \\
      IRAS 07454$-$7112    & $6.7\times10^{-8}$ & $9.5\times10^{-8}$ & $30^{+22}_{-13}$    & $21.0^{+15.7}_{-9.0}$  \\
      AFGL 3068 & $5.5\times10^{-8}$ & $6.0\times10^{-8}$ & $65^{+31}_{-31}$$^{\dagger}$    & $60.0^{+28.3}_{-28.3}$$^{\dagger}$ \\
      IRC~+10\,216    & $3.8\times10^{-8}$ & $7.4\times10^{-8}$ & $37^{+10}_{-18}$    & $19.0^{+5.4}_{-9.2}$   \\
      \hline
      \end{tabular}
      \tablefoot{$^{\dagger}$ The CS isotopic ratios for AFGL 3068 are poorly constrained due to the degeneracy between the fit parameters in the $^{12}$CS modelling (see text).}
\end{table*}
}

Models that reproduce the observed CS emission reasonably well were obtained for all five sources in the sample, yielding well-constrained abundance profiles (see Fig.~\ref{fig:CM_RT_detailed_comparison}). Figure~\ref{fig:15194_CS_line_fits} and Figures~\ref{fig:15082_and_07454_CS_line_fits} - \ref{fig:10216_CS_line_fits} show the best-fit models and the observations, where we have chosen to present only a selection of the many extracted ALMA spectra. The comparison between modelled and observed integrated intensities for all extracted ALMA spectra is shown in Fig~\ref{fig:ALMA_radial_intensity_plots}. For the five modelled stars, we find $f_0$ in the range $1 - 4 \times 10^{-6}$ and $R_\mathrm{e}$ in the range $1.8 - 6.8 \times 10^{16}$ cm (see Table~\ref{tab:RT_modelling_results}).

For each grid of models, maps showing $\chi^2$ as a function of the two adjustable parameters, $f_0$ and $R_\text{e}$, with the best-fit model and corresponding 1$\sigma$, 2$\sigma$, and 3$\sigma$ uncertainty levels marked are shown in Fig.~\ref{fig:chi_sq_maps}. We verified that the best-fit models obtained by considering only the spatial information from the different ALMA apertures, and only the excitation information from the different transitions (the largest aperture ALMA spectrum and the SD lines), are consistent within their respective uncertainties, and show no signs of significant divergence (see Figs.~\ref{fig:10216_chi_sq_maps_comparison},~\ref{fig:chi_sq_contours_comparison_our_stars}). The results for each source are summarised below.
 
Overall, we obtain good fits to the CS line profiles for IRAS 15194$-$5115 (Fig.~\ref{fig:15194_CS_line_fits}), though the quality of the line intensity fits is not as good in comparison to those we get for the other stars in the sample. The precision with which the CS abundance can be constrained in the innermost regions of the CSE is limited by the large uncertainties on the high $J$ HIFI line intensities (Sect.~\ref{subsubsec:HIFI_spectral_survey}). We note that the underlying velocity profile used is also different from that of the other sources for this star (see Sect.~\ref{subsec:CO_Modelling_Results}). Our best-fit model reproduces the line profiles well, and gives a peak abundance of 3.1$\times$10$^{-6}$ and an e-folding radius of 6.8$\times$10$^{16}$ cm. However, this model underestimates the CS $J = 2 - 1$ line intensities from apertures smaller than $\sim$7$\arcsec$ by $\sim$10-50\%, as seen from Fig.~\ref{fig:ALMA_radial_intensity_plots}. Another model with $f_0$ = 4.4$\times$10$^{-6}$ and $R_\text{e}$ = 4.6$\times$10$^{16}$ cm, that falls within the 1$\sigma$ interval of the above-mentioned best-fit model manages to reproduce the line intensities from the inner part of the envelope better, but slightly underestimates the larger aperture ALMA spectra, while also significantly overestimating the higher-$J$ HIFI spectra ($J = 18 - 17$ to $22 - 21$). Our best-fit model also overestimates the APEX CS $J = 7 - 6$ line and a few of the high excitation HIFI lines, mainly $J = 18 - 17$ and $J = 21 - 20$ (see Fig.~\ref{fig:15194_CS_line_fits}). However, we obtain very good fits to the observed line profiles for ALMA apertures from $\sim$7$\arcsec$ onwards, up to around 35$\arcsec$, which encompasses all the detected emission for the $J = 2 - 1$ line (Fig.~\ref{fig:ALMA_radial_intensity_plots}), as well as most of our other APEX and HIFI lines (Fig.~\ref{fig:15194_CS_line_fits}). The complexity in modelling this source is further discussed in light of previous related works in Sect.~\ref{sec:Discussion}.

For IRAS 15082$-$4808 and IRAS 07454$-$7112, we obtain models that reproduce the line profiles from both the various ALMA apertures and the SD lines (see Figs.~\ref{fig:ALMA_radial_intensity_plots}, ~\ref{fig:15082_CS_line_fits},~\ref{fig:07454_CS_line_fits}), with the exception of the $J = 7 - 6$ line for IRAS 07454$-$7112, which our model underestimates (Fig.~\ref{fig:07454_CS_line_fits}). Though our model for IRAS 15082$-$4808 also underestimates the $J = 4 - 3$ line (Fig.~\ref{fig:15082_CS_line_fits}), the modelled spectra for this line falls within the uncertainty range of the observed spectrum (Sect.~\ref{subsubsec:APEX_spectral_survey}). 

For AFGL 3068, we are limited to SD spectra (Fig.~\ref{fig:3068_CS_line_fits}), as we do not have spatially resolved data available. Though we have the CS $J = 7 - 6$ line for this source observed with the ACA, we cannot extract much information about the radial distribution of the emission, as the CS emitting region is similar in size to the ALMA synthesised beam of the observation. We consequently treated this line as an SD observation in this work. The lack of spatial information and the small number of SD lines available imply that we cannot place very rigorous constraints on the abundance profile for this source as we have done in the case of the other four stars. This is reflected in the larger uncertainties of the estimated parameters (Table~\ref{tab:RT_modelling_results}) for this source.

For IRC~+10\,216, we discarded the smallest ALMA aperture (see Fig.~\ref{fig:10216_CS_line_fits}) from the $\chi^2$ minimisation, as it displays an asymmetric, secondary peak around the systemic velocity, which could be due to small scale asymmetries in the emission that cannot be modelled by our smooth, 1D, spherically symmetric RT code. We obtained a good fit to the rest of the spectra and were able to simultaneously fit the cumulative radial emission profile from the ALMA CS $J = 2 - 1$ line (Fig.~\ref{fig:ALMA_radial_intensity_plots}) and the SD line intensities. While the APEX CS $J = 5 - 4$ line from our spectral survey and the IRAM 30 m $J = 3 - 2$, $4 - 3$, $5 - 4$, and $6 - 5$ lines yield modelled spectra that match the observations within the corresponding uncertainties, our model overestimates the Yebes $J = 1 - 0$ line. We also note that the calibration uncertainties for the IRAM 30 m telescope rise significantly with frequency (see Sect.~\ref{subsubsec:Supplementary_CS_observations}).

\subsection{$^{13}$CS and C$^{34}$S}
\label{subsec:isotopologue_modelling_results}
Our models reproduce the observed line profiles for $^{13}$CS and C$^{34}$S for all five stars in the sample. For IRAS 15194$-$5115, IRAS 15082$-$4808, and IRAS 07454$-$7112, we have spatially resolved $J = 2 - 1$ ALMA maps for both isotopologues, so we were able to also model multi-aperture spectra (Sect.~\ref{subsubsec:ALMA_spectral_survey}) for these sources (Figs.~\ref{fig:13CS_ALMA_radial_intensity_plots},~\ref{fig:C34S_ALMA_radial_intensity_plots}), along with the available SD lines. For AFGL 3068 and IRC~+10\,216, we only had SD spectra for the isotopologues. The best-fit models and observed spectra are shown in Figs.~\ref{fig:13CS_line_profiles_15194} - \ref{fig:13CS_line_profiles_3068_and_10216} for $^{13}$CS, and Figs.~\ref{fig:C34S_line_profiles_15194} - \ref{fig:C34S_line_profiles_10216} for C$^{34}$S. The best-fit $f_0$ values obtained are given in Table~\ref{tab:isotopic_peak_abundances_and_ratios}. 

We note that for AFGL 3068, though we get good fits to the isotopologue line profiles using the peak abundance listed in Table~\ref{tab:isotopic_peak_abundances_and_ratios} and the $R_\text{e}$ value from the best-fit $^{12}$CS model (Table~\ref{tab:RT_modelling_results}), there are also other models, around $f_0$ = $2.3\times10^{-8}$ and $R_\text{e}$ = $6.0\times10^{16}$ cm (larger than the best-fit $^{12}$CS $R_\text{e}$ by a factor of $\sim$3), which yield equally good fits. Some models with this $R_\text{e}$ also fall within the 1$\sigma$ uncertainty range from our best-fit $^{12}$CS model (see Fig.~\ref{subfig:3068_chi_sq_map}), but the quality of our $^{12}$CS fits there is considerably lower than that of our best-fit model. This indicates that our $R_\text{e}$ estimate for AFGL 3068 may be underestimated, and the $f_0$ value correspondingly overestimated, which is also evident from the rather degenerate $\chi^2$ contours in Fig.~\ref{subfig:3068_chi_sq_map}.

We also calculated the $^{12}$CS/$^{13}$CS and C$^{32}$S/C$^{34}$S ratios from the peak abundances of the best-fit models (Table~\ref{tab:isotopic_peak_abundances_and_ratios}). The estimated isotopic ratios for AFGL 3068 should be regarded with caution as they suffer from the fact that the models are substantially less constrained compared to the other sources (Sect.~\ref{subsec:CS_Modelling_Results}).

\section{Discussion}
\label{sec:Discussion}

\subsection{Dust and CO}
\label{subsec:Dust_and_CO_discussion}
As noted in Sect.~\ref{subsec:Dust_Modelling}, we fixed the dust condensation radius ($R_\text{c}$) at 3$R_\star$ in our dust modelling, since, if left as a free parameter, its value becomes unreasonably large for a dust-driven wind. This issue can be seen in several previous works as well. \citet{Ramstedt_and_Olofsson_2014} and \citet{Ramstedt_et_al_2008} report values up to $\sim$46 au for M- and S-type stars and up to $\sim$20 au for C-type stars from dust RT models where the inner radius was a free parameter. In general, using a grain size distribution instead of a single fixed grain size may help alleviate this problem, while also approximating the circumstellar dust more realistically. This is, however, outside the scope of this work. Our interest in the dust here amounts to only finding a reasonable estimate of the dust radiation field to serve as input to our CO and CS RT models, rather than undertaking a comprehensive analysis focussed on the dust properties themselves. We obtained reasonable dust condensation temperatures in the range of $\sim$1500 - 1800 K, which is also in line with the $\sim$300 K uncertainty in $T_\text{c}$ estimated by \citet{Nanni_et_al_2016} for amorphous carbon dust.

Our estimates of the gas MLRs from CO RT modelling show very good agreement with those from the literature for all five sources, in most cases within a factor of ~$\sim$2-2.5 \citep[e.g.][]{Ramstedt_and_Olofsson_2014, Danilovich_et_al_2015, Schoier_et_al_2007, Woods_et_al_2003}. Considering this and the general predictions on the uncertainties of AGB MLR estimates \citep[factor of $\sim$3, see e.g.][]{Ramstedt_et_al_2008} from the literature, we estimate an uncertainty of a factor of $\sim$2-3 in our reported MLR values, assuming the input dust properties and CO abundance. The gas kinetic temperature profiles obtained are also well constrained, with all reasonable models in the grid falling within $\pm$1 step size ($\sim$250 K for $T_0$, 0.2 dex for $\alpha$, see Sect.~\ref{subsec:CO_Modelling}) of the best-fit model for all sources except IRAS 15194$-$5115, and within $\pm$2 step sizes of the best-fit model for IRAS 15194$-$5115. We note that these uncertainties on the MLRs and gas temperature profiles do not take into account the errors in the underlying dust properties and assumed canonical CO abundance, which cannot be realistically quantified.

As was noted in Sect.~\ref{subsec:CO_Modelling_Results}, we are unable to strictly constrain the gas velocity profile exponent, $\beta$, for sources other than IRAS 15194$-$5115. Observations of high-$J$ line emission are needed to properly characterise the gas velocity in the innermost regions of the CSE. Likewise, the gas kinetic temperature profiles are also uncertain in the innermost regions of the winds. For IRAS 15194$-$5115 and IRAS 07454$-$7112, where we have sufficiently high-$J$ HIFI observations, up to $J = 16 - 15$ and $J = 14 - 13$, respectively, all otherwise reasonable models in our grids systematically underestimate these lines. For the other sources, no high-excitation data are available to constrain the gas temperature in these regions. Nevertheless, the existing observations provide good constraints on the gas temperature profiles across most of the CSE, except for the innermost regions. High-angular-resolution observations that resolve the inner wind \citep[e.g.][]{Khouri_et_al_2024, Fonfria_et_al_2014,Fonfria_et_al_2019}, and deep observations of narrow molecular lines from vibrationally excited levels, originating in regions close to the star \citep[e.g.][]{Patel_et_al_2008} are needed to constrain the physical properties of the innermost parts of these CSEs.

\citet{Andriantsaralaza_et_al_2021} measured the extent of the CO emission using ALMA observations for several carbon stars including IRAS 07454$-$7112 and AFGL 3068. The measured extent of CO emission and the CO photodissociation radius (see Sect.~\ref{subsec:CO_Modelling}) can differ based on the excitation conditions in the CSE. These quantities will be compared for a number of carbon stars from the DEATHSTAR sample \citep{Ramstedt_et_al_2020, Andriantsaralaza_et_al_2021} in an upcoming paper (Andriantsaralaza et al., in prep.).

\subsection{CS}
\label{subsec:CS_discussion}
In this work, we employ the combined modelling of interferometric and SD data, to put rigorous constraints on the CS models using information about both the radial emission distribution and the conditions traced by different transitions simultaneously, for all sources in our sample except AFGL 3068 for which we have no spatially resolved data available. As was mentioned in Sect.~\ref{subsec:CS_Modelling_Results}, we reproduce the observed CS line profiles for all sources, and are able to robustly constrain their CS abundance profiles, with the exception of AFGL 3068, which presents comparatively large uncertainties. We find that CS maintains a significant gas-phase abundance relatively far out into the CSE, up to $\sim$3-7$\times$10$^{16}$ cm for all five sources. This is consistent with previous predictions from observations of IRC~+10\,216 \citep{Agundez_et_al_2012, Massalkhi_et_al_2019, Velilla-Prieto_et_al_2019}. No significant depletion onto the dust is seen. The e-folding radius increases with MLR as expected. The size of the molecular envelope of CS is set by its photodissociation into C and S by the ambient interstellar UV radiation field. We find no discernible trend in the CS peak abundances of our sources with envelope density beyond the estimated uncertainties. A correlation between CS abundance and CSE density was reported for S-type stars by \citep{Danilovich_et_al_2018}, who also, however, did not detect any such trend for carbon stars.

The CS envelope sizes and peak abundances estimated by \citet{Woods_et_al_2003} for our five sources using SD observations of CS $J = 2 - 1$ and $J = 5 - 4$  and simple photodissociation models match fairly well with those from this work. Our CS peak-abundance values are also consistent with estimates derived from large samples of C-type stars \citep[e.g.][]{Massalkhi_et_al_2019}. We also note that the LTE CS abundances calculated by \citetalias{Unnikrishnan_et_al_2024} using population diagrams are a factor of 2.5-6.5 higher than the non-LTE abundances estimated in this work. This highlights that LTE calculations can serve as simple, useful tools for obtaining first-order abundance estimates, but stricter constraints from non-LTE RT models are required for more precise calculations.

Our best-fit model for IRC$~+10\,216$ has a peak CS abundance approximately double than that calculated by \citet{Massalkhi_et_al_2024}, while the e-folding radii from the two estimates match within the 1$\sigma$ uncertainty level. The difference found in peak abundance is due to the fact that \citet{Massalkhi_et_al_2024} uses an MLR greater than ours by a factor of $\sim$1.6. We have tested the dependence of the derived peak abundances on the input gas MLR for our models, and found that varying the MLR by a factor of two on either side leads to roughly a factor of two change in the best-fit abundance in the opposite direction, indicating an inverse linear relationship as expected. CS in two of our sources (IRAS 07454$-$7112 and IRAS 15194$-$5115) was also modelled by \citet{Danilovich_et_al_2018} using APEX observations. For IRAS 07454$-$7112, we reduce the relative uncertainty in the CS e-folding radius by a factor of $\sim$2.5 with respect to \citet{Danilovich_et_al_2018}, and further provide well-constrained uncertainties on the peak abundance as well.

IRAS 15194$-$5115 is the source with the largest number of observed CS lines among the sample, owing to the lines from the HIFI survey (Sect.~\ref{subsubsec:HIFI_spectral_survey}).  It is known from the literature to be a difficult source to model, possessing a complex circumstellar morphology, with extensive substructure seen in molecular line emission \citepalias[see][]{Unnikrishnan_et_al_2024}, possibly caused by the presence of a binary companion \citep[e.g.][]{Feast_et_al_2003, Lykou_et_al_2018}. \citet{Danilovich_et_al_2018} reported that no satisfactory fit was found for the CS and SiS line profiles of this star using their 1D, constant MLR, smoothly accelerating RT model, which yielded good fits for the same species towards a number of other C-, M-, and S-type AGB stars. Our models with an underlying slow wind acceleration ($\beta$ = 5, instead of 1, see Sect.~\ref{subsec:CO_Modelling_Results}) gave acceptable fits to the CS line profiles of IRAS 15194$-$5115, though as mentioned in Sect.~\ref{subsec:CS_Modelling_Results}, the overall quality of the fits is lower compared to those obtained for the other sources, with difficulties in fitting the inner wind/high-$J$ and extended/low-$J$ emission simultaneously, similar to that reported by \citet{Danilovich_et_al_2018}.

However, owing to the large number of lines available to model for IRAS 15194$-$5115, it is clear that even though a few of the SD spectra are sporadically overestimated (Sect.~\ref{subsec:CS_Modelling_Results}), there appears to be no trends showing deviations consistently correlated with line excitation. Hence, the intermittent mismatches found between observations and models may be attributed, as indicated by \citet{Danilovich_et_al_2018}, to the complex morphology of the gas distribution in the CSE and its effects on line excitation, which cannot be accommodated in our RT model. The availability of spatially resolved observations across a large range of energy levels to constrain RT models alone will not be sufficient to tackle this issue. Also, using more complicated step functions in the inner envelope instead of typical Gaussian abundance profiles, as done by \citet{Danilovich_et_al_2019} for modelling SD observations of M-type stars, might not be completely straightforward in a case where spatially resolved observations sensitive to inner abundance changes are also available. To properly model this source would require RT models employing a more realistic representation of the underlying observed complex gas density structure, and chemical models that take into account the effects due to discrete variations in density and potential binarity \citep[e.g.][]{Cordiner_and_Millar_2009, Van_de_Sande_and_Millar_2022}. The possibility of employing 3D RT modelling \citep[e.g.][]{Homan_et_al_2017, Khouri_et_al_2024}, constrained by morphological information from the ALMA lines also remains to be explored.

\subsubsection{Spatial versus excitation constraints}
\label{subsubsec:spatial_vs_excitation_constraints}

\begin{figure}
    \centering
    \includegraphics[width=0.95\linewidth]{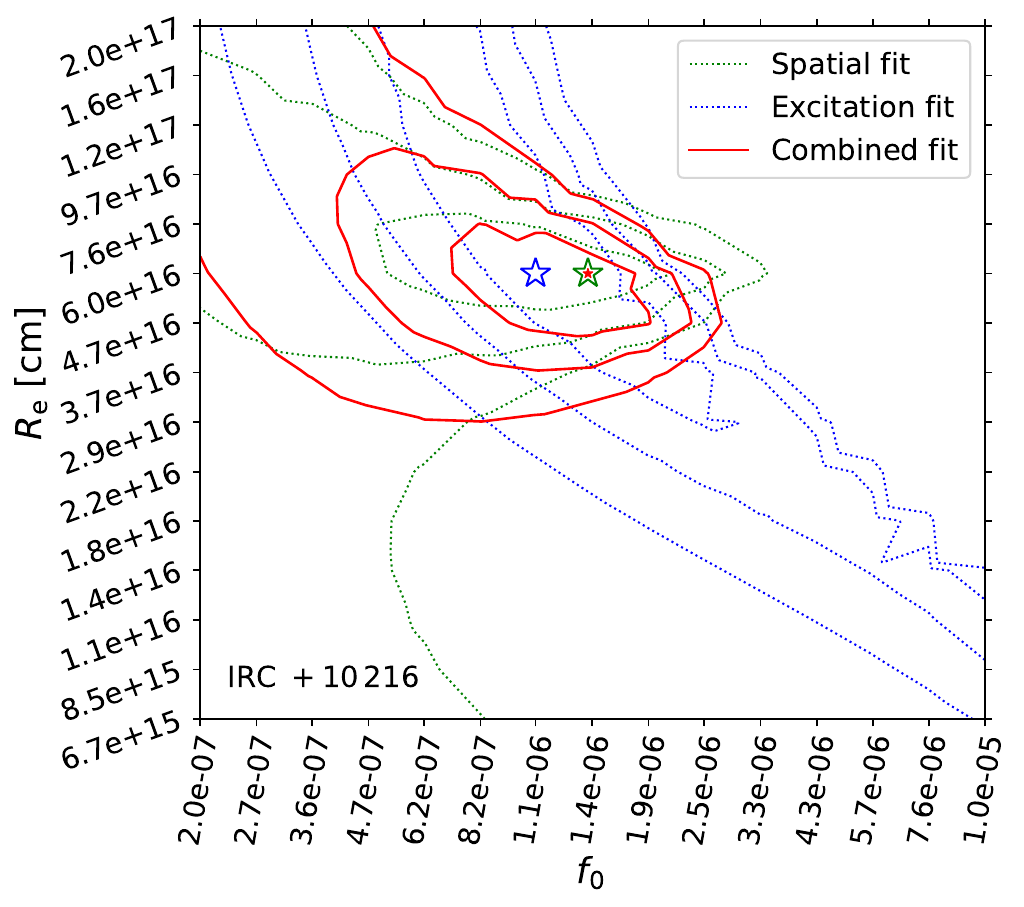}
    \caption{$\chi^2$ values for our grid of CS models for IRC$~+10\,216$. The contours denote 1$\sigma$, 2$\sigma$, and 3$\sigma$ ranges, when constrained using only the spatial information from the CS $J = 2 - 1$ line (green) vs just the excitation information from the different $J$ lines (blue). Also shown are the contours when using both the spatial and excitation information simultaneously (red) to constrain the RT models. The green, blue, and red stars mark the best-fit models estimated from the above three cases, respectively.} 
    \label{fig:10216_chi_sq_maps_comparison}
\end{figure}

\begin{figure}[t]
    \centering
    \begin{subfigure}[t]{0.5\textwidth}
        \centering
        \includegraphics[width=0.95\textwidth]{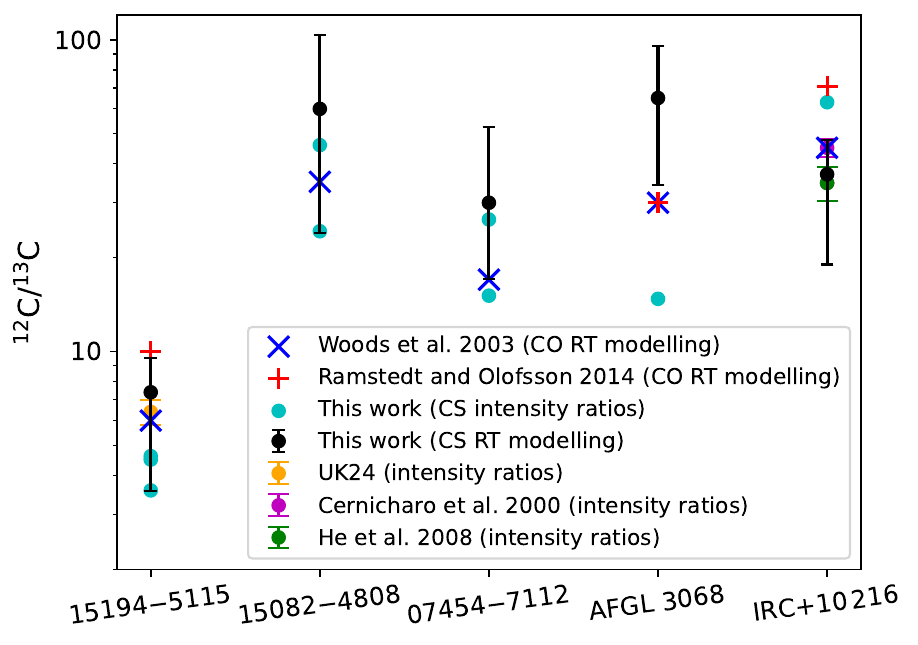}
        \caption{$^{12}$C/$^{13}$C isotopic ratio estimates}
        \label{subfig:C_isotopologue_ratios}
    \end{subfigure}

    \begin{subfigure}[t]{0.5\textwidth}
        \centering
        \includegraphics[width=0.95\textwidth]{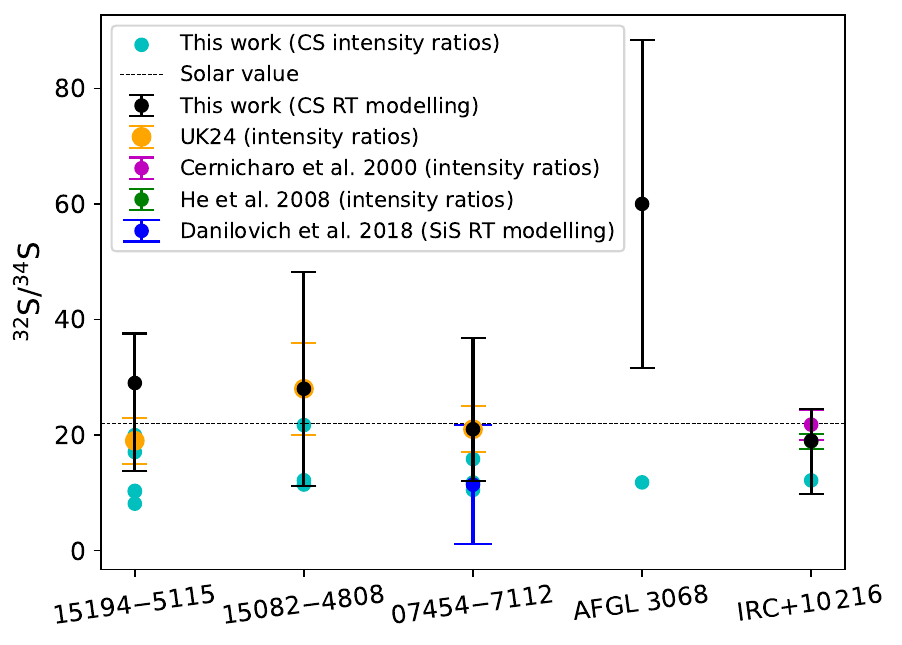}
        \caption{$^{32}$S/$^{34}$S isotopic ratio estimates}
        \label{subfig:S_isotopologue_ratios}
    \end{subfigure}
    \caption{Isotopic ratios from this work compared to estimates from the literature.}
    \label{fig:isotopologue_ratios}
\end{figure}

Fig.~\ref{fig:10216_chi_sq_maps_comparison} shows the difference in $\chi^2$ values obtained for our grid of CS RT models for IRC$~+10\,216$, when constrained using only the spatial information and only the excitation information separately, and also when using both simultaneously. This shows how well the spectra from various transitions and the spectra from different spatial apertures from the interferometric maps manage to independently constrain the CS abundance profiles. Similar figures for the other sources are shown in Fig.~\ref{fig:chi_sq_contours_comparison_our_stars}. The general pattern is similar for all sources that while the best-fit models obtained from both cases fall sufficiently close to each other, the excitation constraints do not manage to strongly break the degeneracy between the two free parameters, resulting in the classic diagonally elongated confidence contours in the $\chi^2$ maps (see Figs.~\ref{fig:10216_chi_sq_maps_comparison},~\ref{subfig:3068_chi_sq_map}). The spatial constraints from the resolved observations, however, limit the overall fit to a narrower range of the parameter space, leading to better estimates of the uncertainties on the best-fit values. This highlights the advantage of incorporating constraints based on the spatial distribution of the line emission to refine RT models.

\begin{figure}[t]
    \centering
    \includegraphics[width=0.95\linewidth]{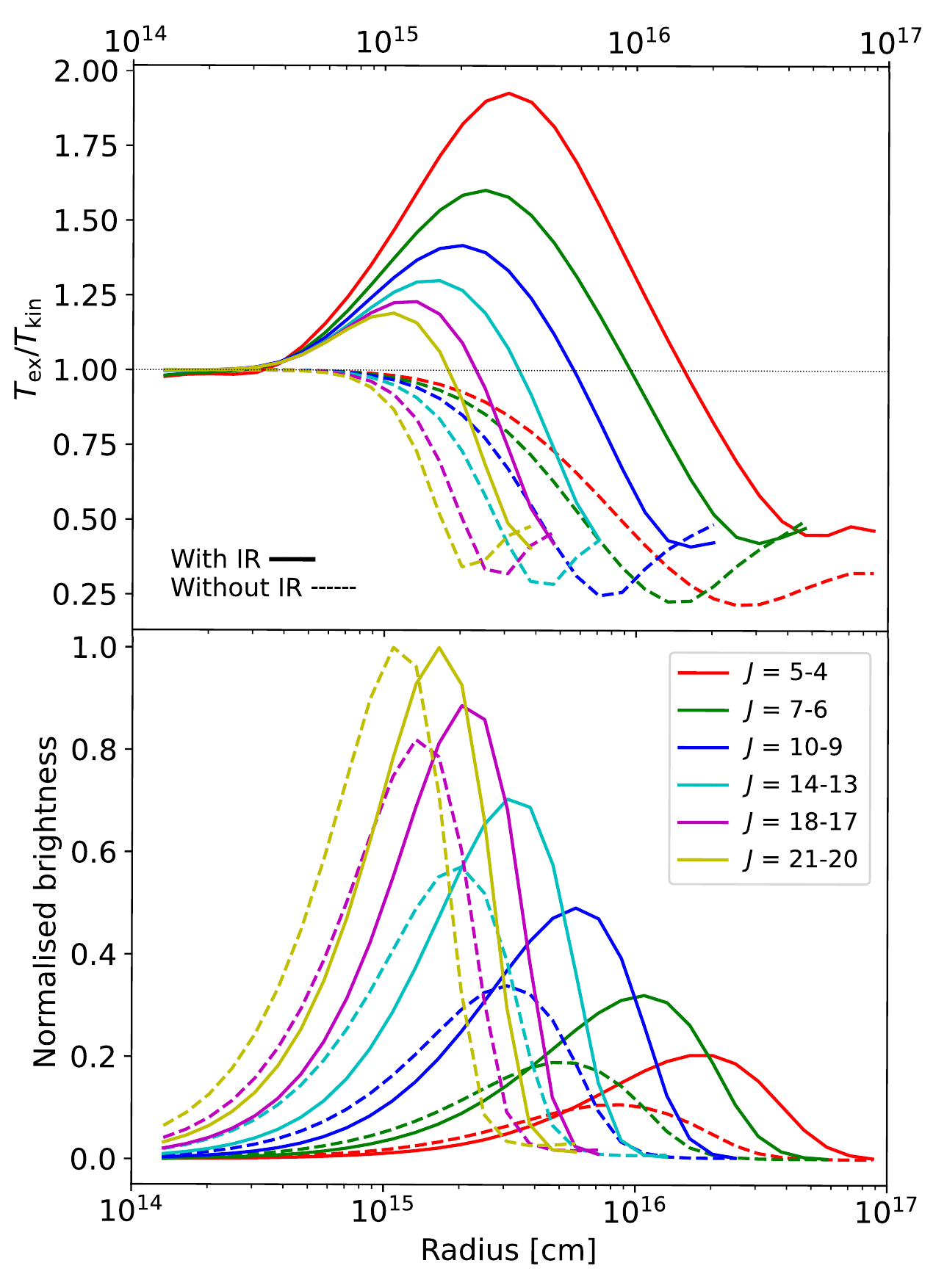}
    \caption{Ratio of the excitation temperature ($T_\mathrm{ex}$) and the kinetic temperature ($T_\mathrm{kin}$) for various low- and high-$J$ CS transitions (top), and the corresponding line emitting regions (bottom), normalised to the peak of the $J = 21 - 20$ emitting region, as functions of radius across the CSE, from the best-fit model for IRAS 15194$-$5115. See text for a more detailed description.}
    \label{fig:CS_temp_ratio_and_emitting_regions_15194}
\end{figure}

\subsubsection{Isotopic ratios}
\label{subsubsec:isotopic_ratios_discussion}
Fig.~\ref{fig:isotopologue_ratios} shows the $^{12}$CS/$^{13}$CS and C$^{32}$S/C$^{34}$S ratios obtained from this work (see Sect.~\ref{subsec:isotopologue_modelling_results} and Table~\ref{tab:isotopic_peak_abundances_and_ratios}) in comparison to various $^{12}$C/$^{13}$C and $^{32}$S/$^{34}$S isotopic ratio estimates from the literature. For all sources except AFGL 3068, the isotopologue abundance ratios we obtain match the isotopic ratios from the literature \citep[\citetalias{Unnikrishnan_et_al_2024}]{Cernicharo_et_al_2000, Woods_et_al_2003, Ramstedt_and_Olofsson_2014, He_et_al_2008, Danilovich_et_al_2018}. The derived $^{12}$CS/$^{13}$CS ratios are hence good measures of the $^{12}$C/$^{13}$C isotopic ratio, suggesting that isotopologue-selective effects are likely not significant in CS chemistry. We likely overestimate the $^{12}$C/$^{13}$C and $^{32}$S/$^{34}$S ratios for AFGL 3068, due to the poorly constrained abundance profiles, which may be overestimating the $^{12}$C$^{32}$S peak-abundance (Sect.~\ref{subsec:isotopologue_modelling_results}).

 IRAS 15194$-$5115 stands out from the rest of the sample with its exceptionally low $^{12}$C/$^{13}$C ratio of $\sim$7 (see Fig.~\ref{fig:isotopologue_ratios}), which is reproduced by our $^{12}$CS/$^{13}$CS abundance ratio as well. \citet{Ramstedt_and_Olofsson_2014} note that it is possible for J-type AGB stars with low MLRs to have $^{12}$C/$^{13}$C ratios as low as 3. Though IRAS 15194$-$5115 is a J-type star \citep{Smith_et_al_2015}, it has a high MLR \citep[see Table~\ref{tab:RT_modelling_results}; and][]{Woods_et_al_2003}. A reason for the very low $^{12}$C/$^{13}$C ratio can be hot-bottom burning (HBB), a process in which the bottom of the convective envelope becomes hot enough to start nuclear burning, destroying $^{12}$C and producing small amounts of $^{13}$C. Stars in the luminosity range -6.4 $<$ $M_\mathrm{bol}$ $<$ -6.3 may retain enough $^{12}$C to remain C-type even after undergoing HBB, resulting in such low $^{12}$C/$^{13}$C ratios \citep{Boothroyd_et_al_1993}. From this work, we however find a bolometric magnitude of -5.3 (Table~\ref{tab:RT_modelling_results}) for IRAS 15194$-$5115, in line with previous estimates \citep[e.g.][]{Woods_et_al_2003, Lykou_et_al_2018} as well. It is hence unclear why IRAS 15194$-$5115 has a peculiarly low $^{12}$C/$^{13}$C ratio, though possible explanations could include differential photodissociation due to the presence of a hot binary companion \citep{Vlemmings_et_al_2013}, extra mixing during the early-AGB phase or during the helium-core flash, or processes like cool bottom processing \citep[see][and references therein]{Ramstedt_and_Olofsson_2014}.

Our estimated C$^{32}$S/C$^{34}$S ratios are very close to the solar $^{32}$S/$^{34}$S ratio \citep[$\sim$22,][]{Asplund_et_al_2009}. This is expected, as $^{32}$S and $^{34}$S are not nucleosynthesised at any stage of low/intermediate-mass stellar evolution up to the AGB, and hence their ratio does not change significantly before or during the AGB phase \citep[e.g.][]{Karakas_and_Lugaro_2016}. These isotopes are mostly produced in supernovae, and also by oxygen-burning \citep[e.g.][]{Hughes_et_al_2008}. It is also interesting to note that the CS line intensity ratios (see Fig.~\ref{fig:isotopologue_ratios}) are always less than the corresponding abundance ratios from RT modelling (with the exception of $^{12}$CS/$^{13}$CS for IRC$+$10$\,$216). This is due to the high optical depths in the $^{12}$C$^{32}$S lines, which lead to suppressed line intensities.

\begin{figure}[t]
    \centering
    \includegraphics[width=\linewidth]{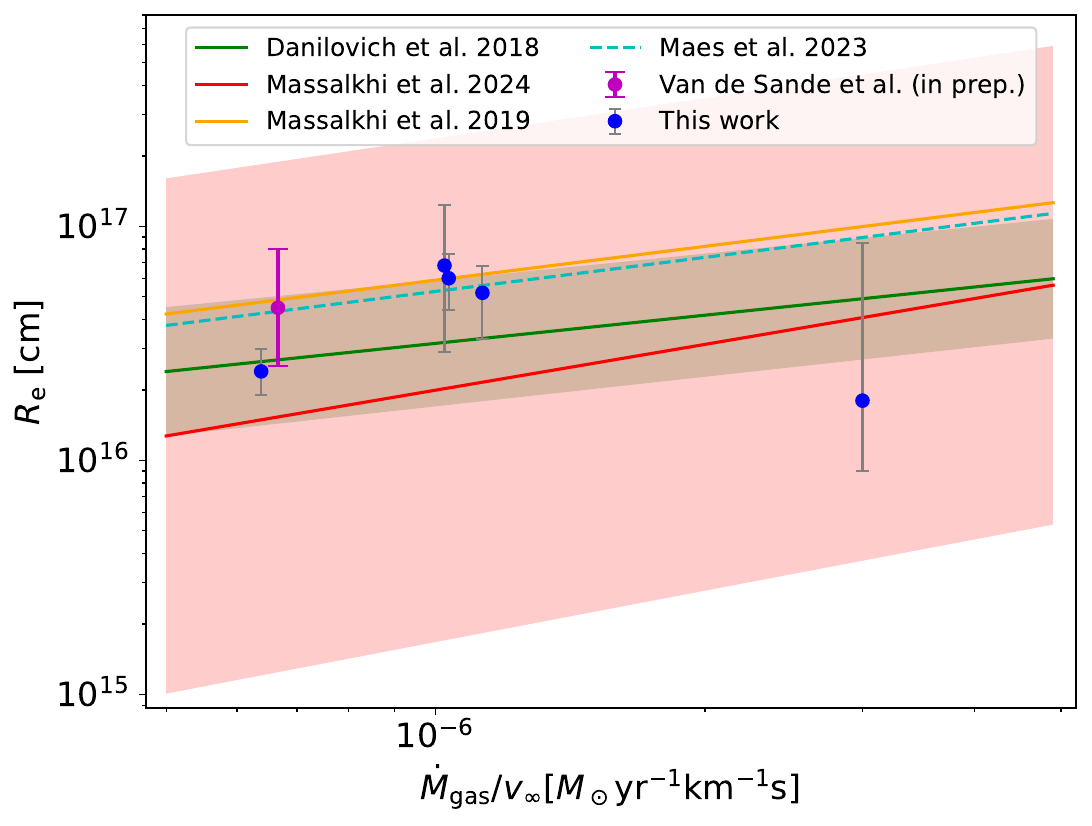}
    \caption{Variation in e-folding radius of the CS abundance profiles with circumstellar gas density. The dashed cyan line \citep{Maes_et_al_2023} is from chemical modelling results, while the other trends plotted are calculated from RT models. The magenta point is obtained from chemical modelling results (Van de Sande et al., in prep.), for a star of MLR 1.0$\times$10$^{-5}$ $M_\odot$ yr$^{-1}$ and a $\varv_\infty$ of 15 km s$^{-1}$, and the corresponding uncertainty range is estimated from sensitivity calculations on chemical models (see text). The rightmost blue circle is for AFGL 3068, and is likely underestimated (see text).}
    \label{fig:CS_Re_empirical_comp}
\end{figure}

\begin{figure*}[t]
    \centering
    \includegraphics[width=0.925\linewidth]{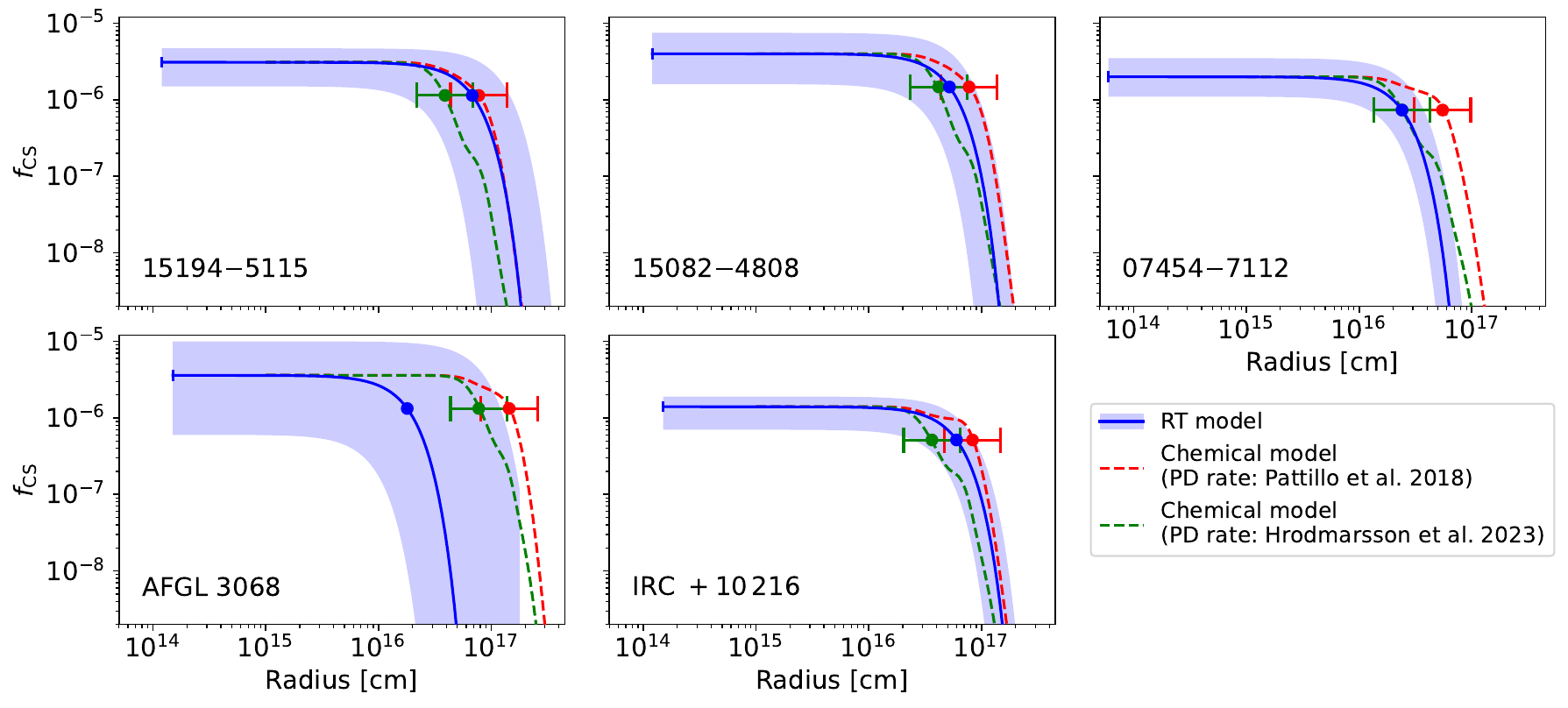}
    \caption{Comparison of CS abundance profiles obtained from RT modelling (blue) and chemical modelling: red - using CS photodissociation rate from \citet{Pattillo_et_al_2018}, green: using CS photodissociation rate calculated from the cross sections from \citet{Hrodmarsson_and_van_Dishoeck_2023}, based on \citet{Xu_et_al_2019}. The shaded blue region represents the 1$\sigma$ uncertainty range in the RT model abundance profiles. The circles denote the $R_\mathrm{e}$ of the respective models. The red and green error bars on the chemical model $R_\mathrm{e}$ denote the estimated uncertainty in the chemical models (Van de Sande et al., in prep.). It is assumed that this uncertainty does not vary significantly with the outflow density and photodissociation rate (see text).}
    \label{fig:CM_RT_detailed_comparison}
\end{figure*}

\subsubsection{A note on CS excitation: IR pumping}
\label{subsubsec:CS_excitation_discussion}
IR pumping refers to the phenomenon where molecules are excited to higher vibrational states by the absorption of IR photons from the dust or the central star, and then spontaneously decay to rotational levels in the ground vibrational state, increasing the populations in these levels. This can increase the line intensities of the  $\varv$ = 0 rotational lines, and has been shown to contribute significantly to the excitation of several species in CSEs \citep[e.g.][]{Agundez_and_Cernicharo_2006, Agundez_et_al_2012}. Using only the $\varv$ = 0 levels is equivalent to not including IR pumping in the models. To check the importance of infrared (IR) pumping in the excitation of CS in our models, we compared the ratio of the excitation temperatures to the gas kinetic temperature and the line emitting regions obtained when both $\varv$ = 0 and 1 levels are considered in the level population calculations, to when only the $\varv$ = 0 levels are used. The $T_\text{ex}/T_\text{kin}$ ratio is shown in Fig.~\ref{fig:CS_temp_ratio_and_emitting_regions_15194} for the best-fit CS model for IRAS 15194$-$5115, for various low- and high-$J$ transitions, along with their respective emitting regions. The emitting region for a line in the RT model is calculated by multiplying the brightness emitted from each radial shell (see Sect.~\ref{subsubsec:CS_modelling_procedure}) by the square of the radius at which the corresponding shell is located. The emitting region moves inwards as we go to higher-$J$ lines as expected, as these lines are excited at higher temperatures and densities compared to their lower-$J$ counterparts.

It is seen that IR pumping affects both the excitation and radial emission distribution in our models. The emitting regions become more compact with their peaks shifted inwards when IR pumping is turned off (Fig.~\ref{fig:CS_temp_ratio_and_emitting_regions_15194}). Not considering IR pumping also leads to a reduction in the observed line intensity for all the modelled CS lines by $\sim$50-70\%. This would lead to overestimating the peak abundances and e-folding radii, showing that it is crucial to include vibrationally excited levels in the calculation of level populations when modelling CS emission. The fact that the modelled line intensities did not change significantly upon testing different collisional rates (see Sect.~\ref{subsubsec:CS_molecular_data}) can also be indicative of the dominant role of radiative excitation in the formation of these lines. In the warm and dense inner part of the envelope, we have $T_\text{ex}$ $\approx$ $T_\text{kin}$, i.e. the rotational levels are thermalised (Fig.~\ref{fig:CS_temp_ratio_and_emitting_regions_15194}). Suprathermal excitation ($T_\text{ex}$ > $T_\text{kin}$) is found for all modelled lines as we move farther away from the star, which can be attributed to IR pumping becoming significant as the IR-emitting dust content builds up. At even larger radii, as the gas density decreases, the excitation eventually becomes subthermal ($T_\text{ex}$ < $T_\text{kin}$) for all the studied lines. These results are in line with the findings of \citet{Massalkhi_et_al_2019} from modelling CS $J = 3 - 2$ line emission for a large set of carbon stars.

\subsubsection{Comparisons with chemical models}
\label{subsubsec:chemical_model_comparisons}
The correlation between the radial extent of molecular abundance profiles and envelope density has been analysed for AGB CSEs by several authors based on RT models \citep{Danilovich_et_al_2018, Massalkhi_et_al_2019, Massalkhi_et_al_2024} and chemical models \citep{Maes_et_al_2023}. This has led to several empirical relations describing the increase in the $R_\mathrm{e}$ of various molecules including CS with increasing circumstellar density, represented by $\dot{M}/\varv_\infty$, where $\dot{M}$ is the gas MLR and $\varv_\infty$ is the terminal expansion velocity of the gas. In Fig.~\ref{fig:CS_Re_empirical_comp}, we explore how the $R_\mathrm{e}$ estimates from our models compare to empirical trends with $\dot{M}/\varv_\infty$ from the literature. We also show the uncertainty in the CS e-folding radii obtained from the chemical model, calculated by performing an uncertainty analysis where the uncertainties of all reaction rates in the \textsc{Rate22} chemical network \citep{Millar_et_al_2024} were propagated through the chemical model using a Monte Carlo sampling method for $\dot{M}$ = 1.0$\times$10$^{-5}$ $M_\odot$ yr$^{-1}$ and $\varv_\infty$ = 15 km s$^{-1}$ (Van de Sande et al., in prep.). Within the overall uncertainties, our results follow the trends from the literature well, and agree with the general trend where denser outflows show larger $R_\mathrm{e}$, indicating larger molecular envelopes. Based on these trends as seen from Fig.~\ref{fig:CS_Re_empirical_comp}, we expect the e-folding radius for the CS abundance profile of AFGL 3068 to increase when constrained better using additional observations, as also noted before (Sect.~\ref{subsec:isotopologue_modelling_results}).

In order to compare our retrieved radial extents of CS abundance profiles directly with predictions from chemical models, we ran chemical kinetic models specific to our sources using the publicly available, state-of-the-art UMIST Database for Astrochemistry \citep[UDfA\footnote{\url{https://umistdatabase.uk}}, ][]{Millar_et_al_2024} circumstellar chemical model\footnote{\url{https://github.com/MarieVdS/rate22_cse_code}}. Previous versions of this chemical model have been extensively employed to model various aspects of AGB circumstellar chemistry \citep[e.g.][]{Van_de_Sande_and_Millar_2019, Van_de_Sande_and_Millar_2022, Van_de_Sande_et_al_2019, Van_de_Sande_et_al_2020, Van_de_Sande_et_al_2021}. In this work, we used the recent \textsc{Rate22} UDfA update \citep{Millar_et_al_2024}, that includes updated rate coefficients for a total of 8767 reactions among 737 species. The 1D chemical kinetics code uses these rates to model the gas-phase chemistry in AGB outflows and derive molecular abundance profiles. It assumes a constant mass-loss rate and gas velocity, leading to a uniformly expanding, spherically symmetric CSE. The gas density in the envelope is assumed to fall as 1/$r^2$, where $r$ is the distance from the centre of the star. For a detailed description of the chemical model, we refer to \citet{Millar_et_al_2000, Cordiner_and_Millar_2009} and \citet{Van_de_Sande_et_al_2018_clumping_and_porosity}.

We adopted parent species abundances compiled by \citet{Agundez_et_al_2020} except for CS, with the addition of metals (Na, Mg, Fe) with initial abundances relative to H$_2$ set to 1.0$\times$10$^{-8}$, as a source of electrons. For each of our sources, we used the peak CS abundance obtained from our best-fit RT models as the initial abundance of CS in the chemical model. Other relevant source properties, including gas MLR and $\varv_\infty$ were taken from Tables~\ref{tab:source_properties} and \ref{tab:RT_modelling_results}. We tested both the \citet{Pattillo_et_al_2018} CS photodissociation rate, included in \textsc{Rate22}, and also the new, lower, rate calculated from the cross sections from \citet{Hrodmarsson_and_van_Dishoeck_2023}, based on \citet{Xu_et_al_2019}. The resulting CS abundance profiles corresponding to the two different photodissociation rates are shown in Fig.~\ref{fig:CM_RT_detailed_comparison}, along with those obtained from RT modelling. We found that varying the input initial CS abundance to the chemical model within the 1$\sigma$ range obtained from the RT models did not cause any significant change to the predicted $R_\mathrm{e}$ values.

We find that the chemical model abundance profiles match our CS abundance profiles from RT modelling within the estimated uncertainties. For three out of the five sources (IRAS 15194$-$5115, IRAS 15082$-$4808, and IRC$+$10$\,216$), both the photodissociation rates match the retrieved RT abundance profiles within 1$\sigma$ intervals. The deviation between the RT and chemical model $R_\mathrm{e}$ values is large in the case of AFGL 3068, owing to its poorly constrained RT models, that likely underestimate $R_\mathrm{e}$ (Sect.~\ref{subsec:isotopologue_modelling_results}). In the case of IRAS 07454$-$7112, we find that the chemical model using the CS photodissociation rate based on \citet{Hrodmarsson_and_van_Dishoeck_2023} matches our retrieved abundance profile, while that with the higher \citet{Pattillo_et_al_2018} rate overestimates the CS radial extent. Fig.~\ref{fig:CM_RT_detailed_comparison} also shows the uncertainty on the chemical model envelope sizes due to uncertainties in the kinetic data, calculated by Van de Sande et al. (in prep.), as described above. We note that this uncertainty on the predicted envelope sizes does not vary significantly with density, though lower densities show slightly larger error bars. This uncertainty calculation was done using the \citet{Pattillo_et_al_2018} rate. Assuming that the error bar on the predicted envelope size does not vary significantly when changing the CS photodissociation rate, we can say that abundance profiles based on the two rates both do a reasonable job overall at reproducing CS abundance profiles obtained by RT modelling, and match within the estimated uncertainty on the chemistry. Further, at high envelope densities, as is the case with our sources, photodissociation is often attenuated by the high dust extinction, irrespective of the employed photodissociation rate \citep[e.g.][]{Massalkhi_et_al_2024}.

The overestimation of CS radial extents by their photodissociation model compared to their RT modelling results described by \citet{Massalkhi_et_al_2024} using the \citet{Pattillo_et_al_2018} rate could atleast partly be due to the fact that their empirical fit to the CS e-folding radii includes both C-rich and O-rich outflows together. CS may have a slightly larger radial extent in C-rich envelopes compared to the O-rich ones, due to additional self-shielding against photodissociation due to its higher abundance in C-rich CSEs \citep[e.g.][]{Danilovich_et_al_2018}. We note that while the location of $R_\mathrm{e}$ is determined by photodissociation, its uncertainty also depends on the chemistry reforming the parent after photodissociation (Van de Sande et al., in prep.). To conclude, this implies that in the case of a parent species such as CS, where different photodissociation rates exist in the literature \citep[e.g.][]{Pattillo_et_al_2018, Xu_et_al_2019}, comparisons between the different photodissociation models should be done with caution, until proper uncertainties have been estimated.

Finally, we note that this study utilises observations from a large number of telescopes, each of which has its own calibration accuracy and uncertainties. This heterogeneous origin of the data used poses an inherent limitation to the precision with which conclusions can be drawn from the presented analysis. We have attempted to accommodate this as far as possible, by taking into account the estimated calibration uncertainties of the respective telescopes when assigning error bars on the line intensities. These uncertainties indeed propagate to the final estimates of the physical parameters and molecular abundances as well. The only way to make these calculations more precise is to avoid uncertainties in relative calibration by using the same telescope as far as possible, though this comes with obvious practical limitations, particularly given the crucial need for spatially resolved data.

\section{Summary and conclusions}
\label{sec:Summary_and_Conclusions}
We modelled the SEDs and CO line emission of our sample of five carbon-rich AGB stars, IRAS 15194$-$5115, IRAS 15082$-$4808, IRAS 07454$-$7112, AFGL 3068, and IRC~+10\,216, to consistently estimate their stellar parameters, dust radiation fields and mass-loss rates. With these results as inputs, we employed detailed non-LTE radiative transfer modelling to constrain the abundance profiles of CS across the five sources. We utilised both spatially resolved ALMA observations of the CS $J = 2 - 1$ line as well as a broad range of SD lines from APEX, \textit{Herschel}/HIFI, and the IRAM 30 m telescopes to simultaneously set spatial and excitation constraints on our RT models. In general, our models reproduce the observed CS line emission well and yield robust estimates of the CS abundance profiles. The heterogeneous origin of the observations used is a major source of uncertainty in the analysis. The calculated peak abundance and e-folding radius have comparatively larger error bars for AFGL 3068, due to the lack of spatially resolved observations, leading to less rigorous constraints on the models. We also determined the $^{12}$CS/$^{13}$CS and C$^{32}$S/C$^{34}$S isotopologue abundance ratios, which are in good agreement with the expected $^{12}$C/$^{13}$C and $^{32}$S/$^{34}$S isotopic ratios for our sources.

Our estimates of the radial extents of the CS abundance profiles follow the expected trends with envelope density. We find no significant difference beyond their respective uncertainties between the calculated CS peak abundances for the studied sources. Our results are also in good agreement with predictions of the radial extent of CS abundance profiles from advanced chemical models, though we cannot discern between the different predicted photodissociation rates of CS beyond the uncertainties involved. These results highlight the importance of leveraging both spatial and excitation information simultaneously in fine-tuning radiative transfer models. A similar analysis of other molecules, especially the parent species, informed by high-resolution observations spanning a large range of excitation conditions is the next step required to advance our understanding of the chemistry in AGB CSEs.


\begin{acknowledgements}
The authors sincerely thank the anonymous referee for their constructive feedback, which improved the quality and clarity of this paper.

We wish to thank P. Bergman for his support in setting up the ALI RT code. We thank M. Ag{\'u}ndez for providing Yebes 40 and IRAM 30 m CS line profiles of IRC~+10\,216, L. Velilla-Prieto for answering our queries regarding the combined ALMA - IRAM 30 m CS $J = 2-1$ data, and J. H. Black for discussions about collisional rates and molecular data files. We are grateful to D. Tafoya for his continuous support in ALMA data related tasks.

RU acknowledges data reduction support from the Nordic ALMA Regional Centre (ARC) node based at Onsala Space Observatory (OSO), Sweden. The Nordic ARC node is funded through Swedish Research Council grant No 2017-00648. 

MA acknowledges support from the Olle Engkvist Foundation
under project 229-0368. 

EDB acknowledges financial support from the Swedish National Space Agency. 

TD is supported in part by the Australian Research Council through a Discovery Early Career Researcher Award (DE230100183).

MVdS acknowledges support from the Oort Fellowship at Leiden Observatory. 

TJM’s research at QUB is supported by grant ST/T000198/1 from the STFC.

The work of MGR is supported by NOIRLab, which is managed by the Association of Universities for Research in Astronomy (AURA) under a cooperative agreement with the National Science Foundation.

This paper makes use of the following ALMA data: ADS/JAO.ALMA\#2013.1.00070.S, ADS/JAO.ALMA\#2015.1.01271.S, ADS/JAO.ALMA\#2017.1.00595.S, ADS/JAO.ALMA\#2013.1.00432.S. ALMA is a partnership of ESO (representing its member states), NSF (USA) and NINS (Japan), together with NRC (Canada), MOST and ASIAA (Taiwan), and KASI (Republic of Korea), in cooperation with the Republic of Chile. The Joint ALMA Observatory is operated by ESO, AUI/NRAO and NAOJ. 

This paper is based on observations with the Atacama Pathfinder EXperiment (APEX) telescope. APEX is a collaboration between the Max Planck Institute for Radio Astronomy, the European Southern Observatory, and the Onsala Space Observatory. Swedish observations on APEX are supported through Swedish Research Council grant No 2017-00648. The APEX observations were obtained under project numbers O-0107.F-9310 (SEPIA/B5), O-0104.F-9305 (PI230), and O-087.F-9319, O-094.F-9318, O-096.F-9336, and O-098.F-9303 (SHeFI). 

HIFI has been designed and built by a consortium of institutes and university departments from across Europe, Canada, and the United States (NASA) under the leadership of SRON, Netherlands Institute for Space Research, Groningen, The Netherlands, and with major contributions from Germany, France and the US. Consortium members are Canada: CSA, U. Waterloo; France: CESR, LAB, LERMA, IRAM; Germany: KOSMA, MPIfR, MPS; Ireland: NUI Maynooth; Italy: ASI, IFSI-INAF, Osservatorio Astrofísico di Arcetri-INAF; The Netherlands: SRON, TUD; Poland: CAMK, CBK; Spain: Observatorio Astronómico Nacional (IGN), Centro de Astrobiología (INTA-CSIC); Sweden: Chalmers University of Technology – MC2, RSS \& GARD, Onsala Space Observatory, Swedish National Space Board, Stockholm University – Stockholm Observatory; Switzerland: ETH Zurich, FHNW; USA: CalTech, JPL, NHSC. 

This work is based on observations carried out with the IRAM 30 m and the Yebes 40 m telescopes. IRAM is supported by INSU/CNRS (France), MPG (Germany) and IGN (Spain). The Yebes 40 m telescope at Yebes Observatory is operated by the Spanish Geographic Institute (IGN, Ministerio de Transportes, Movilidad y Agenda Urbana).

This work has made use of data from the European Space Agency (ESA) mission Gaia\footnote{\url{https://www. cosmos.esa.int/gaia/}}, processed by the Gaia Data Processing and Analysis Consortium\footnote{\url{https://www.cosmos.esa.int/web/gaia/dpac/consortium/}} (DPAC). Funding for the DPAC has been provided by national institutions, in particular the institutions participating in the Gaia Multilateral Agreement.

This publication makes use of data products from the Two Micron All Sky Survey\footnote{\url{https://irsa.ipac.caltech.edu/Missions/2mass.html}} (2MASS), which is a joint project of the University of Massachusetts and the Infrared Processing and Analysis Center/California Institute of Technology, funded by the National Aeronautics and Space Administration and the National Science Foundation.

This work has made use of observations with AKARI\footnote{\url{https://www.isas.jaxa.jp/en/missions/spacecraft/past/akari.html}}, a JAXA project with the participation of ESA.

This research uses data from the Diffuse Infrared Background Experiment\footnote{\url{https://lambda.gsfc.nasa.gov/product/cobe/about_dirbe.html}} (DIRBE) instrument on the Cosmic Background Explorer (COBE). NASA Goddard Space Flight Center developed the DIRBE data sets under the direction of the COBE Science Working Group.

This publication makes use of data from the IRAS point source catalogue\footnote{\url{https://cdsarc.cds.unistra.fr/viz-bin/cat/II/125/}}. The Infrared Astronomical Satellite (IRAS) was a joint project of the US, UK and the Netherlands.

This research has made use of the International Variable Star Index\footnote{\url{https://www.aavso.org/vsx/}} (VSX) database, operated at AAVSO, Cambridge, Massachusetts, USA. 

This research has made use of NASA’s Astrophysics Data System\footnote{\url{https://ui.adsabs.harvard.edu}} (ADS). 

This project has made use of the VizieR\footnote{\url{https://vizier.cds.unistra.fr}} catalogue access tool and the SIMBAD\footnote{\url{https://simbad.cds.unistra.fr/simbad/}} database, operated at CDS, Strasbourg, France. 

This work made use of Astropy:\footnote{\url{https://www.astropy.org}} a community-developed core Python package and an ecosystem of tools and resources for astronomy \citep{astropy:2013, astropy:2018, astropy:2022}. 

This work has made use of GILDAS\footnote{\url{http://www.iram.fr/IRAMFR/GILDAS/}} and CASA\footnote{\url{https://casa.nrao.edu}} software to reduce and analyse data. 

The computations in this work were run in the Vera HPC cluster using resources provided by C3SE, the Chalmers e-Commons e-Infrastructure group at Chalmers University of Technology, Gothenburg, Sweden.
\end{acknowledgements}

\bibliographystyle{aa}
\bibliography{aa54996-25}

\begin{appendix}
\onecolumn
\section{Photometric fluxes and best-fit SEDs}
\label{app:appendix_A}
The photometric flux densities (see Sect.~\ref{subsec:SED_Observations}) for the five stars used for SED fitting are listed in Table~\ref{tab:photometric_fluxes_SED}. The dust radiative transfer modelling procedure adopted is explained in Sect.~\ref{subsec:Dust_Modelling}, and the results are presented in Sect.~\ref{subsec:Dust_Modelling_Results}. See Table~\ref{tab:RT_modelling_results} for the stellar and dust parameters calculated from the SED fitting. The best-fit SEDs obtained from the modelling are shown below in Fig.~\ref{fig:best_fit_SEDs}.

{\renewcommand{\arraystretch}{1.1}%
 \begin{table*}[h]
    \caption{{Photometric flux measurements used in SED fitting}}
    \label{tab:photometric_fluxes_SED}
    \centering
    \begin{tabular}{rccccc}
    \hline \hline
    $\lambda$ [$\mu$m] & \multicolumn{5}{c}{Flux [Jy]} \\
    \cline{2-6} & \\[-2ex]
   & 15194$-$5115 & 15082$-$4808 & 07454$-$7112 & AFGL 3068 & IRC$~+10\,216$ \\
    \hline & \\[-2ex]  
0.5$^{(a)}$ & 1.4$\times 10 ^{-4}$ & 1.1$\times 10 ^{-4}$ & 2.09$\times 10 ^{-3}$ & - & 1.3$\times 10 ^{-4}$ \\
0.58$^{(a)}$ & 6.99$\times 10 ^{-3}$ & 1.19$\times 10 ^{-4}$ & 5.30$\times 10 ^{-2}$ & - & 5.97$\times 10 ^{-4}$ \\
0.72$^{(a)}$ & 2.98$\times 10 ^{-2}$ & 2.15$\times 10 ^{-2}$ & 17.40$\times 10 ^{-2}$ & - & 2.73$\times 10 ^{-2}$ \\
1.24$^{(b)}$ & 3.69 & 8.19$\times 10 ^{-2}$ & 2.54 & 2.15$\times 10 ^{-4}$ & 2.67 \\
1.65$^{(b)}$ & 26.1 & 8.33$\times 10 ^{-1}$ & 21 & 7.26$\times 10 ^{-4}$ & 76.9 \\
2.16$^{(b)}$ & 100 & 5.86 & 95.6 & 4.76$\times 10 ^{-2}$ & 475 \\
3.5$^{(c)}$ & 295 & 139 & 283 & 12 & - \\
4.9$^{(c)}$ & 693 & 405 & 583 & 61 & - \\
11.6$^{(d)}$ & 1.32$\times 10 ^{3}$ & 793 & 613 & 707 & 4.75$\times 10 ^{4}$  \\
18.4$^{(e)}$ & 526 & 764 & 402 & 620 & - \\
23.9$^{(d)}$ & 562 & 424 & 308 & 776 & 2.31$\times 10 ^{4}$  \\
61.9$^{(d)}$ & 145 & 96.2 & 65.7 & 248 & 5.65$\times 10 ^{3}$  \\
65$^{(e)}$ & 108 & 94.8 & 52.74 & 201.3 & - \\
90$^{(e)}$ & 62.2 & 53.6 & 29.34 & 114.6 & - \\
100.2$^{(d)}$ & 51.0 & 27.8 & 21.7 & 73.7 & 922 \\
140$^{(e)}$ & 16.1 & 14.8 & 9.06 & 33 & - \\
160$^{(e)}$ & 15.0 & 10.6& - & 24.4 & - \\
350$^{(f)}$ & - & - & - & - & 63.8 \\
447$^{(g)}$ & - & - & - & - & 17.8 \\
550$^{(f)}$ & - & - & - & - & 24.4 \\
855$^{(g)}$ & - & - & - & - & 9.28 \\    
 \hline
    \end{tabular}
    \tablefoot{$^{(a)}$Gaia, $^{(b)}$2MASS, $^{(c)}$DIRBE, $^{(d)}$IRAS, $^{(e)}$AKARI, $^{(f)}$Planck, $^{(g)}$SCUBA. See Sect.~\ref{subsec:SED_Observations} for references.}
 \end{table*}
}

\begin{figure}[H]
    \centering
    \includegraphics[width=0.95\linewidth]{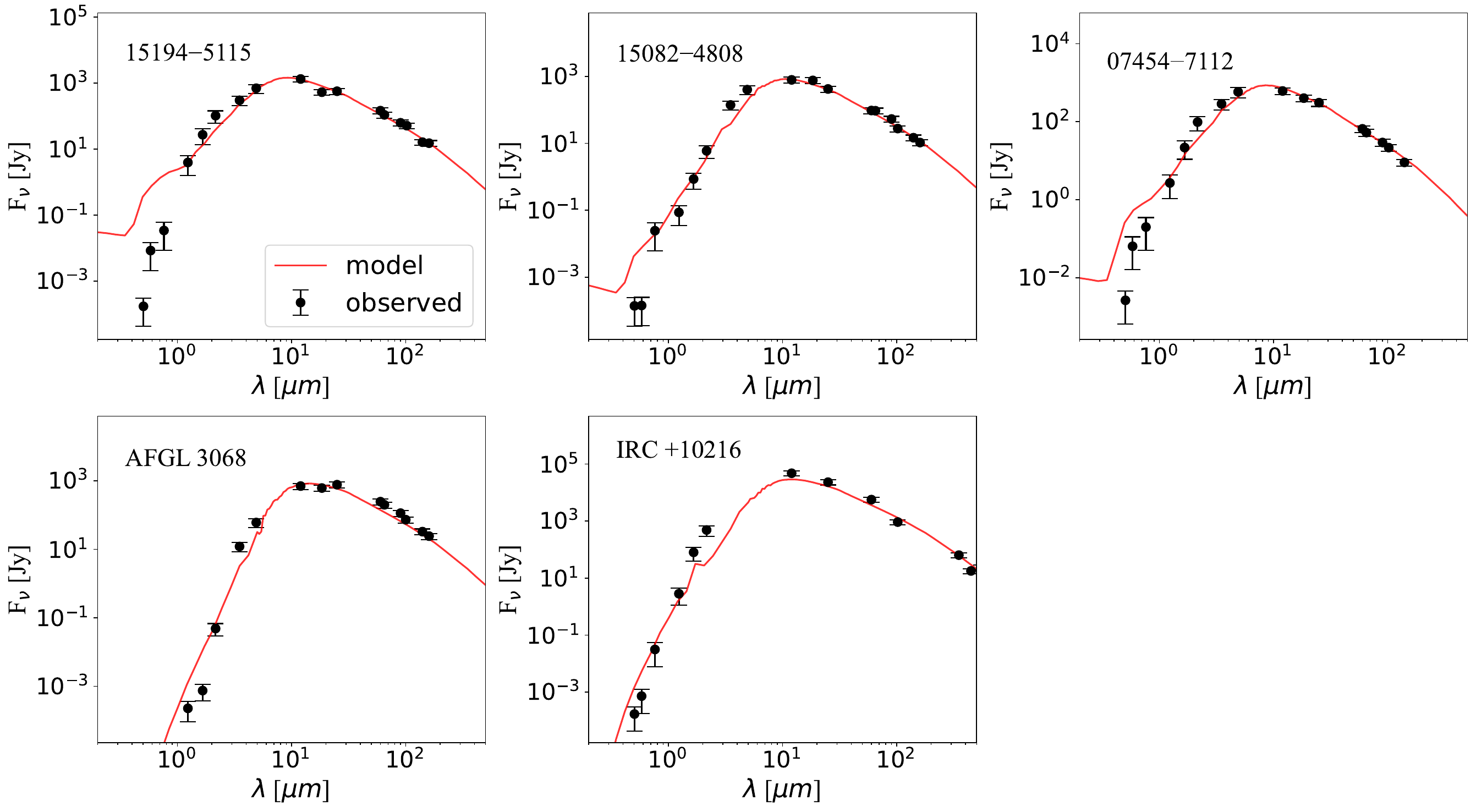}
    \caption{Photometric data (black, Sect.~\ref{subsec:SED_Observations}) and modelled SEDs (red, Sect.~\ref{subsec:Dust_Modelling}).}
    \label{fig:best_fit_SEDs}
\end{figure}

\section{CO line intensities and best-fit models}
\label{app:appendix_B}
This appendix presents the velocity-integrated CO line intensities used to constrain our gas radiative transfer models. The line intensities from the SD observations (Sect.~\ref{subsec:CO_Observations}) used are listed in Table~\ref{tab:CO_int_ints_SD} for all five sources, and the integrated intensities for the spectra from different apertures extracted from the ACA lines (Sect.~\ref{subsec:CO_Observations}) for IRAS 07454$-$7112 and AFGL 3068 are given in Table~\ref{tab:CO_int_ints_ACA}. The CO modelling is described in Sect.~\ref{subsec:CO_Modelling}, and the corresponding results are presented in Sect.~\ref{subsec:CO_Modelling_Results}.

 \begin{table*}[h]
    \caption{{SD CO integrated intensities used to constrain RT models}}
    \label{tab:CO_int_ints_SD}
    \centering
    \begin{adjustbox}{width=18cm}
    \begin{tabular}{lccccccc}
    \hline \hline
    Line & Telescope & Beam size ["] & \multicolumn{5}{c}{$\int T_\mathrm{MB}~d\varv$ [K km s$^{-1}$]} \\
    \cline{4-8} & \\[-2ex]
   & & & 15194$-$5115 & 15082$-$4808 & 07454$-$7112 & AFGL 3068 & IRC$~+10\,216$ \\
    \hline & \\[-2ex]  
        1-0 & IRAM & 21.4 & - & - & - & 101.41$^4$ & 411.94$^4$ \\
        1-0 & OSO & 33 & - & - & - & 26.97$^4$; 48$^8$ & 386$^8$ \\
        1-0 & SEST & 45 & 61$^1$ & 42.4$^{10}$; 36.71$^{11}$ & 25.57$^{11}$ & - & - \\
        1-0 & NRAO & 55 & - & - & - & - & 222.79$^{12}$ \\
        2-1 & IRAM & 10.7 & - & - & - & 118.99$^4$ & 830.58$^4$ \\
        2-1 & JCMT & 21 & - & - & - & 72$^8$ & 689$^8$ \\
        2-1 & SEST & 23 & 151$^1$ & 69.4$^{11}$ & 50$^{11}$ & - & 491.83$^4$ \\
        2-1 & APEX & 27 & 116.32$^2$; 128.12$^3$ & 81.91$^2$ & 55.83$^2$; 56.5$^3$ & 59.69$^2$ & 608.12$^2$ \\
        3-2 & JCMT & 14 & - & - & - & 72$^8$ & 1070$^8$; 852.24$^{12}$ \\
        3-2 & SEST & 15 & 129.3$^1$ & - & - & - & 854.8$^4$ \\
        3-2 & APEX & 18 & 126.96$^1$, 122.71$^3$, 144.4$^4$; 163$^5$ & 90.34$^2$ & 57.41$^2$ & 65.82$^2$; 76.16$^4$ & 752.92$^2$ \\
        3-2 & CSO & 20 & - & - & - & 40.23$^9$ & 672.09$^9$ \\
        4-3 & JCMT & 12 & - & - & - & 93$^8$ & 1230$^8$ \\
        4-3 & APEX & 14 & 134.05$^2$; 155.9$^4$; 128.4$^4$ & 85.67$^2$ & 53.63$^2$ & 72.32$^2$; 87.58$^4$ & 711.92$^2$ \\
        4-3 & CSO & 15.5 & - & - & - & 22.4$^9$ & 639.17$^9$ \\
        5-4 & HIFI & 36.1 & 16.8$^6$ & - & 6.87$^6$ & - & - \\
        6-5 & JCMT & 8 & - & - & - & 55$^8$ & - \\
        6-5 & CSO & 10.3 & - & - & - & 22.4$^9$ & 1130.61$^9$ \\
        6-5 & HIFI & 30.4 & 20.2$^6$ & - & - & - & - \\
        7-6 & APEX & 7.7 & 82.1$^4$; 153.43$^4$ & - & - & 152.69$^4$ & - \\
        9-8 & HIFI & 20.1 & 17.6$^6$ & - & 8.12$^6$ & - & - \\
        10-9 & HIFI & 18.2 & 19.4$^7$ & - & - & - & - \\
        14-13 & HIFI & 12.9 & - & - & 7.62$^6$ & - & - \\
        16-15 & HIFI & 11.5 & 17.2$^7$ & - & - & - & - \\
 \hline
    \end{tabular}
    \end{adjustbox}
    \tablefoot{$^1$\citet{Ryde1999}, $^2$APEX pointing catalogue,$^3$APEX Archive\footnote{\url{https://archive.eso.org/wdb/wdb/eso/apex/form}}, $^4$\citet{De_Beck_et_al_2010}, $^5$\citet{Ramstedt_and_Olofsson_2014}, $^6$\citet{Danilovich_et_al_2015}, $^7$HIFI Archive\footnote{\url{https://archives.esac.esa.int/hsa/whsa/}}, $^8$\citet{Ramstedt_et_al_2008}, $^9$\citet{Teyssier2006}, $^{10}$\citet{Nyman1992}, $^{11}$\citet{Woods_et_al_2003}, $^{12}$\citet{DeBeck2012}. The integrated intensities reported are in the main-beam temperature ($T_\mathrm{MB}$) scale.}
 \end{table*}

 {\renewcommand{\arraystretch}{0.985}
 \begin{table*}[h]
    \caption{{ACA CO integrated intensities extracted from different apertures used to constrain RT models}}
    \label{tab:CO_int_ints_ACA}
    \centering
    \begin{tabular}{cc@{\extracolsep{10pt}}cc}
    \hline \hline & \\[-2ex]
    \multicolumn{2}{c}{IRAS 07454$-$7112} & \multicolumn{2}{c}{AFGL 3068} \\
    \cline{1-2} \cline{3-4} & \\[-2ex]
   \makecell{Aperture size \\\ ["]} & \makecell{$\int I~d\varv$  \\\ [Jy km s$^{-1}$]} & \makecell{Aperture size \\\ ["]} & \makecell{$\int I~d\varv$ \\\ [Jy km s$^{-1}$]} \\ & \\ [-2ex]
    \hline & \\[-2ex]  
       \multicolumn{4}{c}{CO $J = 2 - 1$} \\
        5.93 & 494.61 & 7.30 & 600.10 \\
        8.90 & 765.27 & 11.00 & 822.63 \\
        11.87 & 966.41 & 14.70 & 992.45 \\
        14.84 & 1104.18 & 18.40 & 1121.71 \\
        17.81 & 1202.16 & 22.10 & 1212.02 \\
        20.78 & 1242.66 & 25.80 & 1269.86 \\
        23.75 & 1265.43 & 29.50 & 1304.40 \\
        26.72 & 1261.46 & 33.20 & 1323.68 \\
        -&- & 36.90 & 1333.50 \\
        \multicolumn{4}{c}{CO $J = 3 - 2$} \\
        4.13 & 785.51 & 5.00 & 908.45 \\
        6.20 & 1154.57 &  7.50 & 1149.89\\
        8.27 & 1384.62 & 10.00 & 1282.87 \\
        10.34 & 1453.46 & 12.50 & 1344.00 \\
        12.41 & 1451.51 & 15.00 & 1360.26\\

 \hline
    \end{tabular}
 \end{table*}}

\begin{figure}[H]
    \centering
    \includegraphics[width=0.75\linewidth]{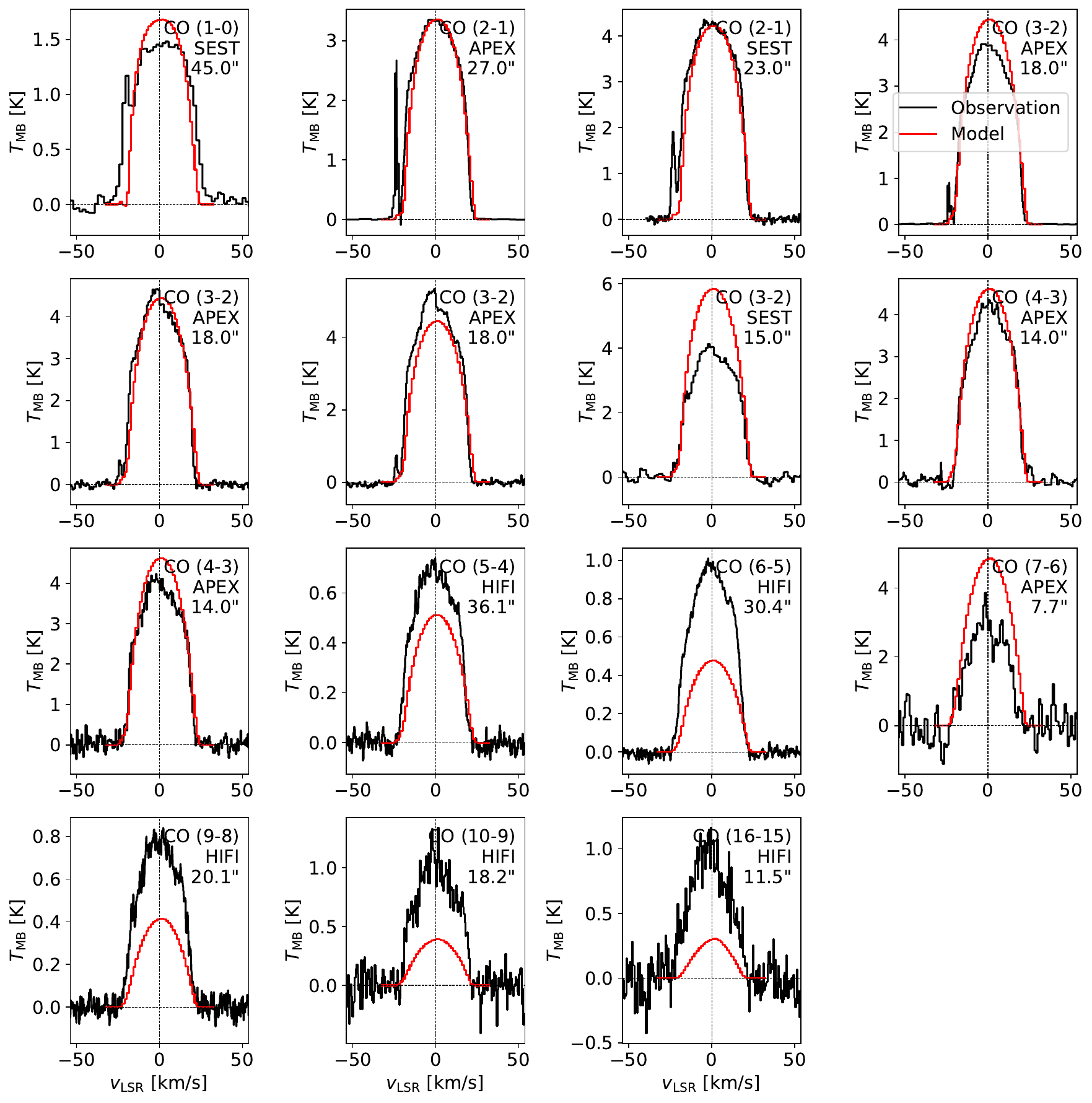}
    \caption{Observed (black, Sect.~\ref{subsec:CO_Observations}) and modelled (red, Sect.~\ref{subsec:CO_Modelling}) CO line profiles for IRAS 15194$-$5115. The transition quantum numbers, telescope used for the observation, and the corresponding beam size in arcseconds are listed at the top right corner of each panel. The lines are in the main-beam temperature ($T_\mathrm{MB}$ [K]) scale for all SD spectra shown, whereas the flux is given in units of Jansky for the ALMA lines where available.}
    \label{fig:CO_line_profiles_15194}
\end{figure}

\begin{figure}[h]
    \centering
    \includegraphics[width=0.575\linewidth]{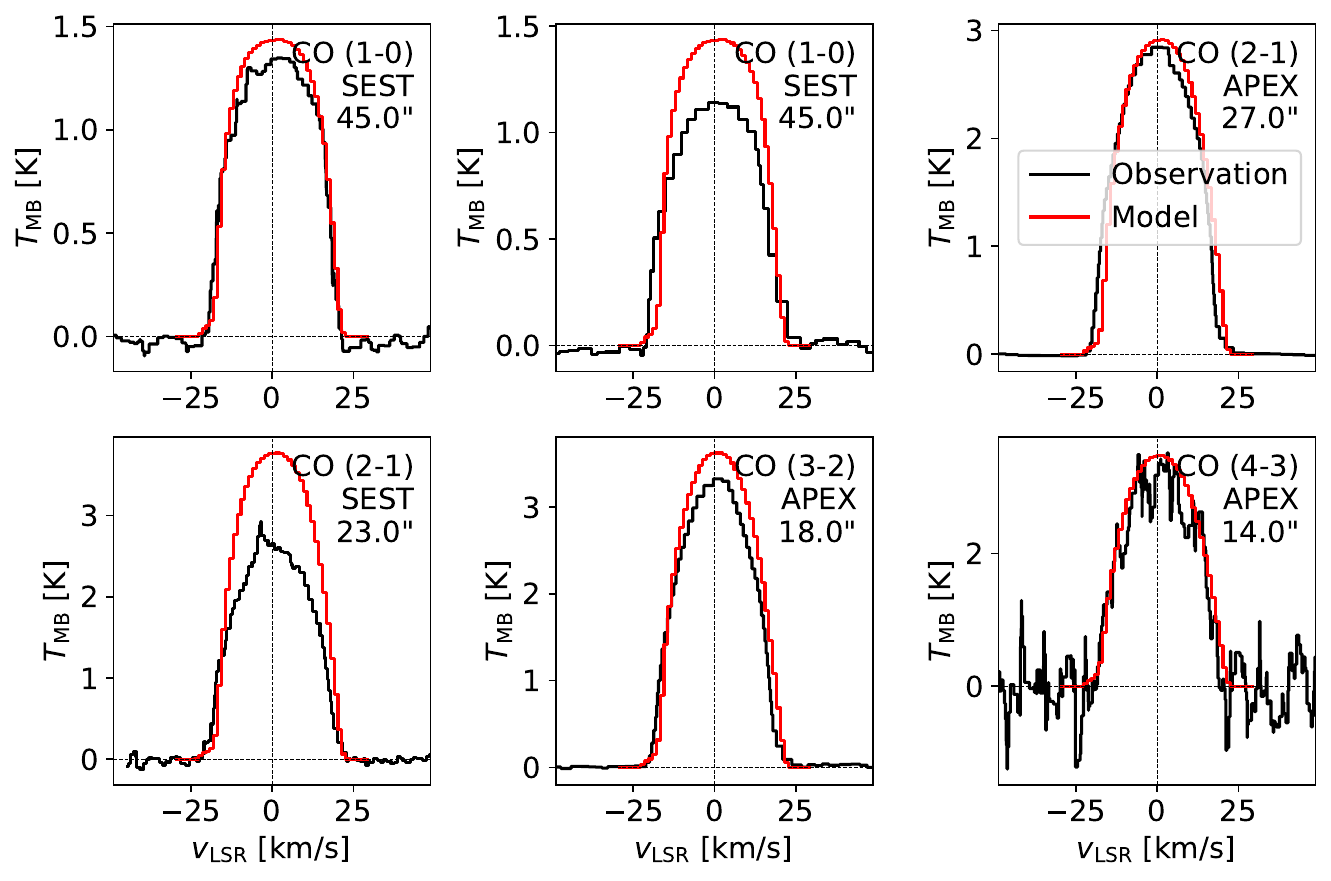}
    \caption{Same as Fig.~\ref{fig:CO_line_profiles_15194}, but for IRAS 15082$-$4808.}
    \label{fig:CO_line_profiles_15082}
\end{figure}

\begin{figure}[h]
    \centering
    \includegraphics[width=0.77\linewidth]{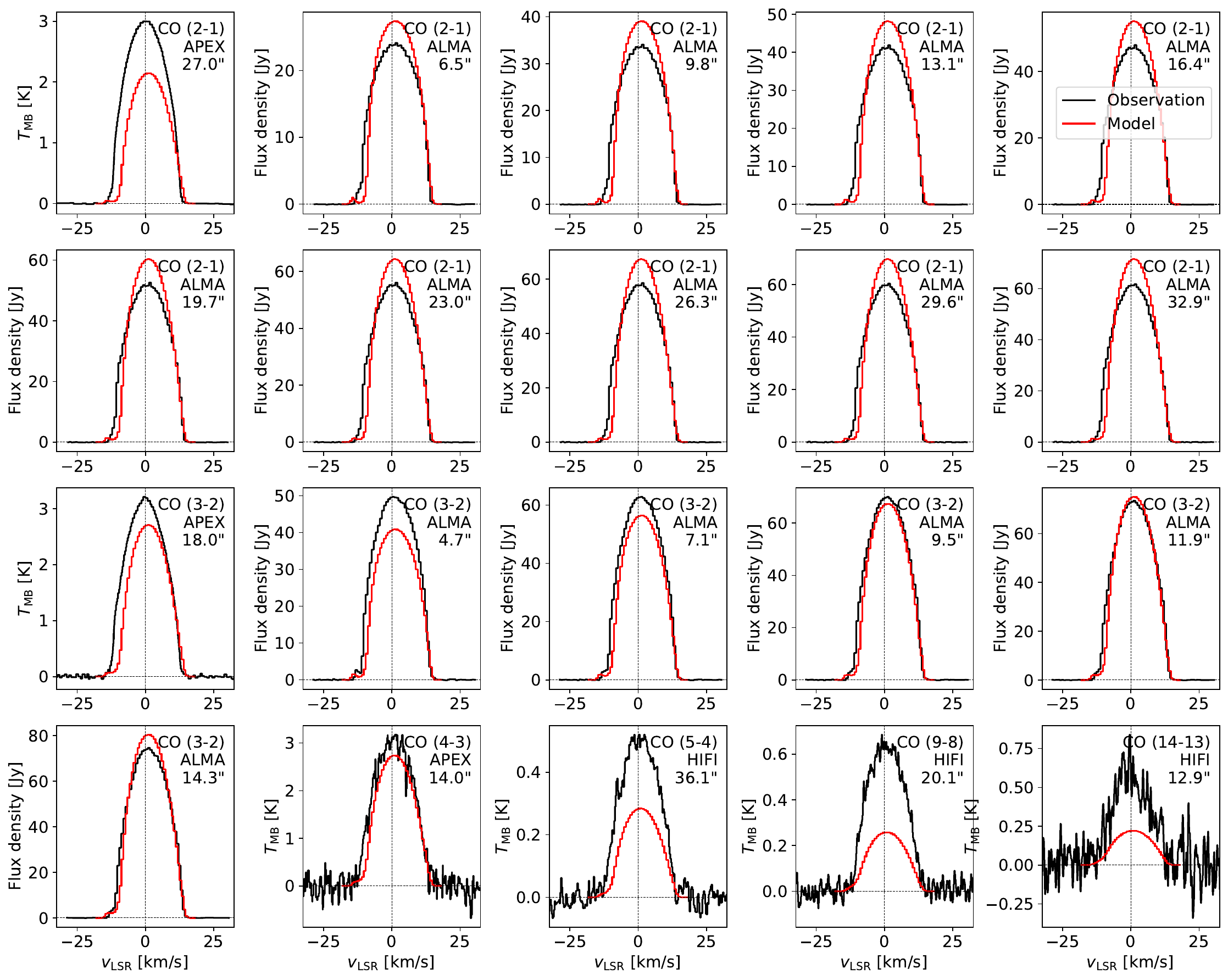}
    \caption{Same as Fig.~\ref{fig:CO_line_profiles_15194}, but for IRAS 07454$-$7112.}
    \label{fig:CO_line_profiles_07454}
\end{figure}

\begin{figure}[h]
    \centering
    \includegraphics[width=0.67\linewidth]{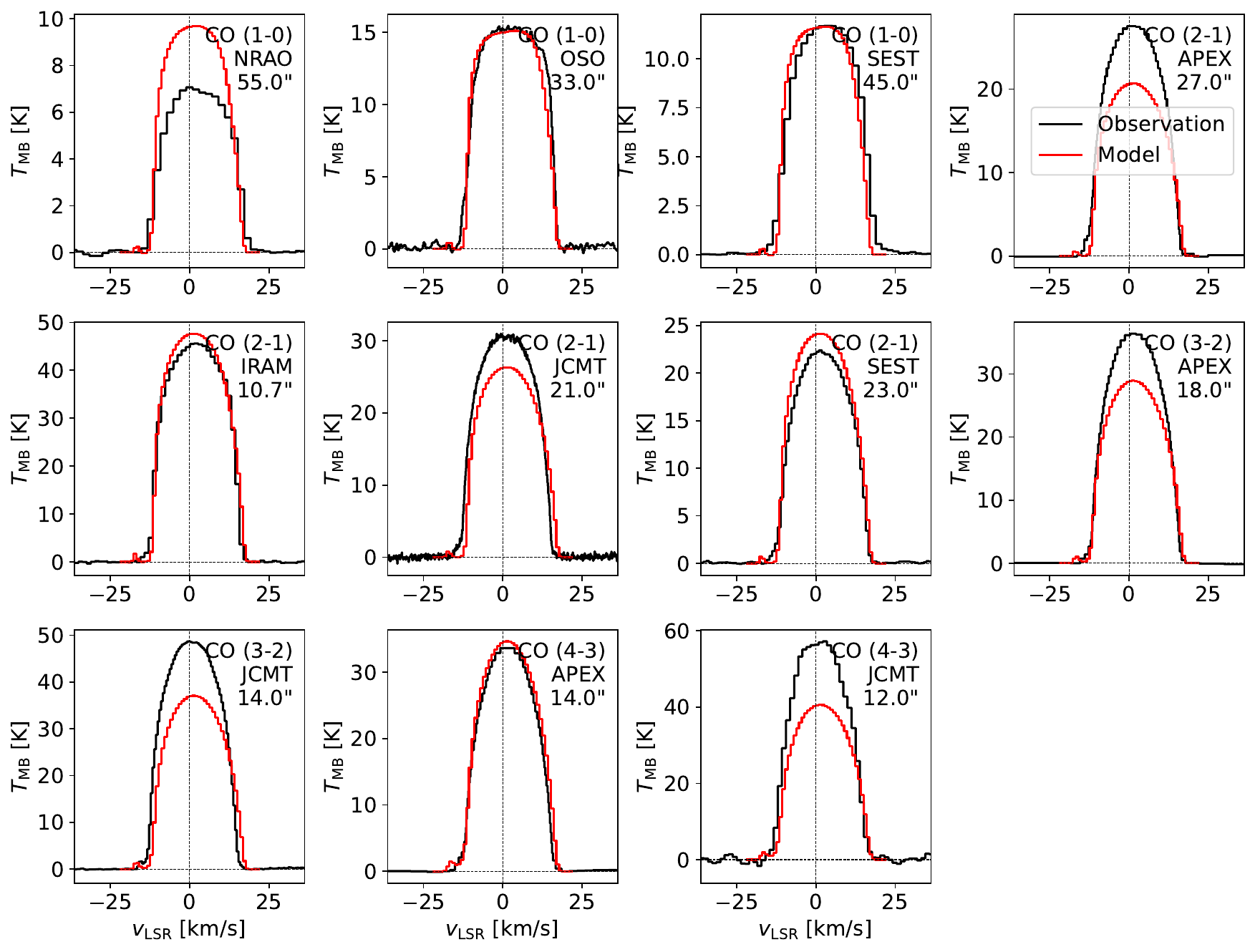}
    \caption{Same as Fig.~\ref{fig:CO_line_profiles_15194}, but for IRC~+10\,216.}
    \label{fig:CO_line_profiles_10216}
\end{figure}

\begin{figure}[h]
    \centering
    \includegraphics[width=\linewidth]{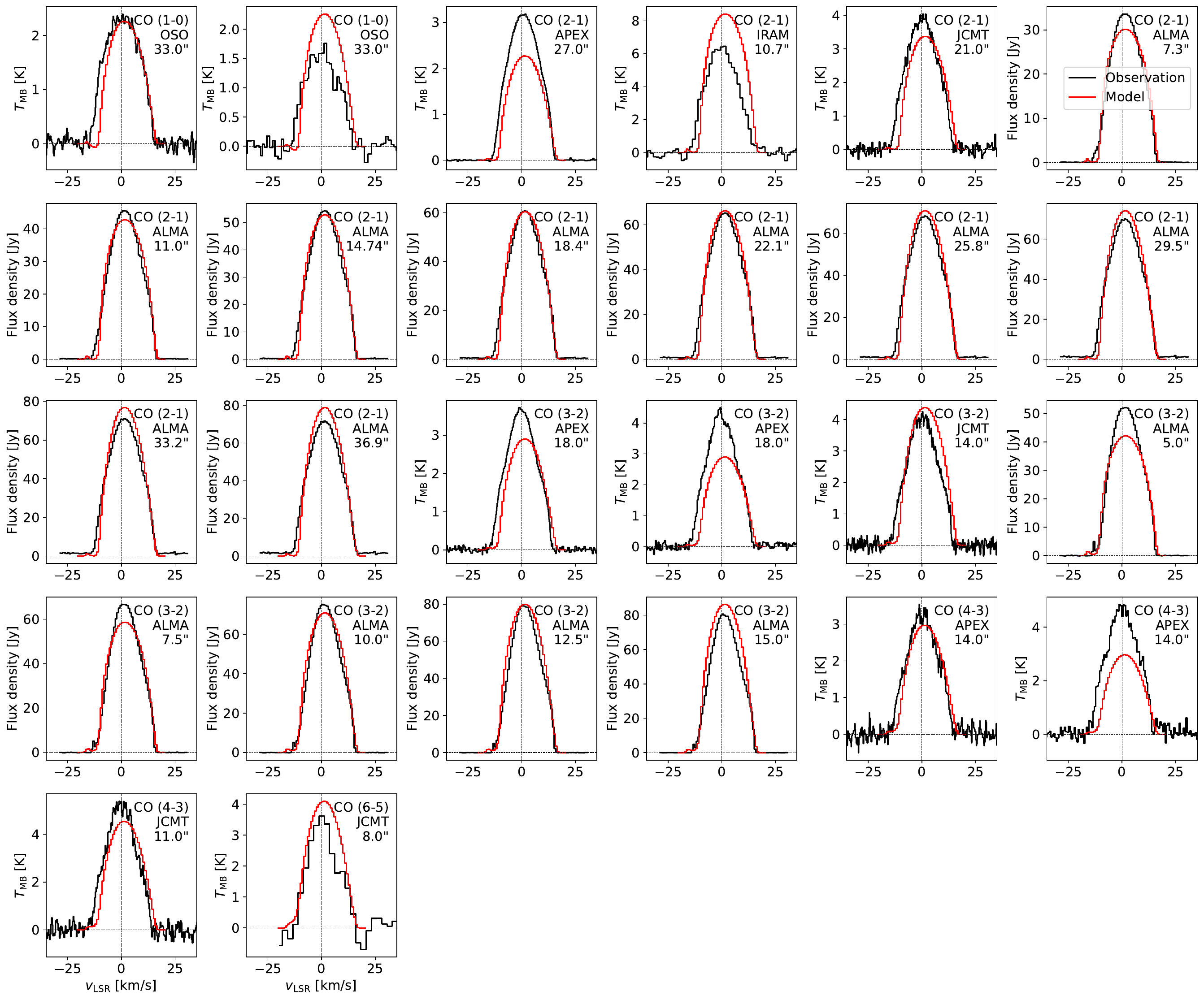}
    \caption{Same as Fig.~\ref{fig:CO_line_profiles_15194}, but for AFGL 3068.}
    \label{fig:CO_line_profiles_3068}
\end{figure}
\FloatBarrier

\section{CS line fits and $\chi^2$ maps}
\label{app:appendix_C}
\begin{figure}[htbp]
    \centering
    \begin{subfigure}[t]{0.48\linewidth}
        \centering
        \includegraphics[width=\linewidth]{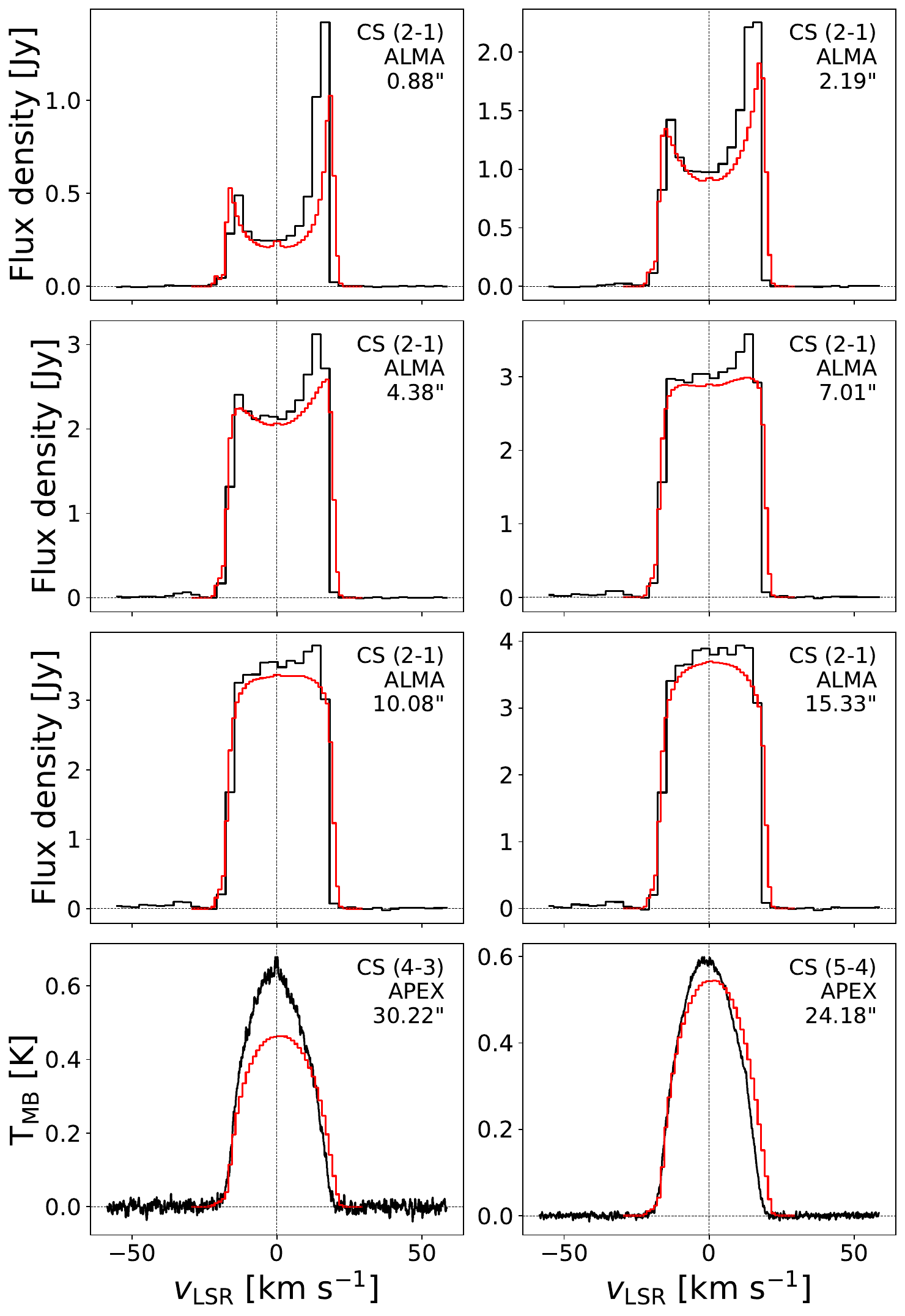}
        \caption{IRAS 15082$-$4808}
        \label{fig:15082_CS_line_fits}
    \end{subfigure}%
    \hfill
    \begin{subfigure}[t]{0.48\linewidth}
        \centering
        \includegraphics[width=\linewidth]{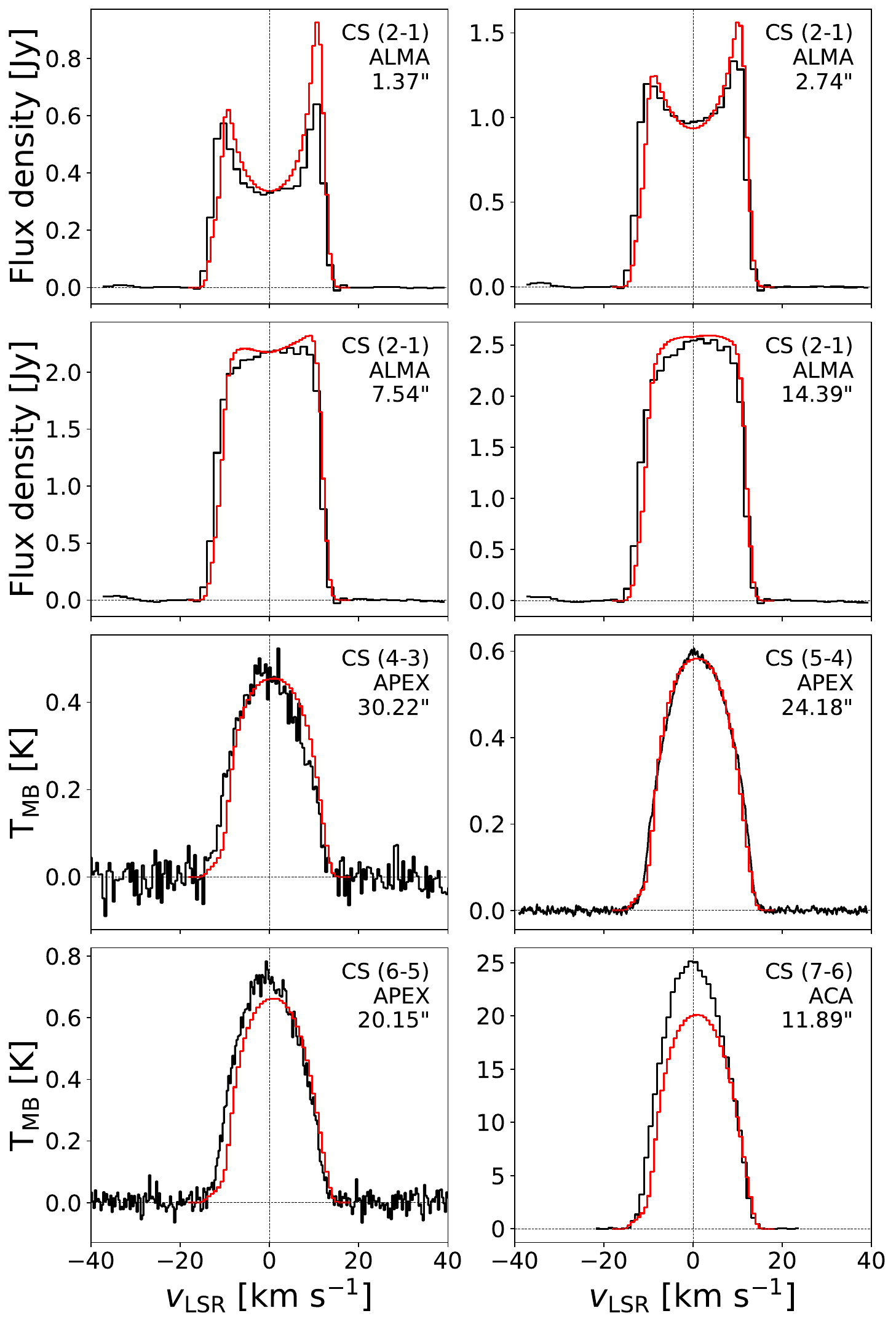}
        \caption{IRAS 07454$-$7112}
        \label{fig:07454_CS_line_fits}
    \end{subfigure}
    \caption{Observed (black) and modelled (red) CS line profiles for  (a) IRAS 15082$-$4808 and (b) IRAS 07454$-$7112. The transition quantum numbers, telescope used for the observation, and the corresponding beam size in arcseconds are listed at the top right corner of each panel. The line fluxes are in units of Jansky for the ALMA spectra, whereas they are given in the main-beam temperature ($T_\mathrm{MB}$ [K]) scale for all SD spectra shown.}
    \label{fig:15082_and_07454_CS_line_fits}
\end{figure}

\begin{figure}[htbp]
    \centering
    \includegraphics[width=0.55\linewidth]{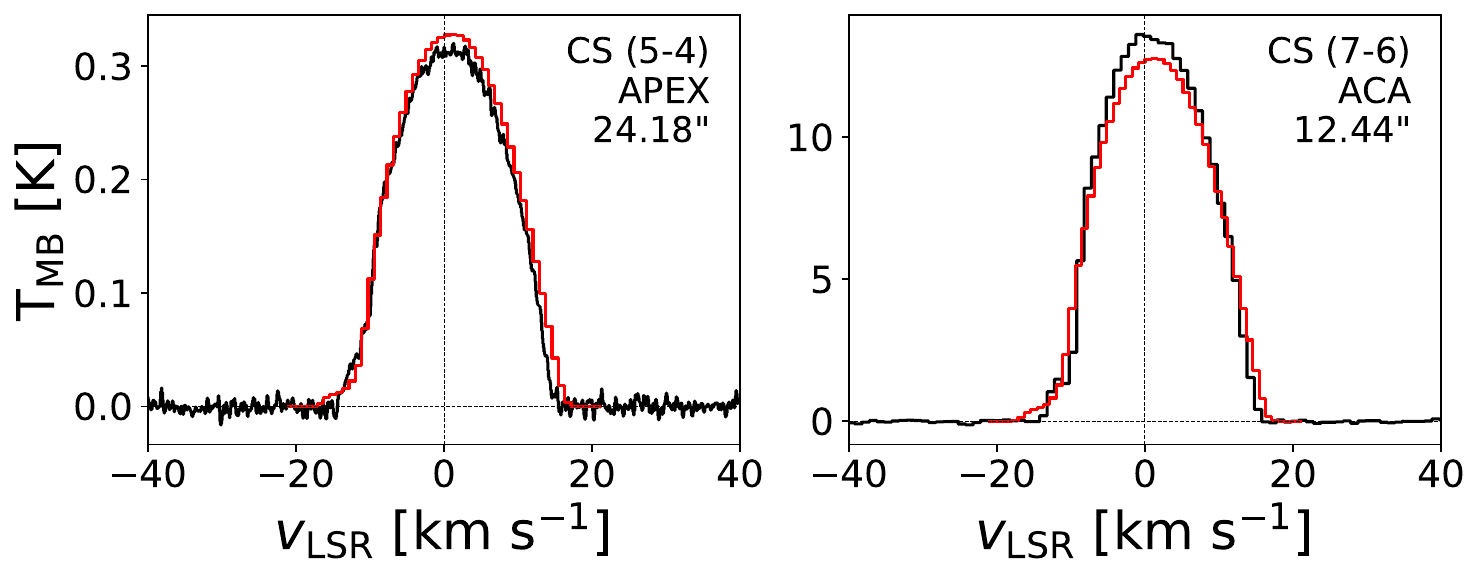}
    \caption{Same as Fig.~\ref{fig:15082_and_07454_CS_line_fits}, but for AFGL 3068.}
    \label{fig:3068_CS_line_fits}
\end{figure}

\begin{figure*}
    \centering
    \includegraphics[width=\linewidth]{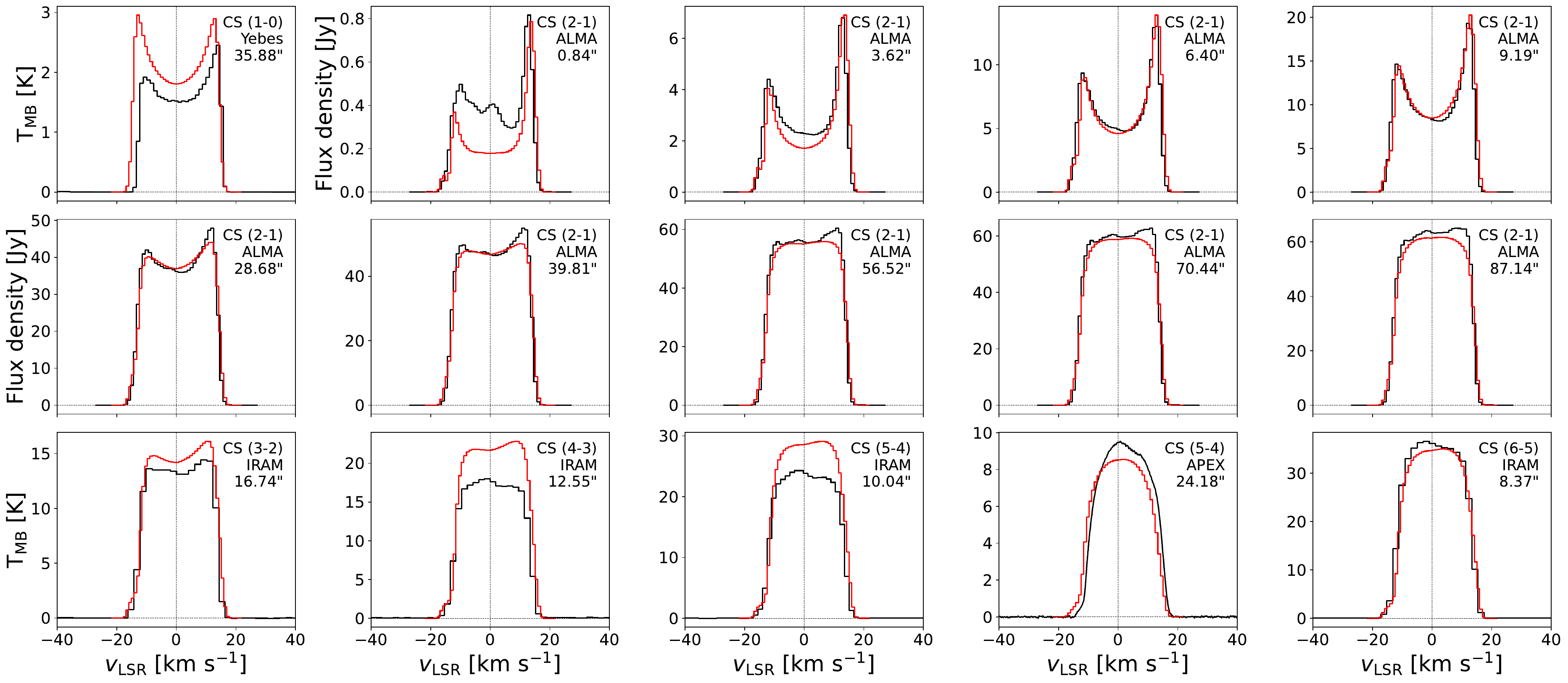}
    \caption{Same as Fig.~\ref{fig:15082_and_07454_CS_line_fits}, but for IRC~+10\,216.}
    \label{fig:10216_CS_line_fits}
\end{figure*}

\begin{figure}[htbp]
    \centering
    \begin{subfigure}[t]{0.32\textwidth}
        \centering
        \includegraphics[width=\textwidth]{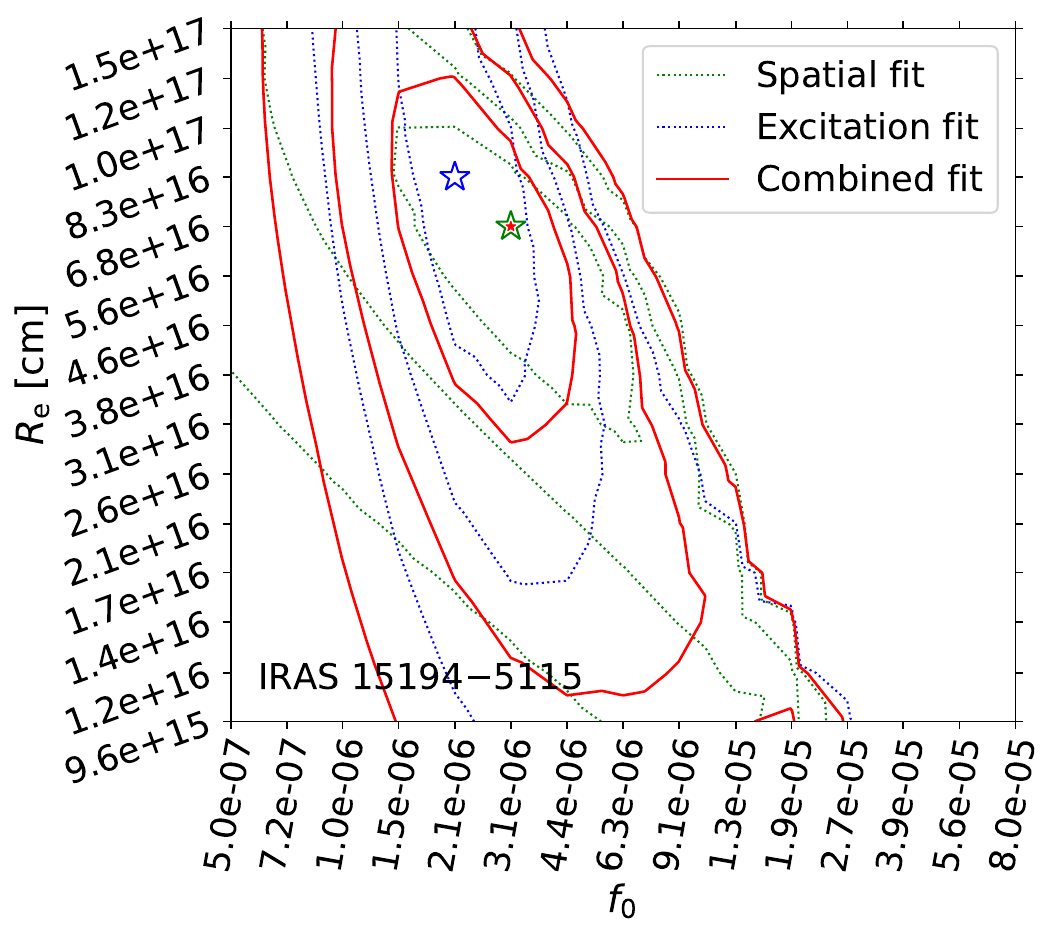}
        \caption{IRAS 15194$-$5115}
        \label{subfig:15194_chi_sq_contours_comparison}
    \end{subfigure}
    \hfill
    \begin{subfigure}[t]{0.32\textwidth}
        \centering
        \includegraphics[width=\textwidth]{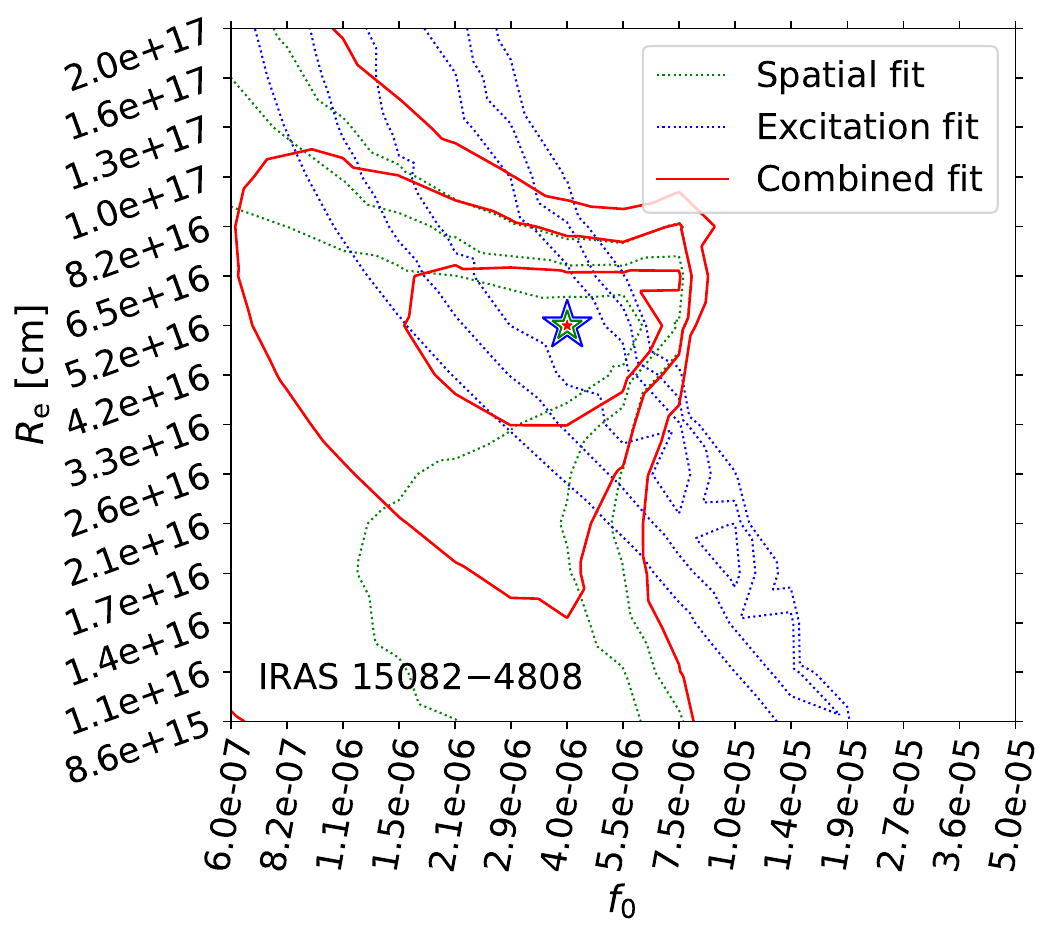}
        \caption{IRAS 15082$-$4808}
        \label{subfig:15082_chi_sq_contours_comparison}
    \end{subfigure}
    \hfill
    \begin{subfigure}[t]{0.32\textwidth}
        \centering
        \includegraphics[width=\textwidth]{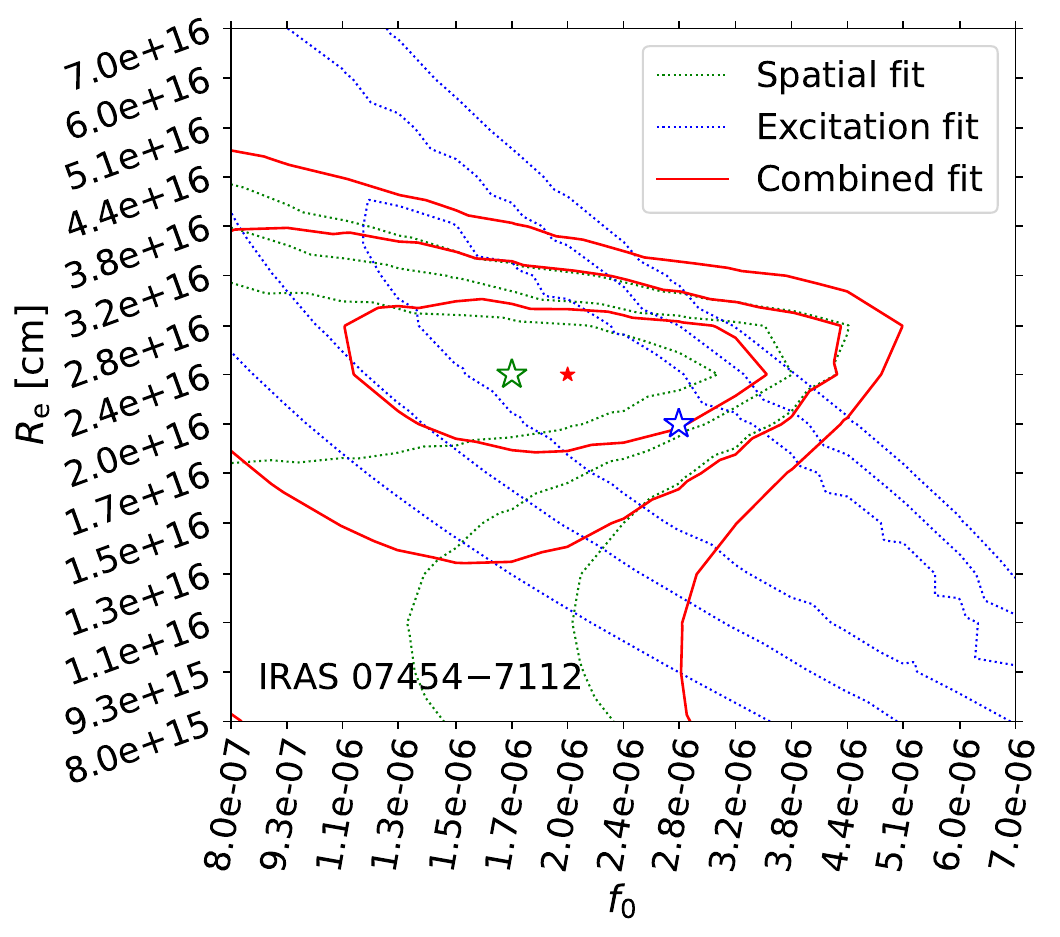}
        \caption{IRAS 07454$-$7112}
        \label{subfig:07454_chi_sq_contours_comparison}
    \end{subfigure}
    \caption{Same as Fig.~\ref{fig:10216_chi_sq_maps_comparison}, but for (a) IRAS 15194$-$5115, (b) IRAS 15082$-$4808, and (c) IRAS 07454$-$7112. The figures show $\chi^2$ values for our grid of CS models for the above sources. The contours denote 1$\sigma$, 2$\sigma$, and 3$\sigma$ ranges, when constrained using only the spatial information from the CS $J = 2 - 1$ line (green) vs just the excitation information from the different $J$ lines (blue). Also shown are the contours when using both the spatial and excitation information simultaneously (red) to constrain the RT models. The green, blue and red stars mark the best-fit models as estimated from the above three cases, respectively.}
    \label{fig:chi_sq_contours_comparison_our_stars}
\end{figure}

\begin{figure}[h]
    \centering
    \begin{subfigure}{0.45\textwidth}
        \centering
        \includegraphics[width=\textwidth]{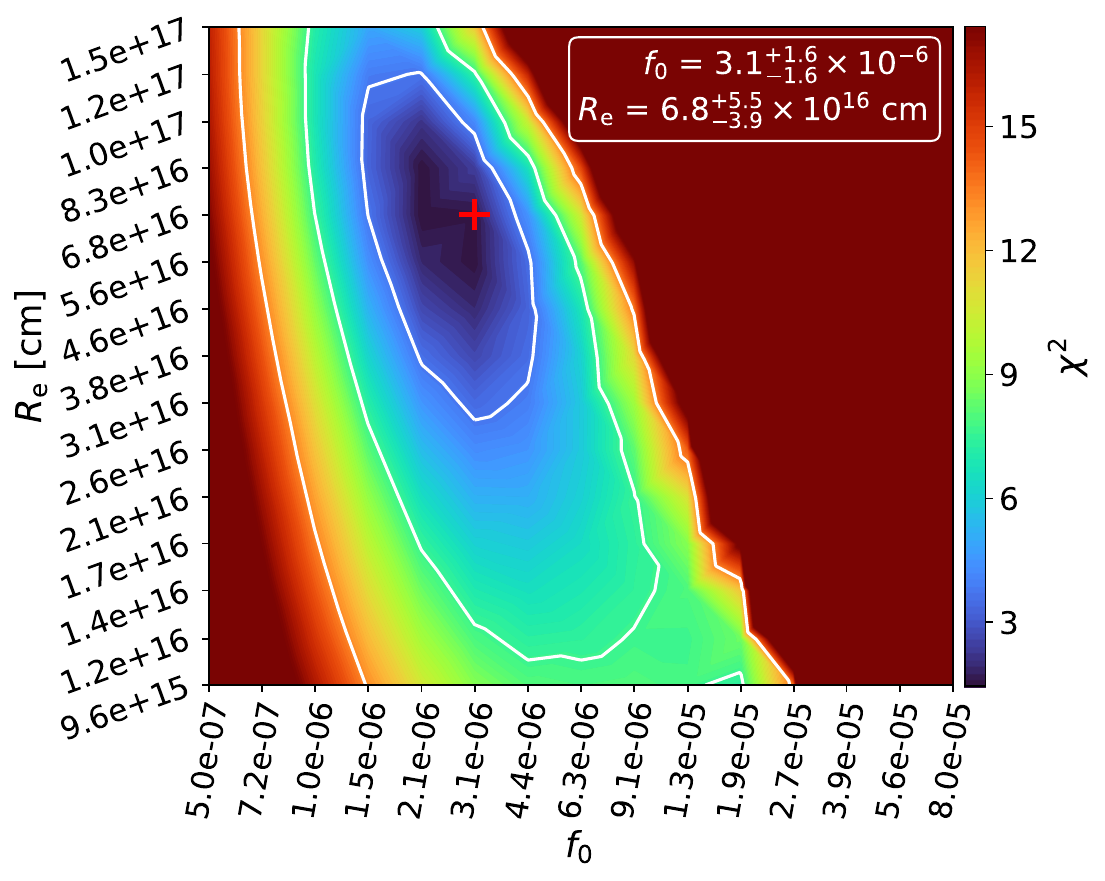}
        \caption{IRAS 15194$-$5115}
        \label{subfig:15194_chi_sq_map}
    \end{subfigure}
    \begin{subfigure}{0.45\textwidth}
        \centering
        \includegraphics[width=\textwidth]{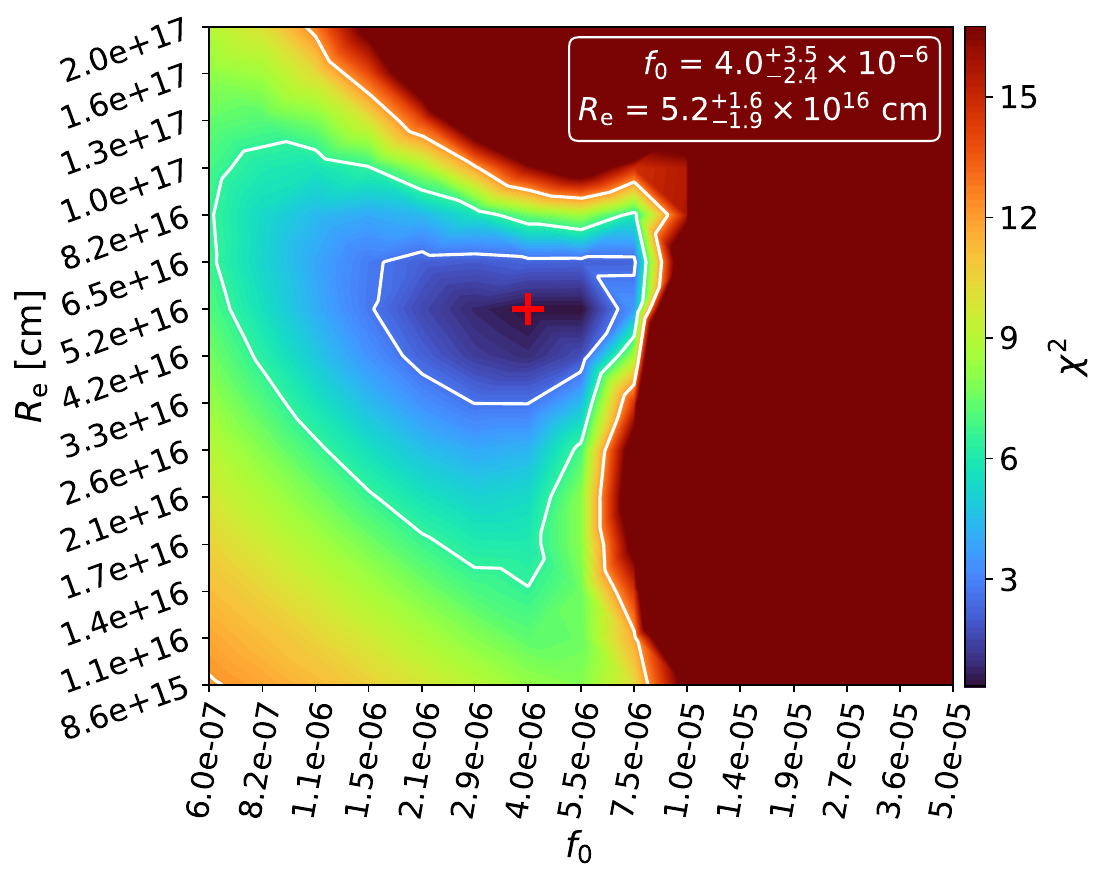}
        \caption{IRAS 15082$-$4808}
        \label{subfig:15082_chi_sq_map}
    \end{subfigure}

    \vspace{0.25cm}
    
    \begin{subfigure}{0.45\textwidth}
        \centering
        \includegraphics[width=\textwidth]{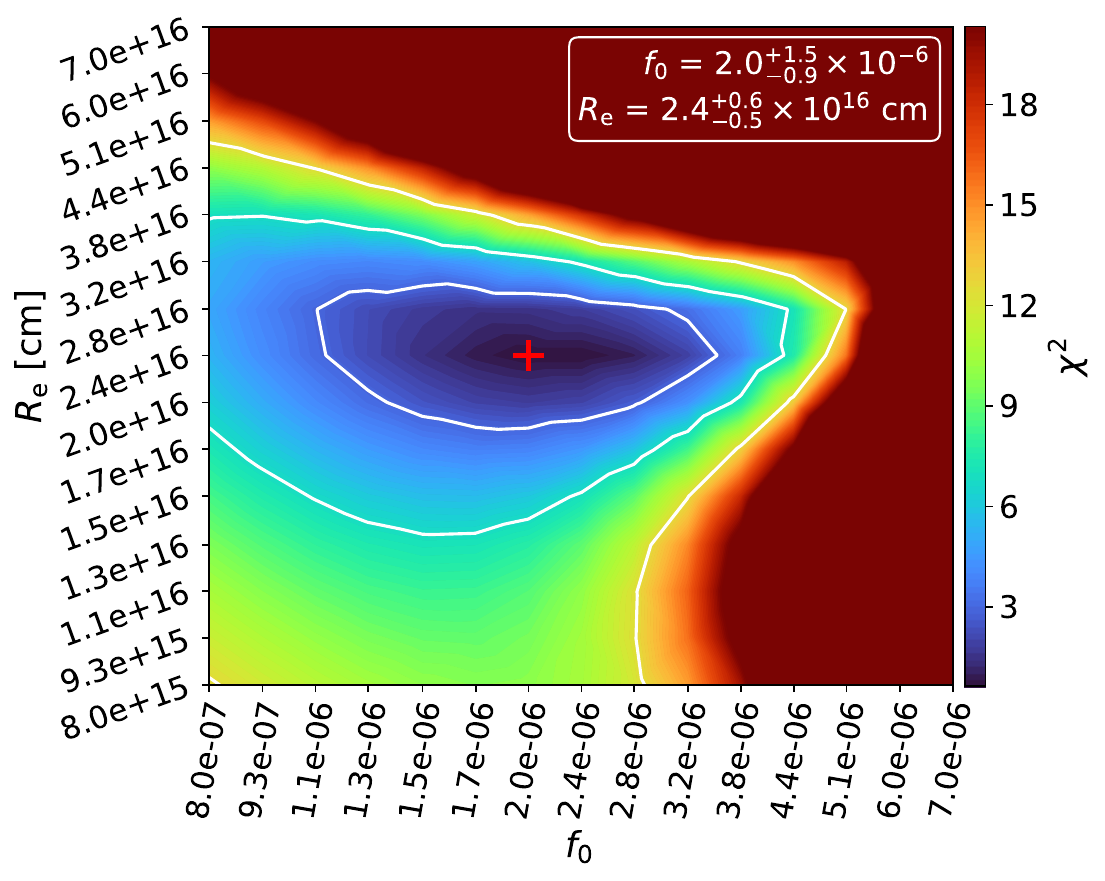}
        \caption{IRAS 07454$-$7112}
        \label{subfig:07454_chi_sq_map}
    \end{subfigure}
    \begin{subfigure}{0.45\textwidth}
        \centering
        \includegraphics[width=\textwidth]{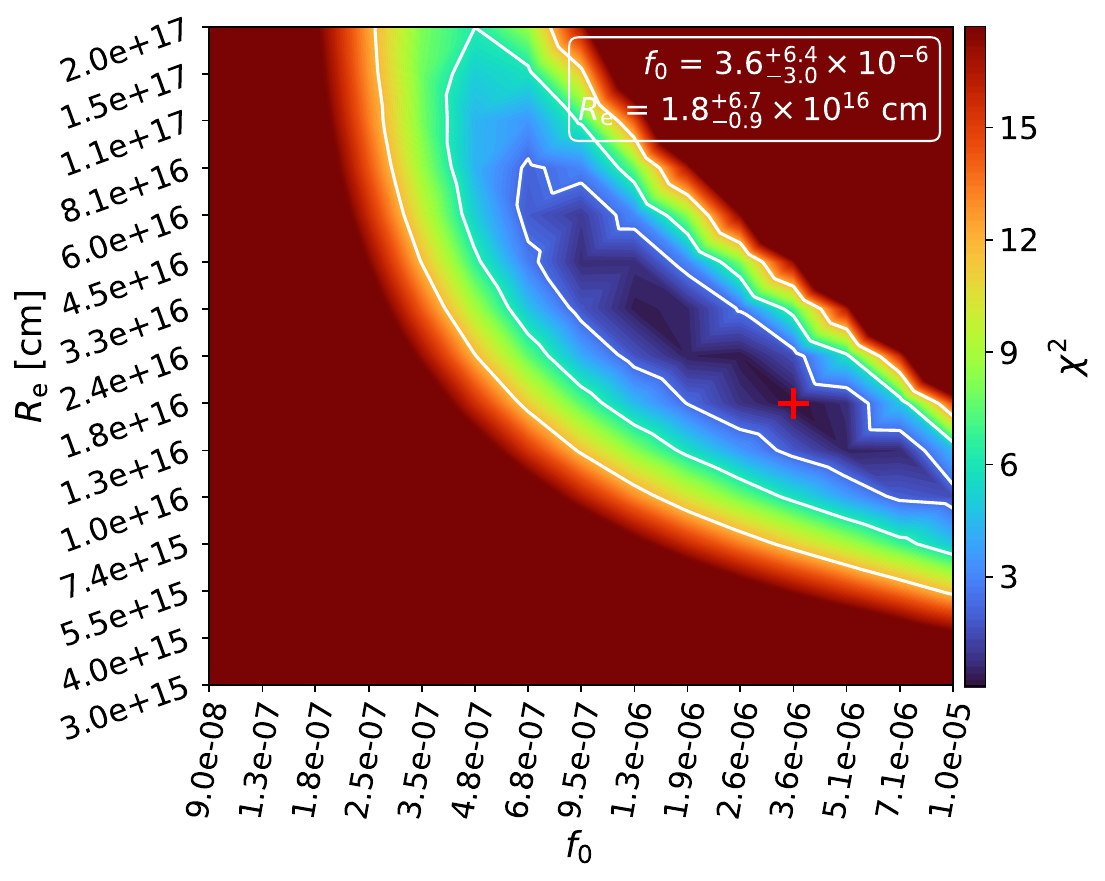}
        \caption{AFGL 3068}
        \label{subfig:3068_chi_sq_map}
    \end{subfigure}

    \vspace{0.25cm}

    \begin{subfigure}{0.45\textwidth}
        \hspace*{1.1cm}
        \centering
        \includegraphics[width=\textwidth]{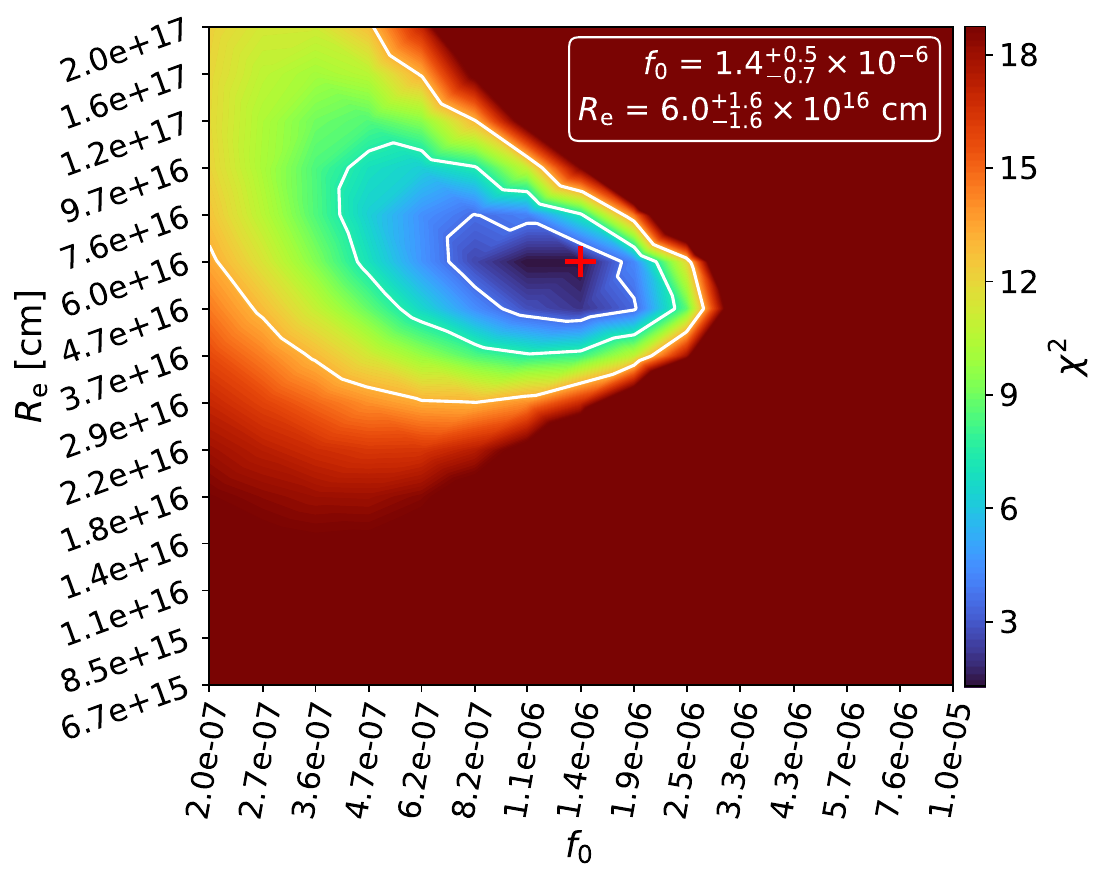}
        \caption{IRC~+10\,216}
        \label{subfig:10216_chi_sq_map}
    \end{subfigure}
    \hfill
    \hspace*{0pt}
    \caption{$\chi^2$ maps for the CS RT models. The free parameters $f_0$ and $R_\mathrm{e}$ are on the x$-$ and y$-$axes, respectively. The red $+$ sign denotes the best-fit model, and the white contours correspond to the 1$\sigma$, 2$\sigma$, and 3$\sigma$ ranges.}
    \label{fig:chi_sq_maps}
\end{figure}

This appendix shows the observed and modelled best-fit CS line profiles for the stars in the sample other than IRAS 15194$-$5115 (Figs.~\ref{fig:15082_and_07454_CS_line_fits} - \ref{fig:10216_CS_line_fits}). The best-fit model for IRAS 15194$-$5115 is shown in Fig.~\ref{fig:15194_CS_line_fits}. The $\chi^2$ maps for the grid of models we ran for each of the five sources, with the CS peak abundance ($f_0$) and the e-folding radii ($R_\mathrm{e}$) of the CS abundance profile as free parameters are also shown (Figs.~\ref{fig:chi_sq_contours_comparison_our_stars}, \ref{fig:chi_sq_maps}). The modelling is described in detail in Sect.~\ref{subsec:CS_Modelling} and the results obtained are presented in Sect.~\ref{subsec:CS_Modelling_Results}.

\FloatBarrier
\section{$^{13}$CS and C$^{34}$S line profile fits and intensity profiles}
\label{app:appendix_D}
Tables~\ref{tab:13CS_line_intensities} and \ref{tab:C34S_line_intensities} list the observed integrated intensities of the $^{13}$CS and C$^{34}$S lines modelled in this work respectively. The line profile fits for $^{13}$CS and C$^{34}$S are also shown in this appendix. The lines of these molecules are modelled by only varying the peak abundance, keeping the e-folding radius fixed to the value obtained from CS modelling for each source. See Sect.~\ref{subsubsec:CS_modelling_procedure} for details.

\begin{table*}[ht]
   \caption{$^{13}$CS lines used in this work.}
   \label{tab:13CS_line_intensities}
   \centering
      \begin{adjustbox}{width=14.85cm}
      \begin{tabular}{c l c c c c c c c}
      \hline\hline & \\[-2ex]
      \makecell{Transition} & \makecell{Rest frequency\\\ [GHz]} & \makecell{E$_{up}$\\\ [K]} & \makecell{Telescope} & \multicolumn{5}{c}{\makecell{Integrated intensity [Jy km s$^{-1}$ for ALMA; K km s$^{-1}$ for others]}} \\
      \cline{5-9} & \\[-2ex]  
      & & & & 15194$-$5115 & 15082$-$4808 & 07454$-$7112 & AFGL 3068 & IRC~+10\,216 \\
      \hline & \\[-2ex]
      $2 - 1$ & 92.494308 & 6.66 & ALMA & 44.4 & 2.8 & 2.2 & $-$ & $-$\\
      $4 - 3$ & 184.981772 & 22.19 & APEX & 4.7 & $-$ & $-$ & $-$ & $-$\\
      $5 - 4$ & 231.2206852 & 33.29 & APEX & 7.2 & 0.6 & 0.7 & 0.4 & 9.3\\
      $6 - 5$ & 277.455405 & 46.61 & APEX & 7.7 & $-$ & $-$ & $-$ & $-$\\
      $11 - 10$ & 508.5347391 & 146.46 & HIFI & 0.5 & $-$ & $-$ & $-$ & $-$\\
      $12 - 11$ & 554.7257657 & 173.08 & HIFI & 0.5 & $-$ & $-$ & $-$ & $-$\\
      $13 - 12$ & 600.90648 & 201.92 & HIFI & 1.0 & $-$ & $-$ & $-$ & $-$\\
      $14 - 13$ & 647.07615 & 232.97 & HIFI & 0.7 & $-$ & $-$ & $-$ & $-$\\
      $15 - 14$ & 693.2337468 & 266.24 & HIFI & 0.9 & $-$ & $-$ & $-$ & $-$\\
      $16 - 15$ & 739.37852 & 301.73 & HIFI & 0.5 & $-$ & $-$ & $-$ & $-$\\
      $17 - 16$ & 785.5095893, & 339.43 & HIFI & 0.5 & $-$ & $-$ & $-$ & $-$\\
      $19 - 18$ & 877.7271914 & 421.46 & HIFI & 0.7 & $-$ & $-$ & $-$ & $-$\\
      $20 - 19$ & 923.8120123 & 465.80 & HIFI & 0.5 & $-$ & $-$ & $-$ & $-$\\
      \hline
      \end{tabular}
      \end{adjustbox}
      \tablefoot{The ALMA and APEX lines reported are from \citetalias{Unnikrishnan_et_al_2024}, unless otherwise specified. All HIFI lines listed are from our \textit{Herschel}/HIFI spectral survey of IRAS 15194$-$5115 (see Sect.~\ref{subsubsec:HIFI_spectral_survey}). For the ALMA lines, the integrated intensity values reported are in units of Jy km s$^{-1}$. For all other SD observations (APEX, HIFI), the integrated intensities are given in units of K km s$^{-1}$, in the main beam ($T_\mathrm{MB}$) temperature scale. The relevant beam size ranges and main beam efficiencies for the different telescopes are described in Sect.~\ref{subsec:CS_Observations}, and the individual beam sizes at each transition frequency are indicated in Figs.~\ref{fig:13CS_line_profiles_15194} - \ref{fig:13CS_line_profiles_3068_and_10216}, alongside the corresponding line spectra.}
\end{table*}

\begin{table*}[ht]
   \caption{C$^{34}$S lines used in this work.}
   \label{tab:C34S_line_intensities}
   \centering
      \begin{adjustbox}{width=14.85cm}
      \begin{tabular}{c l c c c c c c c}
      \hline\hline & \\[-2ex]
      \makecell{Transition} & \makecell{Rest frequency\\\ [GHz]} & \makecell{E$_{up}$\\\ [K]} & \makecell{Telescope} & \multicolumn{5}{c}{\makecell{Integrated intensity [Jy km s$^{-1}$ for ALMA; K km s$^{-1}$ for others]}} \\
      \cline{5-9} & \\[-2ex]  
      & & & & 15194$-$5115 & 15082$-$4808 & 07454$-$7112 & AFGL 3068 & IRC~+10\,216 \\
      \hline & \\[-2ex]
      $2 - 1$ & 96.4129495 & 6.94 & ALMA & 14.1 & 6.0 & 3.7 & $-$ & $-$\\
      $4 - 3$ & 192.8184566 & 23.13 & APEX & 2.1 & 1.4 & 0.8 & 0.2 & 14.2\\
      $5 - 4$ & 241.0160892 & 34.70 & APEX & 1.9 & 1.2 & 0.9 & 0.5 & 17.2\\
      $6 - 5$ & 289.2090684 & 48.58 & APEX & 2.7 & $-$ & $-$ & $-$ & $-$\\
      $7 - 6$ & 337.396459 & 64.77 & APEX & 2.5 & $-$ & $-$ & $-$ & $-$\\
      $10 - 9$ & 481.9158105 & 127.22 & HIFI & 0.3 & $-$ & $-$ & $-$ & $-$\\
      $11 - 10$ & 530.0715537 & 152.66 & HIFI & 0.1 & $-$ & $-$ & $-$ & $-$\\
      $12 - 11$ & 578.217069 & 180.41 & HIFI & 0.2 & $-$ & $-$ & $-$ & $-$\\
      $14 - 13$ & 674.4736177 & 242.84 & HIFI & 0.1 & $-$ & $-$ & $-$ & $-$\\
      $15 - 14$ & 722.5828172 & 277.52 & HIFI & 0.2 & $-$ & $-$ & $-$ & $-$\\
      $16 - 15$ & 770.6780547 & 314.51 & HIFI & 0.2 & $-$ & $-$ & $-$ & $-$\\
      \hline
      \end{tabular}
      \end{adjustbox}
      \tablefoot{The ALMA and APEX lines reported are from \citetalias{Unnikrishnan_et_al_2024}, unless otherwise specified. All HIFI lines listed are from our \textit{Herschel}/HIFI spectral survey of IRAS 15194$-$5115 (see Sect.~\ref{subsubsec:HIFI_spectral_survey}). For the ALMA lines, the integrated intensity values reported are in units of Jy km s$^{-1}$. For all other SD observations (APEX, HIFI), the integrated intensities are given in units of K km s$^{-1}$, in the main beam ($T_\mathrm{MB}$) temperature scale. The relevant beam size ranges and main beam efficiencies for the different telescopes are described in Sect.~\ref{subsec:CS_Observations}, and the individual beam sizes at each transition frequency are indicated in Figs.~\ref{fig:C34S_line_profiles_15194} - \ref{fig:C34S_line_profiles_10216}, alongside the corresponding line spectra.}
\end{table*}

\begin{figure*}
    \centering
    \includegraphics[width=0.9\textwidth]{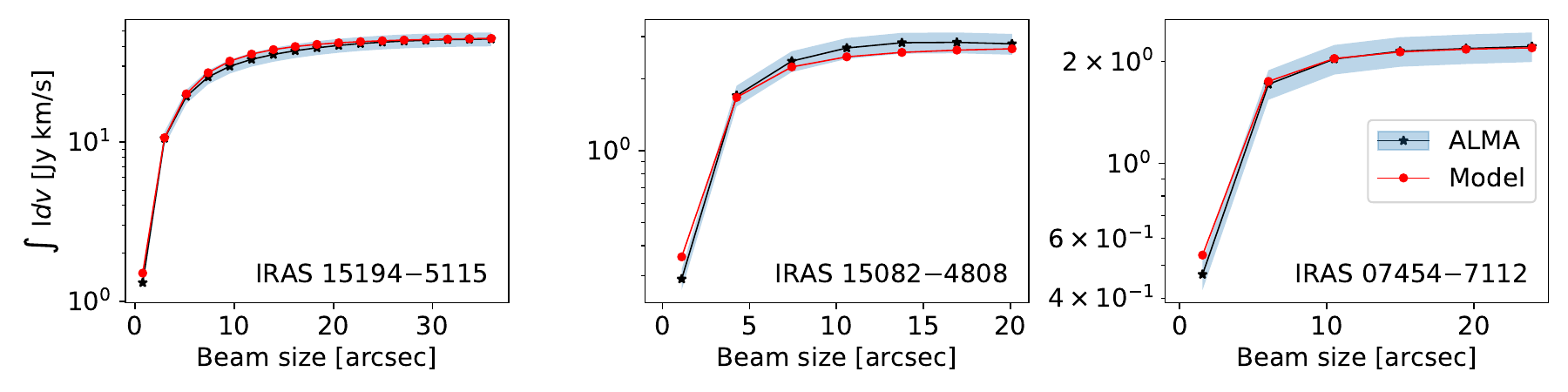}
    \caption{Comparison of modelled (red) and observed (black) integrated intensities of $^{13}$CS $J = 2 - 1$ spectra extracted from successively larger beams (see Sect.~\ref{subsubsec:CS_modelling_procedure}). The blue-shaded region represents the uncertainty in the observed line intensities.}
    \label{fig:13CS_ALMA_radial_intensity_plots}
\end{figure*}

\begin{figure}[h]
    \centering
    \includegraphics[width=0.785\linewidth]{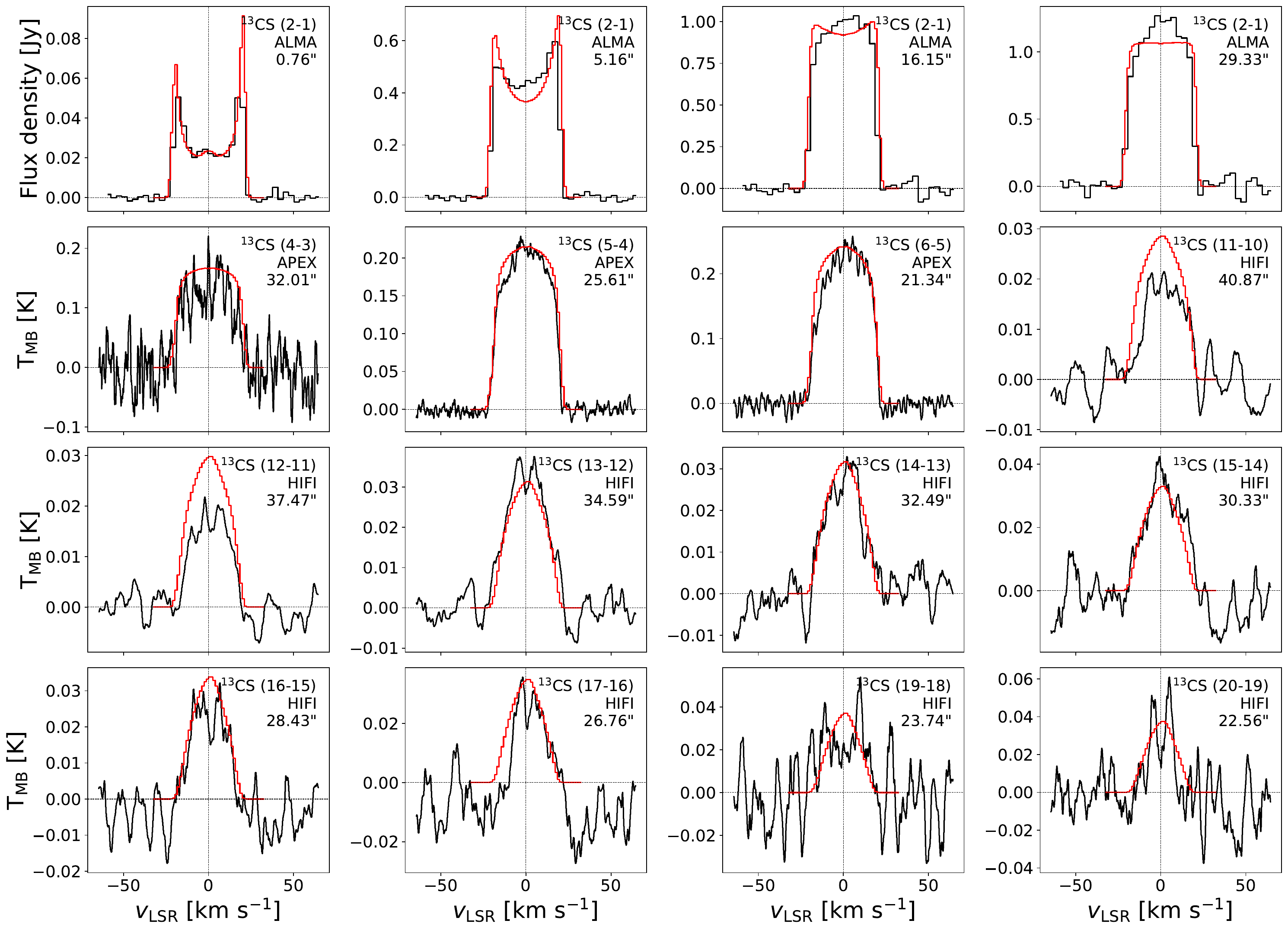}
    \caption{Observed (black, Sect.~\ref{subsec:CS_Observations}) and modelled (red, Sect.~\ref{subsec:CS_Modelling}) $^{13}$CS line profiles for IRAS 15194$-$5115. The transition quantum numbers, telescope used for the observation, and the corresponding beam size in arcseconds are listed at the top right corner of each panel. The line fluxes are in units of Jansky for the ALMA spectra, and in main-beam temperature ($T_\mathrm{MB}$ [K]) scale for all SD spectra.}
    \label{fig:13CS_line_profiles_15194}
\end{figure}

\begin{figure}[h]
    \centering
    \includegraphics[width=0.785\linewidth]{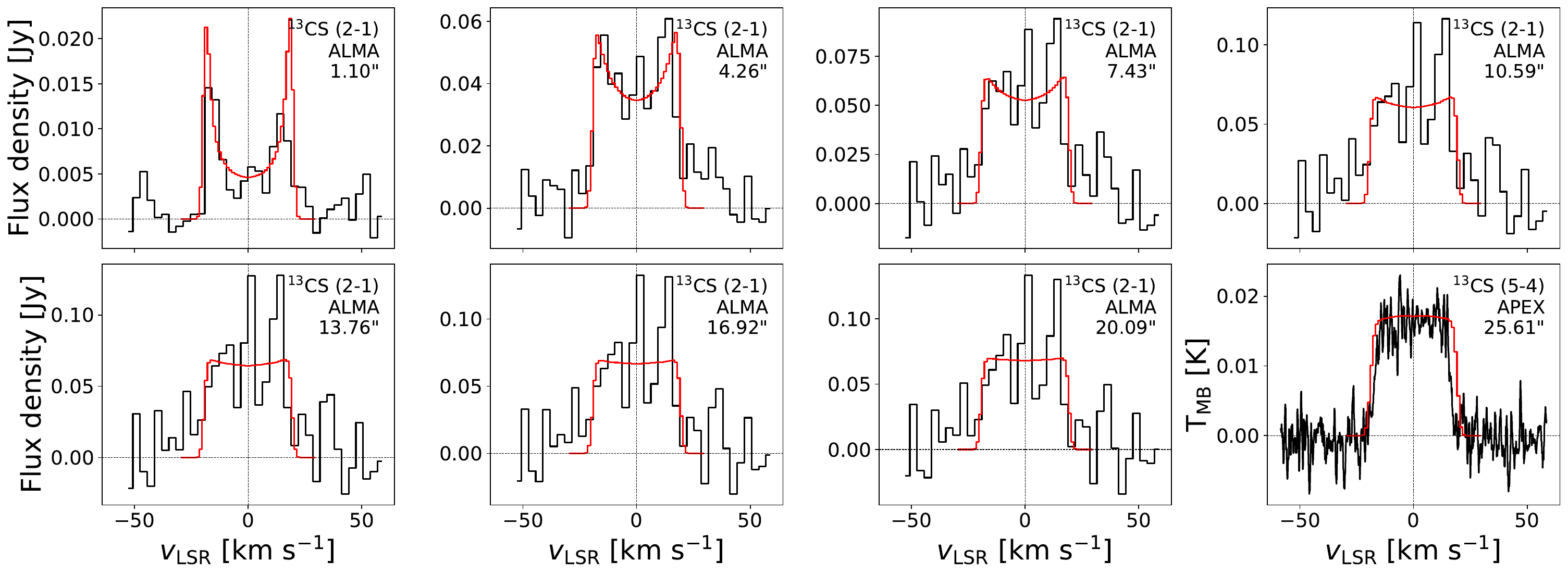}
    \caption{Same as Fig.~\ref{fig:13CS_line_profiles_15194}, but for IRAS 15082$-$4808.}
    \label{fig:13CS_line_profiles_15082}
\end{figure}

\begin{figure}[h]
    \centering
    \includegraphics[width=0.785\linewidth]{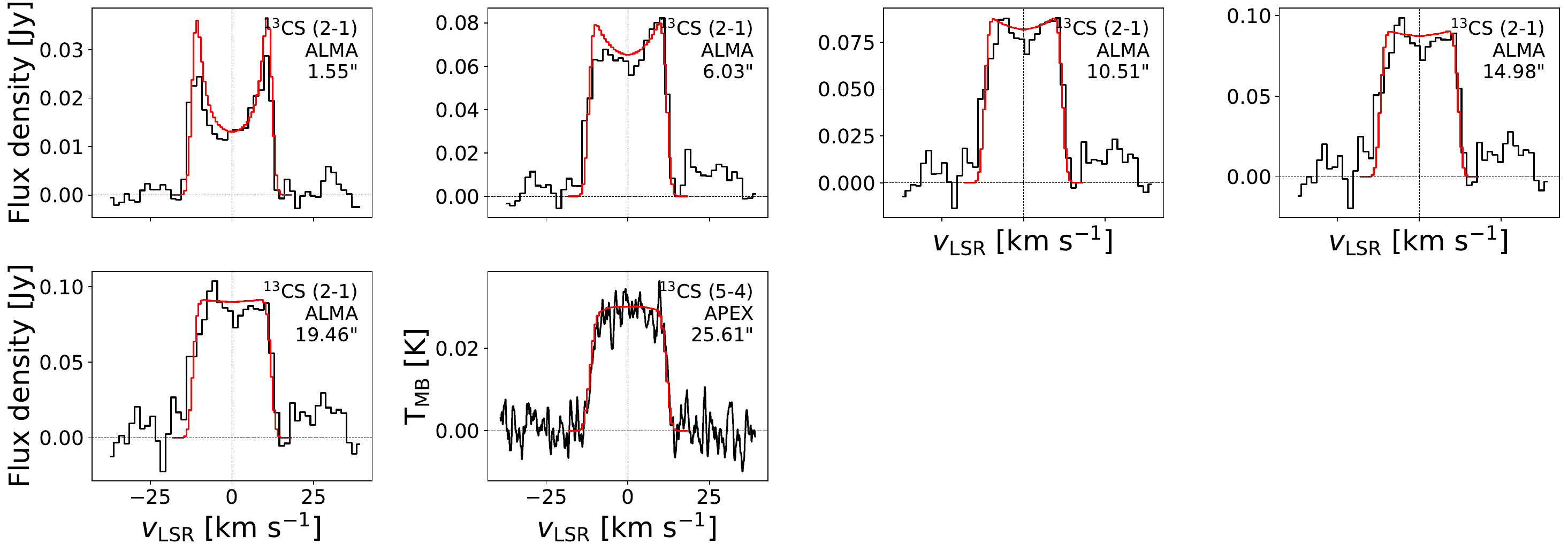}
    \caption{Same as Fig.~\ref{fig:13CS_line_profiles_15194}, but for IRAS 07454$-$7112.}
    \label{fig:13CS_line_profiles_07454}
\end{figure}

\begin{figure}[h]
    \centering
    \includegraphics[width=0.3\linewidth]{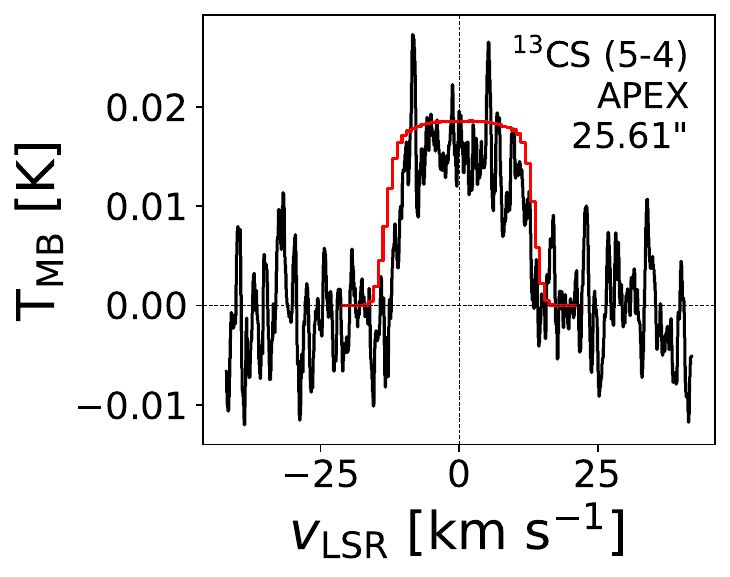}
    \includegraphics[width=0.3\linewidth]{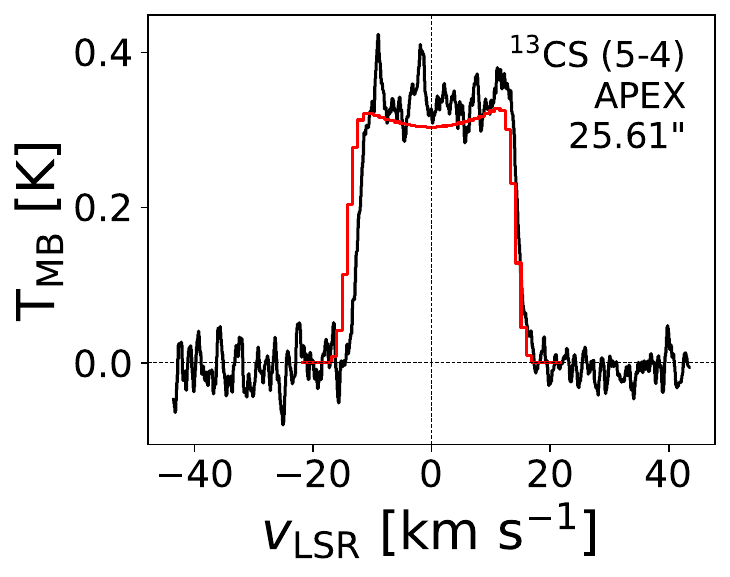}
    \caption{Same as Fig.~\ref{fig:13CS_line_profiles_15194}, but for AFGL 3068 (left) and IRC~+10\,216 (right).}
    \label{fig:13CS_line_profiles_3068_and_10216}
\end{figure}

\begin{figure*}[h]
    \centering
    \includegraphics[width=0.9\textwidth]{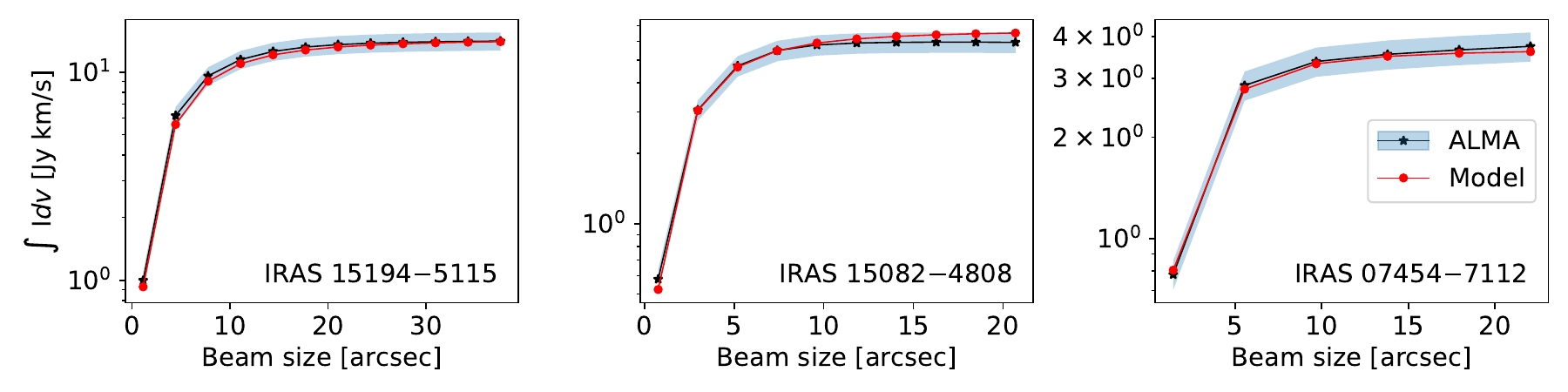}
    \caption{Same as Fig.~\ref{fig:13CS_ALMA_radial_intensity_plots}, but for C$^{34}$S.}
    \label{fig:C34S_ALMA_radial_intensity_plots}
\end{figure*}

\begin{figure}[h]
    \centering
    \includegraphics[width=0.8\linewidth]{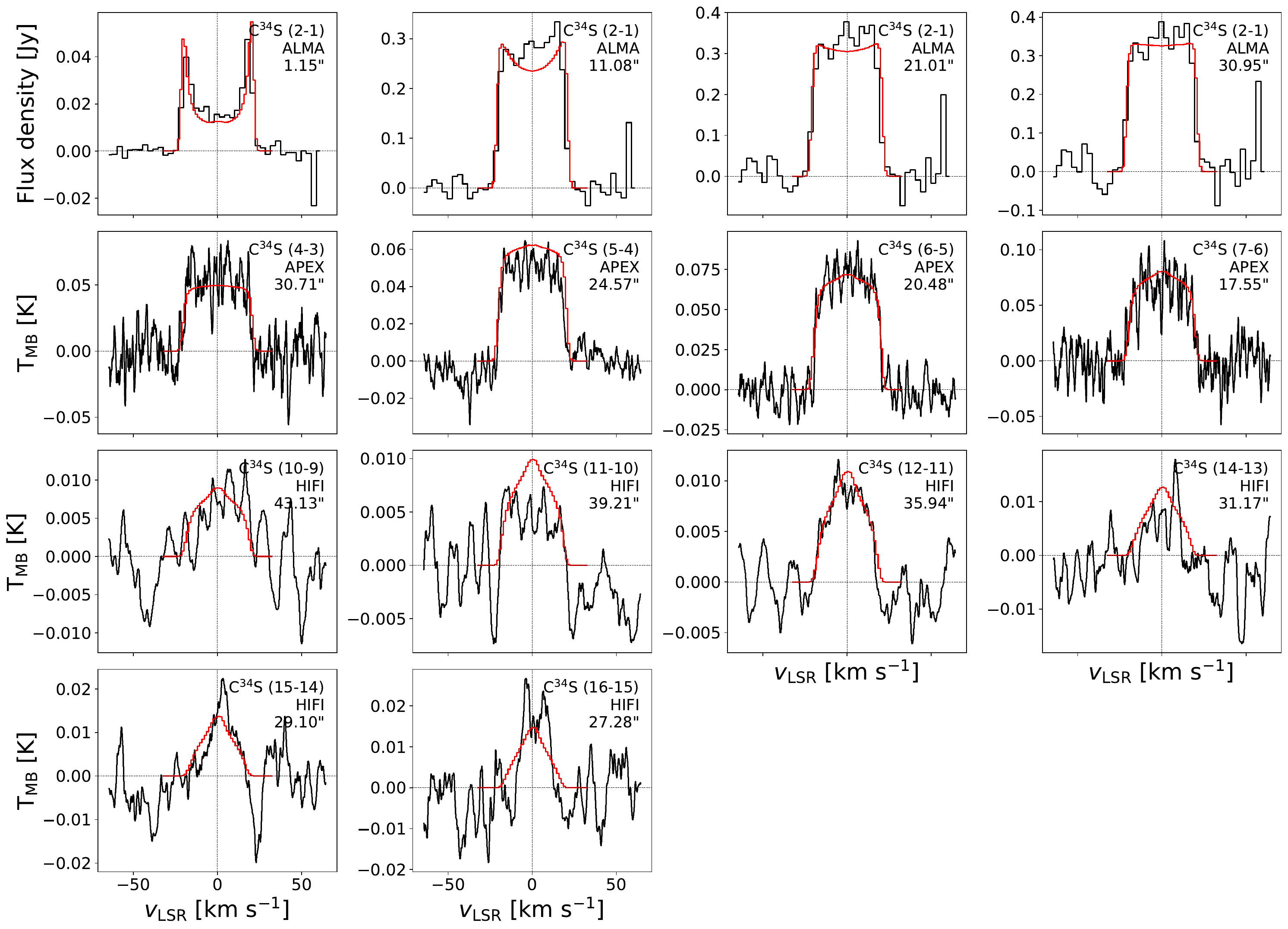}
    \caption{Observed (black, Sect.~\ref{subsec:CS_Observations}) and modelled (red, Sect.~\ref{subsec:CS_Modelling}) C$^{34}$S line profiles for IRAS 15194$-$5115. The transition quantum numbers, telescope used for the observation, and the corresponding beam size in arcseconds are listed at the top right corner of each panel. The line fluxes are in units of Jansky for the ALMA spectra, whereas they are given in the main-beam temperature ($T_\mathrm{MB}$ [K]) scale for all SD spectra shown.}
    \label{fig:C34S_line_profiles_15194}
\end{figure}

\begin{figure}[h]
    \centering
    \includegraphics[width=0.8\linewidth]{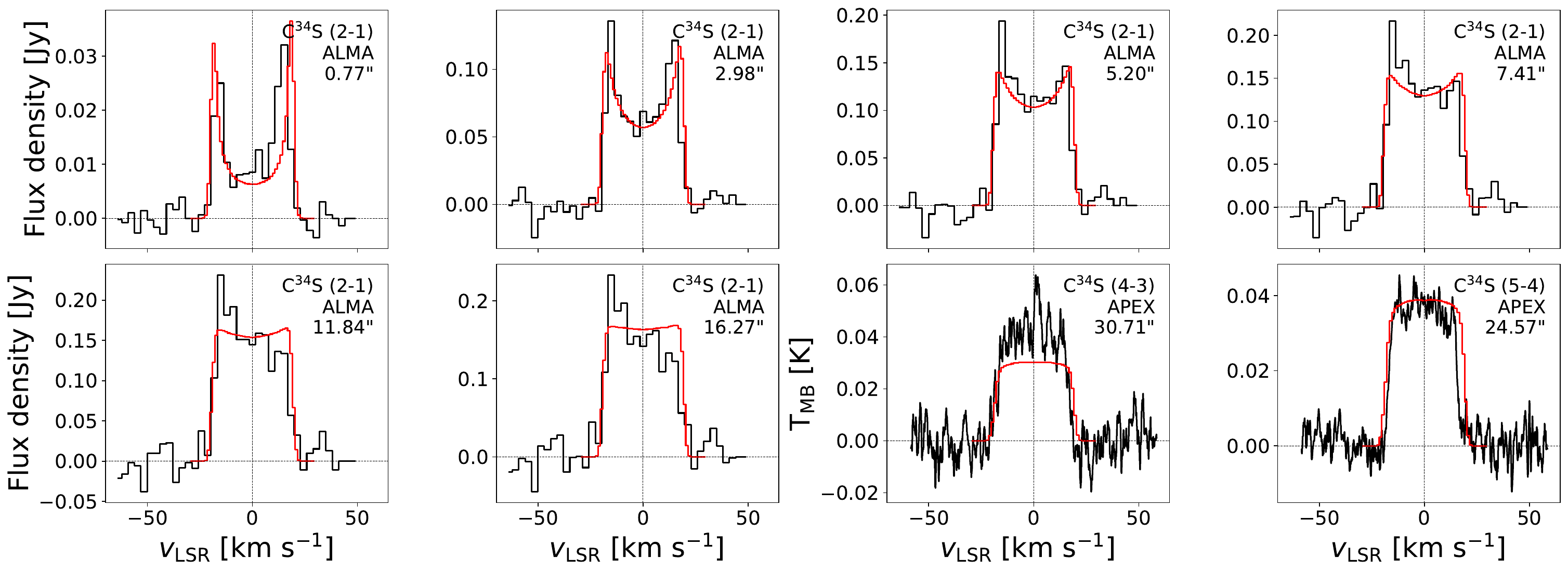}
    \caption{Same as Fig.~\ref{fig:C34S_line_profiles_15194}, but for IRAS 15082$-$4808.}
    \label{fig:C34S_line_profiles_15082}
\end{figure}

\begin{figure}[h]
    \centering
    \includegraphics[width=0.8\linewidth]{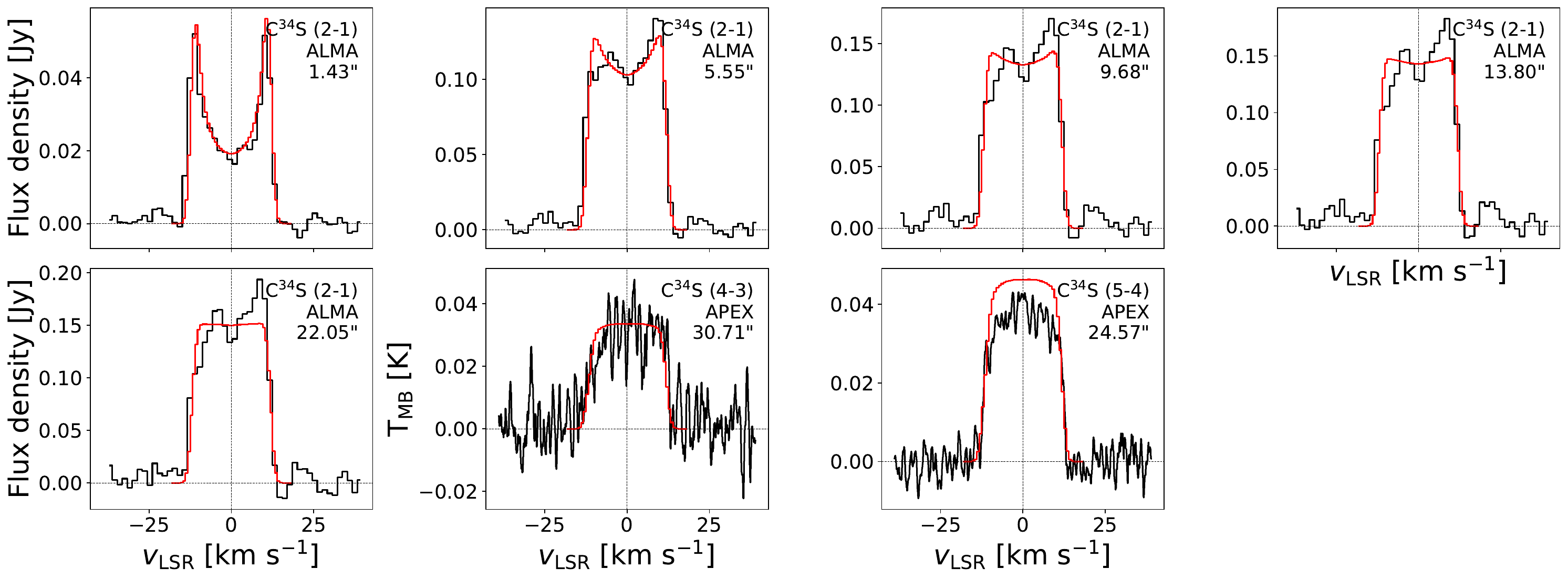}
    \caption{Same as Fig.~\ref{fig:C34S_line_profiles_15194}, but for IRAS 07454$-$7112.}
    \label{fig:C34S_line_profiles_07454}
\end{figure}

\begin{figure}[h]
    \centering
    \includegraphics[width=0.575\linewidth]{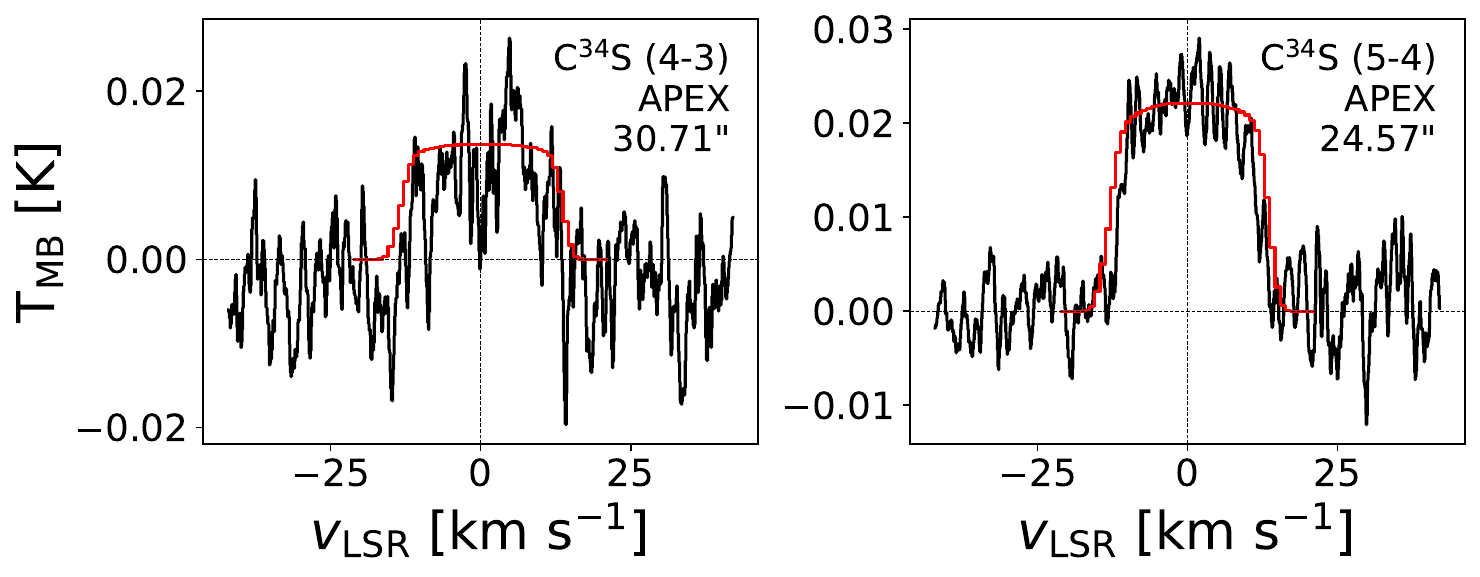}
    \caption{Same as Fig.~\ref{fig:C34S_line_profiles_15194}, but for AFGL 3068.}
    \label{fig:C34S_line_profiles_3068}
\end{figure}

\begin{figure}[h]
    \centering
    \includegraphics[width=0.575\linewidth]{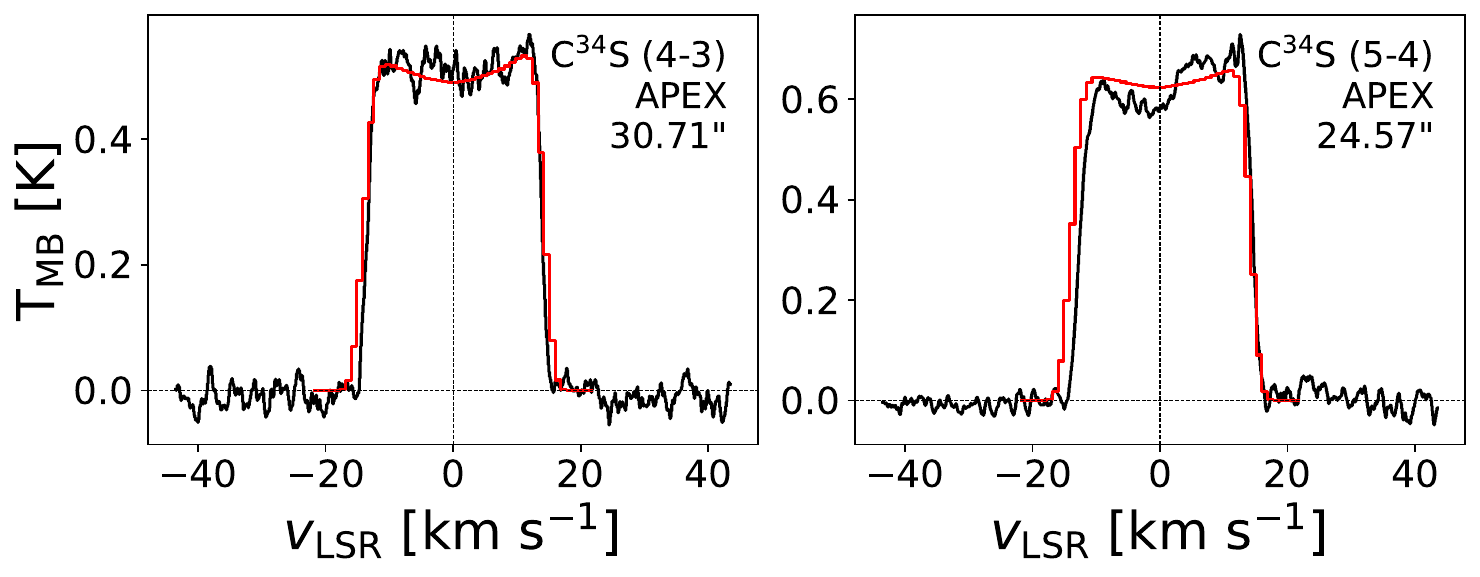}
    \caption{Same as Fig.~\ref{fig:C34S_line_profiles_15194}, but for IRC~+10\,216.}
    \label{fig:C34S_line_profiles_10216}
\end{figure}
\end{appendix}

\end{document}